\documentclass[a4paper,11pt]{article}
\pdfoutput=1 

\usepackage{jheppub} 

\usepackage[T1]{fontenc} 

\usepackage{placeins}
\usepackage[titletoc,title]{appendix}
\usepackage{hyperref}
\usepackage{calc}
\usepackage[utf8]{inputenc}
\usepackage{graphicx}
\usepackage{titlesec}
\usepackage{xspace}
\usepackage{etex}
\usepackage[export]{adjustbox}
\usepackage{setspace}

\newcommand{\FIGWIDTH}{0.31\linewidth}
\newcommand{\FIGHEIGHT}{0.32\textheight}
\newcommand{\FIGWIDTHTWO}{0.33\linewidth}
\newcommand{\FIGHEIGHTTWO}{0.5\linewidth}
\newcommand{\FIGWIDTHTHREE}{0.6\linewidth}

\title{\boldmath High precision determination of $\as$ from a global fit of jet rates}
\author[a,1]{Andrii Verbytskyi\note{Corresponding author.},}
\author[b]{Andrea Banfi,}
\author[c]{Adam Kardos,}
\author[d]{Pier Francesco Monni,}
\author[a]{Stefan Kluth,}
\author[e]{G\'abor Somogyi,}
\author[f]{Zolt\'an Sz\H{o}r,}
\author[e,g]{Zolt\'an Tr\'ocs\'anyi,}
\author[e]{Zolt\'an Tulip\'ant}
\author[a]{and Giulia Zanderighi}
\affiliation[a]{{\scriptsize Max-Planck-Institut f\"{u}r Physik, D-80805 Munich, Germany}}
\affiliation[b]{{\scriptsize University of Sussex, Brighton, BN1 9RH United Kingdom }}
\affiliation[c]{{\scriptsize University of Debrecen, 4010 Debrecen, PO Box 105, Hungary}}
\affiliation[d]{{\scriptsize CERN, Theory Department, CH-1211 Geneva 23, Switzerland}}
\affiliation[e]{{\scriptsize MTA-DE Particle Physics Research Group, University of Debrecen, 4010 Debrecen, PO Box 105, Hungary}}
\affiliation[f]{{\scriptsize PRISMA Cluster of Excellence, Institut f\"{u}r Physik, Universit\"{a}t Mainz, D-55099 Mainz, Germany}}
\affiliation[g]{{\scriptsize Institute for Theoretical Physics,
    E\"{o}tv\"{o}s Lor\'and University, P\'azm\'any P\'eter 1/A,
    H-1117 Budapest, Hungary}}
\preprint{MPP-2019-37, CERN-TH-2019-015, MITP/19-012}

\emailAdd{andrii.verbytskyi@mpp.mpg.de}
\emailAdd{a.banfi@sussex.ac.uk}
\emailAdd{kardos.adam@science.unideb.hu}
\emailAdd{pier.monni@cern.ch}
\emailAdd{stefan.kluth@mpp.mpg.de}
\emailAdd{gabor.somogyi@cern.ch}
\emailAdd{zoltanszoer@uni-mainz.de}
\emailAdd{zoltant@cern.ch}
\emailAdd{tulipant.zoltan@science.unideb.hu}
\emailAdd{zanderi@mpp.mpg.de}

\newcommand{\JRTfitrangeoneRtwoRthree}{$[-1.75+\LQM,-1][-1.5+\LQM,-1]$\xspace}
\newcommand{\JRTfitrangeoneRtwo}{$[-1.75+\LQM,-1]$\xspace}
\newcommand{\JRTfitrangeoneRthree}{$[-1.5+\LQM,-1]$\xspace}
\newcommand{\JRTfitrangetwoRtwoRthree}{$[ -2+\LQM,-1][-1.75+\LQM,-1]$\xspace}
\newcommand{\JRTfitrangetwoRtwo}{$[ -2+\LQM,-1]$\xspace}
\newcommand{\JRTfitrangetwoRthree}{$[-1.75+\LQM,-1]$\xspace}
\newcommand{\JRTfitrangethreeRtwoRthree}{$[-2.25+\LQM,-1][-2+\LQM,-1]$\xspace}
\newcommand{\JRTfitrangethreeRtwo}{$[-2.25+\LQM,-1]$\xspace}
\newcommand{\JRTfitrangethreeRthree}{$[ -2+\LQM,-1]$\xspace}
\def\rhonominal{0.006}
\def\Knominal{0.5}
\def\ydistance{0.25}

\newcommand{\JRTfitrangefourRtwoRthree}{$[-2.5+\LQM,-1][-2.25+\LQM,-1]$\xspace}
\newcommand{\JRTfitrangefourRtwo}{$[-2.5+\LQM,-1]$\xspace}
\newcommand{\JRTfitrangefourRthree}{$[-2.25+\LQM,-1]$\xspace}

\newcommand{\epjcbreak}[1]{}
\newcommand{\draftbreak}[1]{\\#1}
\newcommand{\arxivbreak}[1]{}
\newcommand{\epjconly}[1]{}
\newcommand{\draftonly}[1]{#1}
\newcommand{\arxivonly}[1]{}
\newcommand{\JRTpredictions}{
0.000249 &  64.853 &   2226.9 &    52659 &    -2276.3 &    -18059 &    -52708 \\
0.000304 &  61.462 &   1981.9 &    43613 &    -1951.9 &    -9434&    -49312 \\
0.000371 &  57.969 &   1745.7 &    35503 &    -1644.5 &    -2392&    -45862 \\
0.000453 &  54.589 &   1531.6 &    28680 &    -1370.8 &    2875.1 &    -41808 \\
0.000553 &  51.315 &   1337.6 &    22984 &    -1126.4 &    6726.2 &    -38197 \\
0.000676 &  48.149 &   1162.6 &    18248 &    -911.38 &    9293.8 &    -33891 \\
0.000825 &  45.081 &   1005.5 &    14362 &    -722.69 &    10902&    -29909 \\
0.001008 &  42.125 &   864.92 &    11185 &    -558.20 &    11665&    -25860 \\
0.001231 &  39.267 &   739.71 &    8612&    -416.39 &    11905&    -22437 \\
0.001503 &  36.514 &   628.56 &    6548&    -295.16 &    11595&    -18881 \\
0.001836 &  33.865 &   530.39 &    4910&    -192.34 &    10949&    -15681 \\
0.002243 &  31.318 &   444.20 &    3627&    -106.70 &    10084&    -12771 \\
0.002739 &  28.874 &   368.93 &    2634&    -36.132 &    9080.6 &    -10373 \\
0.003346 &  26.532 &   303.65 &    1875&    20.7909 &    8018.6 &    -8011\\
0.004087 &  24.292 &   247.42 &    1307&    65.2347 &    6936.3 &    -6144\\
0.004992 &  22.155 &   199.36 &    888.5 &    98.9217 &    5886.0 &    -4478\\
0.006097 &  20.118 &   158.62 &    586.9 &    123.332 &    4897.6 &    -3090\\
0.007447 &  18.183 &   124.43 &    374.8 &    139.397 &    3982.9 &    -1863\\
0.009095 &  16.354 &   96.058 &    230.1 &    148.398 &    3173.8 &    -944.6 \\
0.011109 &  14.621 &   72.787 &    134.4 &    151.554 &    2469.0 &    -218.7 \\
0.013569 &  12.988 &   53.978 &    74.07 &    149.800 &    1868.4 &    250.42 \\
0.016573 &  11.455 &   39.019 &    37.89 &    144.067 &    1373.2 &    580.39 \\
0.020242 &  10.019 &   27.368 &    17.63 &    135.256 &    974.01 &    724.38 \\
0.024724 &  8.6830 &   18.495 &    7.213 &    124.148 &    662.70 &    917.46 \\
0.030197 &  7.4420 &   11.934 &    2.471 &    111.328 &    428.24 &    950.16 \\
0.036883 &  6.2962 &   7.2589 &    0.647 &    97.6721 &    260.70 &    897.90 \\
0.045049 &  5.2435 &   4.0822 &    0.110 &    83.5664 &    145.88 &    812.72 \\
0.055023 &  4.2842 &   2.0589 &    0.008 &    69.5476 &    73.141 &    700.85 \\
0.067206 &  3.4173 &   0.8849 &    0.000 &    56.1021 &    31.186 &    571.07 \\
0.082085 &  2.6424 &   0.2950 &    -0.00 &    43.4943 &    10.302 &    449.83 \\
0.100259 &  1.9584 &   0.0630 &    -0.00 &    32.1309 &    2.1788 &    326.17 \\
0.122456 &  1.3647 &   0.0056 &    0.000 &    22.1977 &    0.1915 &    218.79 \\
0.149569 &  0.8612 &   0.0000 &    0.000 &    13.8521 &    0.0005 &    133.14 \\
0.182684 &  0.4500 &   -0.000 &    0.000 &    7.16729 &    -0.000 &    73.647 \\
0.223130 &  0.1366 &   0.0000 &    -0.00 &    2.15434 &    0.0000 &    24.828 \\
0.272532 &  0.1320 &   -0.000 &    0.000 &    2.08088 &    0.0000 &    23.734 \\
}

\newcommand{\JRTtabularinput}{
OPAL~\protect\cite{Pfeifenschneider:1999rz}&$91.2 (91.2)$&$ 91.2 $ &$1508031 $\\
OPAL~\protect\cite{Pfeifenschneider:1999rz}&$189.0 (189.0)$&$ 189 $ &$3300 $\\
OPAL~\protect\cite{Pfeifenschneider:1999rz}&$183.0 (183.0)$&$ 183 $ &$1082 $\\
OPAL~\protect\cite{Pfeifenschneider:1999rz}&$172.0 (172.0)$&$ 172 $ &$224 $\\
OPAL~\protect\cite{Pfeifenschneider:1999rz}&$161.0 (161.0)$&$ 161 $ &$281 $\\
OPAL~\protect\cite{Pfeifenschneider:1999rz}&$130.0-136.0 (133.0)$&$ 133 $ &$630 $\\
L3~\protect\cite{Achard:2004sv}&$201.5-209.1 (206.2)$&$ 206 $ &$4146 $\\
L3~\protect\cite{Achard:2004sv}&$199.2-203.8 (200.2)$&$ 200 $ &$2456 $\\
L3~\protect\cite{Achard:2004sv}&$191.4-196.0 (194.4)$&$ 194 $ &$2403 $\\
L3~\protect\cite{Achard:2004sv}&$188.4-189.9 (188.6)$&$ 189 $ &$4479 $\\
L3~\protect\cite{Achard:2004sv}&$180.8-184.2 (182.8)$&$ 183 $ &$1500 $\\
L3~\protect\cite{Achard:2004sv}&$161.2-164.7 (161.3)$&$ 161 $ &$424 $\\
L3~\protect\cite{Achard:2004sv}&$135.9-140.1 (136.1)$&$ 136 $ &$414 $\\
L3~\protect\cite{Achard:2004sv}&$129.9-130.4 (130.1)$&$ 130 $ &$556 $\\
JADE~\protect\cite{Pfeifenschneider:1999rz}&$43.4-44.3 (43.7)$&$ 44 $ &$4110 $\\
JADE~\protect\cite{Pfeifenschneider:1999rz}&$34.5-35.5 (34.9)$&$ 35 $ &$29514 $\\
ALEPH~\protect\cite{Heister:2003aj}&$91.2 (91.2)$&$ 91.2 $ &$3600000 $\\
ALEPH~\protect\cite{Heister:2003aj}&$206.0 (206.0)$&$ 206 $ &$3578 $\\
ALEPH~\protect\cite{Heister:2003aj}&$189.0 (189.0)$&$ 189 $ &$3578 $\\
ALEPH~\protect\cite{Heister:2003aj}&$183.0 (183.0)$&$ 183 $ &$1319 $\\
ALEPH~\protect\cite{Heister:2003aj}&$172.0 (172.0)$&$ 172 $ &$257 $\\
ALEPH~\protect\cite{Heister:2003aj}&$161.0 (161.0)$&$ 161 $ &$319 $\\
ALEPH~\protect\cite{Heister:2003aj}&$133.0 (133.0)$&$ 133 $ &$806 $\\
}

\newcommand{\JRTtabularresultNNLOandNNLORES}{
\hline 
\JRTfitrangeoneRtwoRthree        &$0.12195\pm 0.00072$ &  $0.12078\pm 0.00066$  \\
$S^{C}$              &$120/143 = 0.84$&  $140/143 = 0.98$  \\\hline 
\JRTfitrangetwoRtwoRthree         &$0.12163\pm 0.00061$ &  $0.12065\pm 0.00056$  \\
$S^{C}$              &$153/187 = 0.82$ &  $176/187 = 0.94$  \\\hline 
\JRTfitrangethreeRtwoRthree       &$0.12075\pm 0.00044$ &  $0.11994\pm 0.00041$  \\
$S^{C}$              &$208/251 = 0.83$&  $222/251 = 0.88$  \\\hline 
\JRTfitrangefourRtwoRthree       &$0.12143\pm 0.00043$ &  $0.12089\pm 0.00044$  \\
$S^{C}$              &$321/331 = 0.97$&  $336/331 = 1.01$  \\\hline 
\hline 
\JRTfitrangeoneRtwoRthree        &$0.12068\pm 0.00073$ &  $0.11956\pm 0.00066$  \\
$H^{C}$              &$126/143 = 0.88$&  $147/143 = 1.03$  \\\hline 
\JRTfitrangetwoRtwoRthree         &$0.12006\pm 0.00061$ &  $0.11913\pm 0.00054$  \\
$H^{C}$              &$163/187 = 0.87$ &  $188/187 = 1.01$  \\\hline 
\JRTfitrangethreeRtwoRthree       &$0.11869\pm 0.00043$ &  $0.11793\pm 0.00043$  \\
$H^{C}$              &$221/251 = 0.88$&  $238/251 = 0.95$  \\\hline 
\JRTfitrangefourRtwoRthree       &$0.11845\pm 0.00045$ &  $0.11799\pm 0.00047$  \\
$H^{C}$              &$302/331 = 0.91$&  $310/331 = 0.94$  \\\hline 
\hline 
\JRTfitrangeoneRtwoRthree        &$0.12248\pm 0.00068$ &  $0.12129\pm 0.00063$  \\
$H^{L}$              &$121/143 = 0.85$&  $141/143 = 0.99$  \\\hline 
\JRTfitrangetwoRtwoRthree         &$0.12211\pm 0.00057$ &  $0.12110\pm 0.00053$  \\
$H^{L}$              &$155/187 = 0.83$ &  $180/187 = 0.96$  \\\hline 
\JRTfitrangethreeRtwoRthree       &$0.12071\pm 0.00044$ &  $0.11989\pm 0.00045$  \\
$H^{L}$              &$209/251 = 0.83$&  $227/251 = 0.90$  \\\hline 
\JRTfitrangefourRtwoRthree       &$0.12041\pm 0.00044$ &  $0.11990\pm 0.00044$  \\
$H^{L}$              &$266/331 = 0.80$&  $278/331 = 0.84$  \\\hline 
}
\newcommand{\JRTtabularresultNNLOandNNLORESepjc}{
\hline 
\JRTfitrangeoneRtwo        &$0.12195\pm 0.00072$ &  $0.12078\pm 0.00066$  \\
\JRTfitrangeoneRthree        &$120/143 = 0.84$&  $140/143 = 0.98$  \\ 
$S^{C}$              & &  \\\hline 
\JRTfitrangetwoRtwo         &$0.12163\pm 0.00061$ &  $0.12065\pm 0.00056$  \\
\JRTfitrangetwoRthree         &$153/187 = 0.82$ &  $176/187 = 0.94$  \\ 
$S^{C}$              & &  \\\hline 
\JRTfitrangethreeRtwo       &$0.12075\pm 0.00044$ &  $0.11994\pm 0.00041$  \\
\JRTfitrangethreeRthree       &$208/251 = 0.83$&  $222/251 = 0.88$  \\ 
$S^{C}$              & &  \\\hline 
\JRTfitrangefourRtwo       &$0.12143\pm 0.00043$ &  $0.12089\pm 0.00044$  \\
\JRTfitrangefourRthree       &$321/331 = 0.97$&  $336/331 = 1.01$  \\ 
$S^{C}$              & &  \\\hline 
\hline 
\JRTfitrangeoneRtwo        &$0.12068\pm 0.00073$ &  $0.11956\pm 0.00066$  \\
\JRTfitrangeoneRthree        &$126/143 = 0.88$&  $147/143 = 1.03$  \\ 
$H^{C}$              & &  \\\hline 
\JRTfitrangetwoRtwo         &$0.12006\pm 0.00061$ &  $0.11913\pm 0.00054$  \\
\JRTfitrangetwoRthree         &$163/187 = 0.87$ &  $188/187 = 1.01$  \\ 
$H^{C}$              & &  \\\hline 
\JRTfitrangethreeRtwo       &$0.11869\pm 0.00043$ &  $0.11793\pm 0.00043$  \\
\JRTfitrangethreeRthree       &$221/251 = 0.88$&  $238/251 = 0.95$  \\ 
$H^{C}$              & &  \\\hline 
\JRTfitrangefourRtwo       &$0.11845\pm 0.00045$ &  $0.11799\pm 0.00047$  \\
\JRTfitrangefourRthree       &$302/331 = 0.91$&  $310/331 = 0.94$  \\ 
$H^{C}$              & &  \\\hline 
\hline 
\JRTfitrangeoneRtwo        &$0.12248\pm 0.00068$ &  $0.12129\pm 0.00063$  \\
\JRTfitrangeoneRthree        &$121/143 = 0.85$&  $141/143 = 0.99$  \\ 
$H^{L}$              & &  \\\hline 
\JRTfitrangetwoRtwo         &$0.12211\pm 0.00057$ &  $0.12110\pm 0.00053$  \\
\JRTfitrangetwoRthree         &$155/187 = 0.83$ &  $180/187 = 0.96$  \\ 
$H^{L}$              & &  \\\hline 
\JRTfitrangethreeRtwo       &$0.12071\pm 0.00044$ &  $0.11989\pm 0.00045$  \\
\JRTfitrangethreeRthree       &$209/251 = 0.83$&  $227/251 = 0.90$  \\ 
$H^{L}$              & &  \\\hline 
\JRTfitrangefourRtwo       &$0.12041\pm 0.00044$ &  $0.11990\pm 0.00044$  \\
\JRTfitrangefourRthree       &$266/331 = 0.80$&  $278/331 = 0.84$  \\ 
$H^{L}$              & &  \\\hline 
}
\newcommand{\JRTtabularresultNNLOandNNLOREStwo}{
\hline 
\JRTfitrangeoneRtwo        &$0.12121\pm 0.00095$ &  $0.11849\pm 0.00092$  \\
$S^{C}$              &$20/86 = 0.24$&  $20/86 = 0.24$  \\\hline 
\JRTfitrangetwoRtwo         &$0.12114\pm 0.00081$ &  $0.11864\pm 0.00075$  \\
$S^{C}$              &$26/100 = 0.26$ &  $26/100 = 0.26$  \\\hline 
\JRTfitrangethreeRtwo       &$0.12119\pm 0.00060$ &  $0.11916\pm 0.00063$  \\
$S^{C}$              &$44/150 = 0.29$&  $44/150 = 0.29$  \\\hline 
\JRTfitrangefourRtwo       &$0.12217\pm 0.00052$ &  $0.12075\pm 0.00055$  \\
$S^{C}$              &$89/180 = 0.50$&  $107/180 = 0.59$  \\\hline 
\hline 
\JRTfitrangeoneRtwo        &$0.11957\pm 0.00098$ &  $0.11698\pm 0.00093$  \\
$H^{C}$              &$22/86 = 0.26$&  $22/86 = 0.25$  \\\hline 
\JRTfitrangetwoRtwo         &$0.11923\pm 0.00079$ &  $0.11687\pm 0.00076$  \\
$H^{C}$              &$29/100 = 0.29$ &  $28/100 = 0.28$  \\\hline 
\JRTfitrangethreeRtwo       &$0.11868\pm 0.00068$ &  $0.11679\pm 0.00064$  \\
$H^{C}$              &$43/150 = 0.28$&  $40/150 = 0.27$  \\\hline 
\JRTfitrangefourRtwo       &$0.11849\pm 0.00050$ &  $0.11723\pm 0.00053$  \\
$H^{C}$              &$58/180 = 0.32$&  $58/180 = 0.32$  \\\hline 
\hline 
\JRTfitrangeoneRtwo        &$0.12171\pm 0.00109$ &  $0.11897\pm 0.00092$  \\
$H^{L}$              &$21/86 = 0.25$&  $21/86 = 0.24$  \\\hline 
\JRTfitrangetwoRtwo         &$0.12144\pm 0.00078$ &  $0.11893\pm 0.00075$  \\
$H^{L}$              &$28/100 = 0.28$ &  $26/100 = 0.26$  \\\hline 
\JRTfitrangethreeRtwo       &$0.12080\pm 0.00069$ &  $0.11881\pm 0.00063$  \\
$H^{L}$              &$43/150 = 0.28$&  $39/150 = 0.26$  \\\hline 
\JRTfitrangefourRtwo       &$0.12024\pm 0.00051$ &  $0.11897\pm 0.00053$  \\
$H^{L}$              &$57/180 = 0.32$&  $52/180 = 0.29$  \\\hline 
}
\newcommand{\JRTtabularresultNNLOthree}{
\hline 
\JRTfitrangeoneRthree        &$0.12176\pm 0.00113$  \\
$S^{C}$              &$39/56 = 0.70$ \\\hline 
\JRTfitrangetwoRthree         &$0.12088\pm 0.00088$  \\
$S^{C}$              &$56/86 = 0.65$   \\\hline 
\JRTfitrangethreeRthree       &$0.11996\pm 0.00074$  \\
$S^{C}$              &$74/100 = 0.74$  \\\hline 
\JRTfitrangefourRthree       &$0.11853\pm 0.00068$  \\
$S^{C}$              &$111/150 = 0.74$ \\\hline 
\hline 
\JRTfitrangeoneRthree        &$0.12053\pm 0.00114$  \\
$H^{C}$              &$41/56 = 0.74$ \\\hline 
\JRTfitrangetwoRthree         &$0.11933\pm 0.00084$  \\
$H^{C}$              &$61/86 = 0.70$   \\\hline 
\JRTfitrangethreeRthree       &$0.11810\pm 0.00074$  \\
$H^{C}$              &$83/100 = 0.83$  \\\hline 
\JRTfitrangefourRthree       &$0.11645\pm 0.00061$  \\
$H^{C}$              &$125/150 = 0.83$ \\\hline 
\hline 
\JRTfitrangeoneRthree        &$0.12257\pm 0.00112$  \\
$H^{L}$              &$39/56 = 0.70$ \\\hline 
\JRTfitrangetwoRthree         &$0.12178\pm 0.00088$  \\
$H^{L}$              &$57/86 = 0.66$   \\\hline 
\JRTfitrangethreeRthree       &$0.12076\pm 0.00068$  \\
$H^{L}$              &$75/100 = 0.75$  \\\hline 
\JRTfitrangefourRthree       &$0.11931\pm 0.00067$  \\
$H^{L}$              &$115/150 = 0.77$ \\\hline 
}

\makeatletter
\def\ifabsgreater #1#2{\ifpdfabsnum 
    \dimexpr#1pt>\dimexpr#2pt\relax
    \expandafter\@firstoftwo\else\expandafter\@secondoftwo\fi }
\makeatother

\newcommand{\figdependencesqrtscombinedthree}[2]{

\begin{figure}\centering\boldmath
\adjustbox{width=0.65\linewidth, height=1.0\linewidth}{
\begin{tikzpicture}
\linespread{0.4}
\pgfplotsset{try min ticks=7}
\pgfplotsset{
  errorBars/.style={
    error bars/error bar style={
     thick,
    },
    error bars/y dir=both,
    error bars/y explicit,
  error bars/error mark=none,
  thick,mark size=2.0pt
  }
}
\begin{axis}[ylabel=$\alpha_s(M_Z)$,xlabel={$\sqrt{s}$, $\[\GeV\]$},
     every axis plot/.append style={errorBars},
     xticklabel=\empty,
    xmajorgrids=true,
    xmin=0.2,xmax=9.8,
    ymin=0.1075,ymax=0.1238,
    grid style={ draw=gray!70},
     major grid style={dashed, draw=gray!70},
         mminor grid style={line width=0.0pt,draw=gray!50},
    minor tick num=5,
            legend cell align={left},
    enlarge x limits={abs=0.0},    
    enlarge y limits={abs=0.0},    
    ticklabel style={font=\large,fill=none},
    xtick={1.5,2.5,3.5,4.5,5.5,6.5,7.5,8.5},
    xtick style={grid=major,draw=none},
    yticklabel style={/pgf/number format/fixed, /pgf/number format/precision=3, /pgf/number format/fixed zerofill, font=\large} ,
        xlabel style={at={(ticklabel* cs:1)},anchor=north west, xshift=-3.0cm, yshift=0.6cm, font=\bf\LARGE},
    ylabel style={at={(ticklabel* cs:1)},anchor=south west, xshift=-2.6cm, yshift=0.4cm, font=\LARGE},
    legend style={draw=none},
        extra x ticks={              0.9,              2,            3,          4,             5,                 6,                 7,              8,                     9.1  },
        extra x tick labels={ {$35\textnormal{-}207$}, {$35$}, {$44$}, {$91$}, {$130\textnormal{-}136$}, {$161\textnormal{-}172$},  {$183$}, {$189\textnormal{-}196$}, {$200\textnormal{-}207$} },
        every extra x tick/.style={grid=none,tick label style={font=\Large,rotate=-90,anchor=south},major tick length=0.0cm,yshift=1.0cm,xshift=-0.3cm,draw=none,thick,align=center},
        legend style={legend pos=north east,font={\bf\LARGE}},
        width=0.7\textwidth,
        height=0.6\textheight
    ]        
\addplot[mark=o,color=blue] table [x expr = 1.0-0.25, y expr={1.0*\thisrowno{0}},  y error  expr={1.0*\thisrowno{1}}, col sep=space] {../../share//Results/fit_asmz_0G0D110000.dat};
\addplot[mark=o,color=blue] table [x expr = 2.0-0.25,  y expr={1.0*\thisrowno{0}}, y error  expr={1.0*\thisrowno{1}}, col sep=space] {../../share//Results/fit_asmz_0G0D110010.dat}; 
\addplot[mark=o,color=blue] table [x expr = 3.0-0.25,  y expr={1.0*\thisrowno{0}},y error  expr={1.0*\thisrowno{1}}, col sep=space] {../../share//Results/fit_asmz_0G0D110020.dat}; 
\addplot[mark=o,color=blue] table [x expr = 4.0-0.25,  y expr={1.0*\thisrowno{0}},  y error  expr={1.0*\thisrowno{1}}, col sep=space] {../../share//Results/fit_asmz_0G0D110030.dat}; 
\addplot[mark=o,color=blue] table [x expr = 5.0-0.25,  y expr={1.0*\thisrowno{0}},  y error  expr={1.0*\thisrowno{1}}, col sep=space] {../../share//Results/fit_asmz_0G0D1100J0.dat};
\addplot[mark=o,color=blue] table [x expr = 6.0-0.25,  y expr={1.0*\thisrowno{0}},  y error  expr={1.0*\thisrowno{1}}, col sep=space] {../../share//Results/fit_asmz_0G0D1100K0.dat};
\addplot[mark=o,color=blue] table [x expr = 7.0-0.25,  y expr={1.0*\thisrowno{0}}, y error  expr={1.0*\thisrowno{1}}, col sep=space] {../../share//Results/fit_asmz_0G0D110090.dat};
\addplot[mark=o,color=blue] table [x expr = 8.0-0.25,  y expr={1.0*\thisrowno{0}}, y error  expr={1.0*\thisrowno{1}}, col sep=space] {../../share//Results/fit_asmz_0G0D1100L0.dat};
\addplot[mark=o,color=blue] table [x expr = 9.0-0.25,  y expr={1.0*\thisrowno{0}},  y error  expr={1.0*\thisrowno{1}}, col sep=space] {../../share//Results/fit_asmz_0G0D1100M0.dat};

\addplot[mark=*,color=black] table [x expr = 1.0+0.25, y expr={1.0*\thisrowno{0}},  y error  expr={1.0*\thisrowno{1}}, col sep=space] {../../share//Results/fit_asmz_0H0D110000.dat};
\addplot[mark=*,color=black] table [x expr = 2.0+0.25,  y expr={1.0*\thisrowno{0}}, y error  expr={1.0*\thisrowno{1}}, col sep=space] {../../share//Results/fit_asmz_0H0D110010.dat}; 
\addplot[mark=*,color=black] table [x expr = 3.0+0.25,  y expr={1.0*\thisrowno{0}},y error  expr={1.0*\thisrowno{1}}, col sep=space] {../../share//Results/fit_asmz_0H0D110020.dat}; 
\addplot[mark=*,color=black] table [x expr = 4.0+0.25,  y expr={1.0*\thisrowno{0}},  y error  expr={1.0*\thisrowno{1}}, col sep=space] {../../share//Results/fit_asmz_0H0D110030.dat}; 
\addplot[mark=*,color=black] table [x expr = 5.0+0.25,  y expr={1.0*\thisrowno{0}},  y error  expr={1.0*\thisrowno{1}}, col sep=space] {../../share//Results/fit_asmz_0H0D1100J0.dat};
\addplot[mark=*,color=black] table [x expr = 6.0+0.25,  y expr={1.0*\thisrowno{0}},  y error  expr={1.0*\thisrowno{1}}, col sep=space] {../../share//Results/fit_asmz_0H0D1100K0.dat};
\addplot[mark=*,color=black] table [x expr = 7.0+0.25,  y expr={1.0*\thisrowno{0}}, y error  expr={1.0*\thisrowno{1}}, col sep=space] {../../share//Results/fit_asmz_0H0D110090.dat};
\addplot[mark=*,color=black] table [x expr = 8.0+0.25,  y expr={1.0*\thisrowno{0}}, y error  expr={1.0*\thisrowno{1}}, col sep=space] {../../share//Results/fit_asmz_0H0D1100L0.dat};
\addplot[mark=*,color=black] table [x expr = 9.0+0.25,  y expr={1.0*\thisrowno{0}},  y error  expr={1.0*\thisrowno{1}}, col sep=space] {../../share//Results/fit_asmz_0H0D1100M0.dat};

\addplot[mark=x,color=red] table [x expr = 1.0+0.00, y expr={1.0*\thisrowno{0}},  y error  expr={1.0*\thisrowno{1}}, col sep=space] {../../share//Results/fit_asmz_0C0D110000.dat};
\addplot[mark=x,color=red] table [x expr = 2.0+0.00,  y expr={1.0*\thisrowno{0}}, y error  expr={1.0*\thisrowno{1}}, col sep=space] {../../share//Results/fit_asmz_0C0D110010.dat}; 
\addplot[mark=x,color=red] table [x expr = 3.0+0.00,  y expr={1.0*\thisrowno{0}},y error  expr={1.0*\thisrowno{1}}, col sep=space] {../../share//Results/fit_asmz_0C0D110020.dat}; 
\addplot[mark=x,color=red] table [x expr = 4.0+0.00,  y expr={1.0*\thisrowno{0}},  y error  expr={1.0*\thisrowno{1}}, col sep=space] {../../share//Results/fit_asmz_0C0D110030.dat}; 
\addplot[mark=x,color=red] table [x expr = 5.0+0.00,  y expr={1.0*\thisrowno{0}},  y error  expr={1.0*\thisrowno{1}}, col sep=space] {../../share//Results/fit_asmz_0C0D1100J0.dat};
\addplot[mark=x,color=red] table [x expr = 6.0+0.00,  y expr={1.0*\thisrowno{0}},  y error  expr={1.0*\thisrowno{1}}, col sep=space] {../../share//Results/fit_asmz_0C0D1100K0.dat};
\addplot[mark=x,color=red] table [x expr = 7.0+0.00,  y expr={1.0*\thisrowno{0}}, y error  expr={1.0*\thisrowno{1}}, col sep=space] {../../share//Results/fit_asmz_0C0D110090.dat};
\addplot[mark=x,color=red] table [x expr = 8.0+0.00,  y expr={1.0*\thisrowno{0}}, y error  expr={1.0*\thisrowno{1}}, col sep=space] {../../share//Results/fit_asmz_0C0D1100L0.dat};
\addplot[mark=x,color=red] table [x expr = 9.0+0.00,  y expr={1.0*\thisrowno{0}},  y error  expr={1.0*\thisrowno{1}}, col sep=space] {../../share//Results/fit_asmz_0C0D1100M0.dat};

\end{axis}
\end{tikzpicture}
}
\caption{Dependence of fit results on the used $\sqrt{s}$ range.
The fit range is default.}
\label{fig:dependencesqrtscombined}
\end{figure}
}

\include{JRT-auth}
\newcommand{\JRTresultHNNLORtwoRthree}{ 0.11989\pm  0.00045  {\text( exp.)}\pm 0.00098{\text(hadr.)}\pm 0.00046{\text(ren.)}\pm 0.00017{\text(res.)}}

\newcommand{\JRTresultHNNLORtwo}{ 0.11881\pm  0.00063  {\text( exp.)}\pm 0.00101{\text(hadr.)}\pm 0.00045{\text(ren.)}\pm 0.00034{\text(res.)}}
\newcommand{\JRTresultHNNLORtwocomb}{ 0.11881\pm 0.00131{\text(comb.)}}

\newcommand{\JRTresultHNNLONOJADERtwo}{ 0.11838\pm  0.00077  {\text( exp.)}}
\newcommand{\JRTresultHNNLOOPALRtwo}{ 0.11893\pm  0.00137  {\text( exp.)}}
\newcommand{\JRTresultHNNLOALEPHRthree}{ 0.11905\pm  0.00251  {\text( exp.)}}
\newcommand{\JRTresultHNNLOOPALSHIFTSRtwo}{ 0.11761\pm  0.00179  {\text( exp.)}}
\newcommand{\JRTresultHNNLOMULTRtwo}{ 0.11881\pm  0.00063  {\text( exp.)}\pm 0.00109{\text(hadr.)}}

\newcommand\prog[1]     {{\tt #1}}

\def\beq{\begin{equation}}
\def\eeq{\end{equation}}
\def\bsp#1\esp{\begin{split}#1\end{split}}
\def\bal#1\eal{\begin{align}#1\end{align}}

\newcommand\tsig[2]    {\sigma^{\rm{#1}}_{#2}}

\newcommand\as  	       {\ensuremath{\alpha_{\rm{s}}}}
\newcommand\asbar[1]  	       {\ensuremath{\frac{\alpha_{\rm{S}}(#1)}{2\pi}}}
\newcommand\Oa[1]      {\ensuremath{\mathcal O(\as^{#1})}}

\newcommand{\CF}       {C_{\rm{F}}}
\newcommand{\CA}       {C_{\rm{A}}}
\newcommand{\TR}       {T_{\rm{R}}}

\newcommand{\Nf}       {\ensuremath{n_{\rm{f}}}}

\newcommand{\ndof}       {{\rm ndof}}

\newcommand{\xiR}      {x_{R}}
\newcommand{\xiRsq}      {(x_{R}^2)}

\newcommand{\eVdist}{\kern-0.06667em}
\newcommand{\GeV}{{\,\text{Ge}\eVdist\text{V\/}}}

\newcommand{\NLL}{\text{NLL}}
\newcommand{\NNLL}{\text{NNLL}}

\newcommand{\ycut}{y_{\rm cut}}

\newcommand{\LQM}{{\cal L}}

 \newcommand{\DissertoriNNLOALEPHRthree}{ 0.1175\pm  0.0020  {\text(exp.)}\pm  0.0015  {\text(theo.)}}

\abstract{\\We present state-of-the-art extractions of the strong coupling based
on N$^3$LO+NNLL accurate predictions for the two-jet rate in the
Durham clustering algorithm at $e^+e^-$ collisions, as well as a
simultaneous fit of the two- and three-jet rates taking into account
correlations between the two observables. The fits are performed on a
large range of data sets collected at LEP and PETRA colliders, with
energies spanning from $35$ GeV to $207$ GeV. Owing to the high
accuracy of the predictions used, the perturbative uncertainty is
considerably smaller than that due to hadronization. Our best
determination at the $Z$ mass is $\alpha_s(M_Z) =
\JRTresultHNNLORtwo$, which is in agreement with the latest world
average and has a comparable total uncertainty.
}

\newcommand{\durham}{Durham\xspace}

\begin{document}
\maketitle
\flushbottom
\tableofcontents
\newpage
\clearpage
\pagenumbering{arabic}
\pagestyle{plain}
\newpage
\section{Introduction}
\label{sec:introduction}

Measurements using hadronic final states in $e^{+}e^{-}$ annihilation
provide a unique opportunity to study the strong interaction in a well
controlled environment without strongly interacting particles in the
initial state. Accordingly, Quantum Chromodynamics (QCD) was tested
extensively at
LEP (see e.g.~\cite{Fritzsch:1973pi,Gross:1973id,Politzer:1973fx,Gross:1973ju}).

QCD is a well-established theory by now, and its coupling constant
$\alpha_s$ has been measured in a variety of different processes at
different energies. Still, the last PDG average of $\alpha_s$ has an
uncertainty of about 1\%~\cite{Olive:2016xmw}, which is considerably larger
than errors in other gauge couplings. This uncertainty has an
important impact on the current LHC precision physics program.
Furthermore, a number of outlier fits of $\alpha_s$ exist,
hence any further precise determination of the value of the strong
coupling constant is very valuable.

Many measurements of the strong coupling $\alpha_s$ at $e^+e^-$
colliders are based on comparisons of differential distributions of
event shapes or jet rates to perturbative predictions. Presently
progress in such measurements depends solely on the improvement in the
accuracy of theoretical predictions as new data are not foreseen in
the near future.

Compared to event shapes, jet-rates are known to be less sensitive to
hadronization corrections\footnote{See, for instance, the discussion in section 4.5 of
  Ref.~\cite{Dokshitzer:1995qm}.}, hence more suited for precise determinations
of the strong coupling constant.
Fully differential predictions for the
process $e^{+}e^{-}\to 3$ partonic jets are available to NNLO
($\as^{3}$)
accuracy~\cite{GehrmannDeRidder:2007hr,GehrmannDeRidder:2008ug,Weinzierl:2008iv,Weinzierl:2009ms,DelDuca:2016ily}.
Using the predictions for the three-jet rate at NNLO and the total
cross-section at N$^3$LO~\cite{Gorishnii:1990vf}, the two-jet rate can be deduced at N$^3$LO
accuracy.

For a jet rate with a resolution parameter $y$, the fixed order
predictions have a limited range of validity and are not reliable near
the boundaries of the phase space, dominated by soft and collinear QCD
radiation. In particular, for $y\to 0$, the perturbative prediction at
order ${\cal O}(\alpha_s^n)$ features logarithmic terms of the type
$\alpha_s^n L^{m}$ with $L=\ln(y)$, and $m\leq 2n$.
When $L$ is large, such terms have to be summed up to all orders in
perturbation theory. 

The accuracy of the resummation for observables for which double logarithmic
terms ${\cal O}(\as^n L^{2n})$ exponentiate, is usually defined in
terms of the logarithm of the cumulative distribution $\Sigma$.
For such exponentiating observables, we define leading logarithms (LL)
as terms of the form $\as^n L^{n+1}$, next-to-leading logarithms (NLL)
as $\as^n L^n$, next-to-next-to-leading logarithms (NNLL) as $\as^n
L^{n-1}$, in $\ln\Sigma$.  The state of the art for most event shapes
and jet rates measured at LEP is either NNLL or even
N$^3$LL~\cite{Becher:2008cf,Abbate:2010xh,Monni:2011gb,Becher:2012qc,Hoang:2014wka,deFlorian:2004mp,Banfi:2014sua,Banfi:2016zlc,Tulipant:2017ybb,Moult:2018jzp,Banfi:2018mcq,Bell:2018gce}.
These resummed predictions matched to  fixed-order ones were used for 
precise extractions of
$\alpha_s$~\cite{Becher:2008cf,Abbate:2010xh,Gehrmann:2012sc,Hoang:2014wka,Kardos:2018kqj}.

Since the structure of logarithmic terms of predictions for jet rates
is commonly more involved than that of most event shapes, until very
recently, only next-to-double logarithmic corrections $\alpha_s^n
L^{2n-1}$ to these observables were known~\cite{Catani:1991hj}. In
this study, we focus on jet rates obtained with the Durham clustering
algorithm~\cite{Catani:1991hj} and will use for the first time NNLL
predictions for the two-jet rate $R_2(y)$, which became available in
Ref.~\cite{Banfi:2016zlc}.

The main result of this paper is an extraction of the strong
coupling from the Durham two-jet rate to data measured at the LEP and
PETRA colliders which relies on N$^{3}$LO+NNLL accurate theoretical
predictions. As an additional result, we also present a fit based
simultaneously on the two- and three-jet rates, where the latter is
computed at NNLO accuracy.  For the first time, a fit of the coupling
based on the two-jet rate features perturbative uncertainties that are
considerably smaller than those of hadronization. This is due to the
accurate N$^{3}$LO+NNLL prediction adopted in the extraction.


\section{Observables and predictions}                      
\subsection{Jet rates}                      
A jet clustering algorithm is a procedure to classify final-state events
into different jet multiplicities. This categorisation depends on the
underlying algorithm used. In this paper we adopt the
\durham
clustering~\cite{Catani:1991hj}.  This is a sequential recombination
algorithm, which requires a distance measure in phase space between momenta%
\footnote{These momenta belong to either particles, or pseudo-jets
obtained during the recombination process.} $p_i^\mu$ and $p_j^\mu$,
\begin{equation}
y_{ij} = 2\frac{\min(E_i^{2}, E_j^{2})}{E^2_\text{vis}} (1- \cos\theta_{ij}),
\end{equation}
where $\theta_{ij}$ is the angle between the spatial components of the
pair and $E_\text{vis}$ the visible energy in the event.  If the
smallest of these, $y_{\min} = \min(\{y_{ij}\})$ is below a
pre-defined number, $y_\text{cut}$, then the corresponding pair of momenta is
recombined into a single one.  The procedure continues until all
distance measures become larger than $y_\text{cut}$. The momenta are
recombined using some recombination scheme. Here we adopt the
$E$-scheme~\cite{Catani:1991hj}, according to which the four-momenta
of the two clustering particles are simply added together.
The $n$-jet rate is then defined as 
\begin{equation}
R_n(y) = \frac{\sigma_{n\text{-jet}}(y)}{\sigma_{\text{tot}}}\,,
\end{equation}
where $\sigma_{n\text{-jet}}$ is the cross section for $n$-jet
production in hadronic final states obtained with the above algorithm
and $\sigma_{\rm tot}$ is the total hadronic cross section.

In the following sections we briefly review the fixed-order and
resummed predictions used in this study. The two results can be
combined by means of a matching procedure to obtain the predictions
that we ultimately use in the fit.

\subsection{Fixed-order predictions}
\label{sec:th:fo}
In this work we use predictions obtained with the CoLoRFulNNLO 
method~\cite{Somogyi:2006da,Somogyi:2006db,DelDuca:2016ily}. 
The perturbative expansion of the $n$-jet rate $R_{n}$ 
as function of $\ycut$ at the default renormalization scale $\mu_{\rm ren}=Q$ reads
\begin{align}\begin{split}
R_n(\ycut) = \delta_{2,n}&+
        \asbar{Q}  A_{n}(\ycut)
        +
        \left(\asbar{Q}\right)^2  B_{n}(\ycut)
        +\epjcbreak{&+}
        \left(\asbar{Q}\right)^3  C_{n}(\ycut)
        +
        \Oa{4}\,,\draftbreak{&}
\end{split}\end{align}
where $A$, $B$ and $C$ are the perturbative coefficients computed by
the \prog{MCCSM} code~\cite{Kardos:2016pic}.  For massless quarks, these
coefficients are independent of $Q$. The renormalization scale
dependence of the fixed-order prediction can be restored using the
renormalization group equation for $\as$,
\begin{align}\begin{split}
R_{n}(\ycut,\mu_{\rm ren}){} =\delta_{2,n} &+
        \asbar{\mu_{\rm ren}}  A_{n}(\ycut,\xiR)+
        \epjcbreak{&+}       
        \left(\asbar{\mu_{\rm ren}}\right)^2  B_{n}(\ycut,\xiR) +
        \\&+
        \left(\asbar{\mu_{\rm ren}}\right)^3  C_{n}(\ycut,\xiR)+
         \Oa{4}\,,
\end{split}\end{align}
where
\begin{align}\begin{split}
A_{n}(\ycut,\xiR) &= A_{n}(\ycut)\,,
\\
B_{n}(\ycut,\xiR) &= B_{n}(\ycut) + A_{n}(\ycut) \frac{1}{2}\beta_0 \ln\xiRsq\,,
\\
C_{n}(\ycut,\xiR) &= C_{n}(\ycut) + B_n(\ycut) \beta_0 \ln\xiRsq
        +\epjcbreak{&+} A_{n}(\ycut) \bigg(\frac{1}{4} \beta_1 \ln\xiRsq 
        + \frac{1}{4}\beta_0^2 \ln^2\xiRsq \bigg)\,,
\end{split}
\label{eq:ABC}
\end{align}
with $\xiR = \mu_{\rm ren}/Q$, $\beta_0 = (11\CA - 4\Nf \TR)/3$ and $\beta_1 = (34 \CA^2 - 20 \CA \TR \Nf - 12 \CF \TR \Nf)/3$. 
The numerical values for fixed-order coefficients for $R_3$, $R_4$ and
$R_5$ are reported in Appendix~\ref{app:PTcoeffs},
Tab.~\ref{tab:predictions}. 
These can be used to build up the
necessary fixed-order predictions for $R_2$ that we use in the following.

\subsection{Resummed predictions}
\label{sec:th:res}

The resummation technique adopted here was formulated in
Refs.~\cite{Banfi:2014sua,Banfi:2016zlc,Banfi:2018mcq}, hence we
present only the main features of these results.  The essence of
the procedure described in Ref.~\cite{Banfi:2014sua} is that the NLL
cross section is given by all-order configurations made of partons
independently emitted off the Born legs and widely separated in
angle~\cite{Banfi:2001bz}. The NNLL corrections are obtained by
correcting a {\it single} parton of the above ensemble to account for
all kinematic configurations that give rise to NNLL
effects~\cite{Banfi:2016zlc}.
The two-jet rate at NNLL can be written as
\begin{align}
\label{eq:y3-cs}
   R_2(\ycut)  \draftonly{&}\arxivonly{&} = e^{-R_{\NNLL}(\ycut)} \left[
    \left(1+ \frac{\as(Q)}{2\pi} H^{(1)}+ \right. \right. \notag &\\ 
    \draftonly{&}\arxivonly{&} \left. \left.  + \frac{\as(Q \sqrt{\ycut})}{2\pi} C^{(1)}_{\rm hc}
    \right)
    \mathcal{F}_{\NLL}(\ycut) + \frac{\as(Q)}{\pi}\delta{\mathcal F}_{\rm NNLL}(\ycut)
  \right]\epjconly{&},
  \end{align}
  where the Sudakov radiator $R_{\NNLL}$ and the coefficients
  $H^{(1)}$, $C^{(1)}_{\rm hc}$ are defined in
  Ref.~\cite{Banfi:2018mcq}\footnote{This observable corresponds to
    setting $a=2$ and $b_\ell=0$ in the corresponding equations.}, while
  the functions $\mathcal{F}_{\NLL}$ and
  $\delta{\mathcal F}_{\rm NNLL}$ are given in
  Refs.~\cite{Banfi:2014sua,Banfi:2016zlc}.

For the two-jet rate $R_{2}$ the resummation is performed with the
\prog{ARES} program~\cite{Banfi:2016zlc} and the matching to
fixed-order is done according to the $\ln R$
scheme~\cite{Catani:1992ua}.

The resummation of the three-jet rate is much more involved due to the
extra number of emitting particles. Accordingly, the state-of-the-art
resummed predictions have a much lower logarithmic accuracy and
include only terms ${\cal O}(\as^n L^{2n})$ and
${\cal O}(\as^n L^{2n-1})$ in $R_3(y)$~\cite{Catani:1991hj}, in
contrast to the logarithmic counting that we introduced for $R_2$
which refers to the logarithm of the jet rate. While $R_3$ is more
sensitive to $\as$ than $R_2$, the low theoretical accuracy of the
resummation does not guarantee a good theoretical control in the
region where the logarithms are large. Therefore, for the present
analysis, we do not perform any resummation of $R_3$ and limit the fit
to a range where the fixed order is reliable.


\subsection{Effects of quark masses}
The effect of the non-vanishing $b$-quark mass on the predictions has
also been considered in the literature. In particular, predictions for
$e^{+}e^{-}\to$ partons are known including $\as^2$ corrections with
massive $b$-quarks~\cite{Nason:1997tz}.  The resummed predictions for
the Durham $R_{2}$ and $R_{3}$ observables with non-zero $b$-quark
masses are only known at next-to-double logarithmic
accuracy~\cite{Krauss:2003cr} (i.e. $\as^n L^{2n-1}$ in the jet-rates)
and are not used in this analysis as this does not match our target
accuracy needed to guarantee a robust theoretical control even in the
region where the logarithms become large.

In order to take $b$-quark mass corrections into account, we subtract
the fraction of $b$-quark events $r_b(Q)$ from the massless result and
add back the corresponding massive contribution computed at fixed
order. Hence, we include mass effects directly at the level of the
final distributions according to the formulae
\begin{align}
R_{2}(y) &= 
        R_{2}^{\mathrm{N^3LO+NNLL}}(y)_{m_b=0}(1 - r_b(Q)) \nonumber 
        +\epjcbreak{&+} r_b(Q)R_{2}^{\mathrm{NNLO}}(y)_{m_b \ne 0 }\,, \\        
R_{3}(y) &= 
        R_{3}^{\mathrm{NNLO}}(y)_{m_b=0}(1 - r_b(Q)) 
        +\epjcbreak{&+} r_b(Q)R_{3}^{\mathrm{NLO}}(y)_{m_b \ne 0 }\,.
\label{eq:mb-corr-base}
\end{align}

The latter quantities are obtained by combining the total cross
section at NNLO including mass corrections as obtained from
Ref.~\cite{Chetyrkin:2000zk}, and the three- and four-jet rate
${\cal O}(\as^2)$ predictions as computed with the \prog{Zbb4}
program~\cite{Nason:1997tz,Nason:1997nw}. The strong coupling used in
the prediction of \prog{Zbb4} is then converted into a 5-flavour
coupling by means of the matching relation for
$\alpha_s(m_b)$~\cite{Abbiendi:1999fs,Chetyrkin:2000yt}.

We define the fraction of $b$-quark events as the ratio 
of the total $b$-quark production cross section divided by the total 
hadronic cross section,
\begin{equation}
r_b(Q) \equiv \frac{\tsig{}{m_b \ne 0}(e^+e^- \to b\bar{b})}
        {\tsig{}{m_b \ne 0}(e^+e^- \to \mathrm{hadrons})}\,.
\end{equation}
We evaluate the ratio of these cross sections to approximate ${\cal
O}(\as^3)$ according to Ref.~\cite{Chetyrkin:2000zk}.

For the bottom quark we used a pole mass of $m_b = 4.78\GeV$, which is
consistent with the corresponding world average~\cite{PDG2018}.

\section{Determination of $\as$}
\label{sec:fit}
To extract the strong coupling we compare the theory predictions
described above to the available data, taking into account the
non-perturbative corrections from Monte Carlo (MC) models, as
described in detail in the following.
\subsection{Data sets}
\label{sec:data}
We select experimental data that satisfy the following basic
requirements: (i) measurements are obtained with both charged and
neutral final state particles, (ii) corrections for detector effects
have been taken into account, and (iii) corrections for initial-state
QED radiation have been taken into account.  We found that the data
from Refs.~\cite{Buskulic:1992hq,Ackerstaff:1997kk,Alexander:1996kh,Pfeifenschneider:1999rz,Acton:1992fa,Schieck:2012mp,Achard:2004sv,Abreu:1996mk,Heister:2003aj}
satisfy these basic criteria.

However, the measurements of Ref.~\cite{Ackerstaff:1997kk} and
Ref.~\cite{Alexander:1996kh} are superseded by the measurements in
Ref.~\cite{Pfeifenschneider:1999rz} and hence not included in our
analysis.  The analysis in Ref.~\cite{Acton:1992fa} was excluded as
the provided combined uncertainties are much lower than those from
later refined analyses and are close to expected statistical
uncertainties estimated from earlier analysis in
Ref.~\cite{Akrawy:1990ac}.
We also excluded the measurements from
Ref.~\cite{Schieck:2012mp} as these contain explicit corrections for
the contributions of the process $e^{+}e^{-}\to  b\bar{b}$ from
MC simulations and cannot be treated in the same way as the
data without such corrections. 

From the selected analyses we also excluded the measurements at
$\sqrt{s}=172\GeV$ from Ref.~\cite{Achard:2004sv} as the background
subtraction procedure performed with data of limited statistics has
biased the measurement, e.g.\ introduced a non-monotonous behaviour of
$R_2(y)$, see Ref.~\cite{Acciarri:1997xr} for details.  For similar
reasons we exclude data from Ref.~\cite{Abreu:1996mk}. The sum of the
rates for the data sets in Ref.~\cite{Heister:2003aj} deviates from
unity, and the largest deviation reaches almost $0.03$ for the data
set at $\sqrt{s}=200\GeV$. For this reason we excluded this data from
the fits as well.

We summarise the information on the selected data sets in
Tab.~\ref{tab:input}.  For some data sets the uncertainties were
updated before the fit, as follows.  We added in quadrature the two
available systematic uncertainties for the measurements at
$\sqrt{s}=91\GeV$ from Ref.~\cite{Heister:2003aj}.  The data from
JADE~\cite{Pfeifenschneider:1999rz} does not include systematic
uncertainties related to the choice of MC samples used for the
calculations of detector corrections. Later studies with the same
data~\cite{Schieck:2012mp} indicated that such uncertainties can be at
least as large as the statistical uncertainties.  Therefore, in this
case we added a relative uncertainty of $1.5\%$ as an estimation of
missing systematic uncertainties.

\renewcommand{\arraystretch}{1.0}
\begin{table}[!htbp]\centering\small
\begin{tabular}{|c|c|c|c|}\hline
           &                  Data &          MC  &         \\
Experiment & $\sqrt{s},$            &$\sqrt{s},$    &Events         \\
           & $\GeV$               &    $\GeV$     &   \\
\hline\hline
\JRTtabularinput\hline
\end{tabular}
\caption{
  Data used for the determination of $\as$ in this work.  The ranges of collision 
  energies,  their weighted average value (in brackets) and the number of events for 
  each  experiment are given as quoted in the original publications.}
\label{tab:input}
\end{table}

The measurements of the jet rates selected for the analysis are
provided in the original publications without correlations between the
individual points and without decomposition of the total systematic
uncertainties. To perform an accurate extraction procedure, we
examined the available data and uncertainties and built a covariance
matrix for all measured sets of data. This procedure consists of
multiple steps.

In the first step we estimated the statistical correlation matrix of
the individual points in the fit range as described in
Ref.~\cite{Verbytskyi:2016ymc} from the MC generated samples.  We find
that these are in good agreement with statistical correlation matrices
obtained from data in unpublished Ref.~\cite{Verbytskyi201813}.  In the second
step we built the full covariance matrix for each measurement from the
correlation matrix, statistical and systematic uncertainties.

As the original publications do not contain enough information, the
procedure is based on some assumptions.  Namely, for a pair of
measurements $R_{n}(y_{1})$ and $R_{m}(y_{2})$ we make the following
assumptions on the correlation coefficients of the systematic
uncertainties: (i) ${\rm corr}_{{\rm syst}}[R_{n}(y_{1}),R_{m}(y_{2})]
= {\rm corr}_{{\rm stat}}[R_{n}(y_{1}),R_{m}(y_{2})]\times K \times
\rho^{|\log(y_{1})-\log(y_{2})|}$
for $|\log(y_{1})-\log(y_{2})|<\ydistance$ and (ii) ${\rm corr}_{{\rm
syst}}(R_{n}(y_{1}),R_{m}(y_{2}))=0$ for
$|\log(y_{1})-\log(y_{2})|>\ydistance$.  We selected
$\rho=\rhonominal$ in order to mimic patterns of systematic
uncertainties observed in Ref.~\cite{Verbytskyi201813}.  This
corresponds to ${\rm corr}_{{\rm
syst}}[R_{n}(y_{1}),R_{m}(y_{2})]\approx 0.5 K \, {\rm corr}_{{\rm
stat}}[R_{n}(y_{1}),R_{m}(y_{2})]$ for measurements with
$|\log(y_{1})-\log(y_{2})|=0.125$, i.e.\ in the neighbouring bins for
typical binning used in the measurements.  $K$ is equal to $1$ if $n=m$ and $\Knominal$
otherwise.  This approach approximates correlations between
observables $R_2$ and $R_3$.

\FloatBarrier 
\subsection{Monte Carlo event generation setup}
\label{sec:mcsetup}

In our analysis we model non-perturbative effects in the
$e^{+}e^{-}\to$ hadrons process using state-of-the-art particle level
MC event generators. As usual, we estimate the non-perturbative
corrections of the jet rate distributions by comparing distributions
at hadron and parton level in the simulated samples. In particular, we
used the \prog{Herwig7.1.4}~\cite{Bellm:2015jjp} MC event generator to 
deliver our final results, and the 
\prog{Sherpa2.2.6}~\cite{Gleisberg:2008ta} MC event generator for
cross-checks.

We generated event samples for the process $e^{+}e^{-}\to$ hadrons  at the
centre-of-mass energies listed in Tab.~\ref{tab:input}.  In all cases,
we switched off the simulation of initial state radiation and used
default generator settings, unless stated otherwise. We used $\as(M_{Z})=0.1181$~\cite{PDG2018} as input for the generation. 
Moreover, we
adopted the $G_\mu$ scheme with input parameters $M_{Z}=91.1876\GeV$,
$M_{W}=80.379\GeV$ and $G_{F}=1.1663787\times10^{-5} \GeV^{-2}$. The pole masses of $b$- and
$t$-quarks were set to $4.78\GeV$ and $173\GeV$ respectively.

The \prog{Herwig7.1.4} samples were generated with the unitarised
MENLOPS method~\cite{Platzer:2012bs} using the 
\prog{MadGraph5}~\cite{Alwall:2011uj} matrix element generator and the
\prog{OpenLoops}~\cite{Cascioli:2011va} one-loop library to produce
matrix elements for the 2, 3, 4 and 5 parton final states in the hard
process.  The two- and three-parton final state matrix elements have
again NLO accuracy in perturbative QCD and the QCD matrix elements
were also calculated using massive $b$-quarks.  The merging parameter
was set to $\sqrt{s}\times10^{-1.25}$.

To test the fragmentation and hadronization model dependence, the
events generated with \prog{Herwig7.1.4} were hadronized using either
the cluster fragmentation model~\cite{Webber:1983if} or the Lund
string fragmentation model~\cite{Andersson:1983ia}. The cluster
fragmentation model is natively implemented in \prog{Herwig7.1.4}. The
Lund string fragmentation model is implemented in
\prog{Pythia8.2.35}~\cite{Sjostrand:2007gs} and was used in
\prog{Herwig7.1.4} via the \prog{ThePEG1.6.1++}~\cite{Lonnblad:2006pt}
toolkit.  For both setups the hadron decays were performed by the 
\prog{EvtGen1.7.0}~\cite{Lange:2001uf} package.  We label the first
setup $H^{C}$ and the second one $H^{L}$.

The \prog{Sherpa2.2.6} samples were generated with the MENLOPS method
using the matrix element generators
\prog{AMEGIC}~\cite{Krauss:2001iv}, \prog{COMIX}~\cite{Duhr:2006iq}
and the \prog{OpenLoops}~\cite{Cascioli:2011va} one-loop library to
produce matrix elements of the processes
$e^{+}e^{-}\to Z/\gamma^*\to $ 2, 3, 4, 5 partons.  The two parton
final state matrix elements again have NLO QCD accuracy.  The QCD matrix
elements were calculated assuming massive $b$-quarks.  The merging
parameter $Y_{cut}$ was set to $10^{-2.5}$.  The events were
hadronized using the cluster fragmentation model~\cite{Winter:2003tt}
as implemented in \prog{Sherpa2.2.6}.  We label this setup $S^{C}$.

\subsection{Estimation of hadronization effects from MC models}
\label{sec:mchad}
The estimation of hadronization corrections is an integral part of
comparing the parton-level QCD predictions to the data measured at
hadron (particle) level. While the principle of local parton-hadron
duality leads to close values of observable quantities at parton and
hadron level, the difference between them is not negligible and must
be taken into account if one aims at a precise determination of $\as$.
One possible way to do this is to apply correction factors estimated
from MC simulations to the perturbative predictions.

To obtain the jet rates at parton and hadron level, we processed the
MC generated samples in the same way as for the data, using
undecayed/stable particles for hadron level calculations. To define
jets with the \durham clustering, we used the implementation of the
\prog{FastJet/fjcore3.3.0} package~\cite{Cacciari:2011ma}.  The
selected resulting distributions are shown in
Fig.~\ref{fig:partons:all} in Appendix~\ref{app:MC}.

The predictions obtained with all setups describe the data well for all 
ranges of $y$, with the exception of the regions of small $y$ at
all values of $\sqrt{s}$ (see  Figs.~\ref{fig:hadrons:one},
 \ref{fig:hadrons:two} and \ref{fig:hadrons:three}).

 Among all considered MC setups, $H^{L}$ was selected to be the
 reference.  The selected setup uses the very well tested Lund
 hadronization model that provides stable and physically reliable
 predictions throughout a wide range of centre-of-mass energies in
 $e^{+}e^{-}$ collisions.  Moreover, precisely this hadronization
 model was used for the modelling of $e^{+}e^{-}$ collisions in the
 original
 publications~\cite{Pfeifenschneider:1999rz,Heister:2003aj,Achard:2004sv}.

 To estimate the uncertainty related to the hadronization modelling,
 we consider half the difference between the $H^{L}$ and $H^{C}$
 setups, as explained in more detail in the following.
 
MC generators were already used to model hadronization effects in the
previous QCD analyses of $e^{+}e^{-}$
data~\cite{Kardos:2018kqj,Dissertori:2007xa}.  Typically, hadron
level predictions for every observable were obtained by multiplying
the perturbative predictions by some factor derived from the analysis
of MC generated events.  In the present analysis the approach was
amended to take into account physical constraints on $R_{n}$, namely
that jet rates are positive and that their sum should be one.  These constraints
are implemented by introducing the variables $\xi_1$ and $\xi_2$ such
that at parton level
\begin{equation}
R_{2}^{({\rm p})}=\cos^2\xi_1\,,
\quad
R_{3}^{({\rm p})}=\sin^2\xi_1\cos^{2}\xi_2
\quad
\text{and}
\quad\epjcbreak{\]\[}
R_{\geq 4}^{({\rm p})}=\sin^2\xi_1\sin^{2}\xi_2\,,
\label{eq:xiparton}
\end{equation} 
with the constraint
\begin{equation}
\label{eq:unitarity}
R_{2}^{({\rm p})} + R_{3}^{({\rm p})} + R_{\geq 4}^{({\rm p})} = 1\,.
\end{equation}
The corresponding relations at hadron level are
\begin{align}
R_{2}^{({\rm h})}&=\cos^2(\xi_1+\delta\xi_1)\,,
\quad
R_{3}^{({\rm h})}=\sin^2(\xi_1+\delta\xi_1)\cos^{2}(\xi_2+\delta\xi_2), \nonumber\\
R_{\geq 4}^{({\rm h})}&=\sin^2(\xi_1+\delta\xi_1)\sin^{2}(\xi_2+\delta\xi_2).
\label{eq:xihadron}
\end{align}
The functions $\delta\xi_1(y)$ and $\delta\xi_2(y)$ take into account
the non-perturbative corrections.  We obtained the numerical values
for $\delta\xi_1(y)$ and $\delta\xi_2(y)$ from the MC simulated
samples. In order to extract the hadronization corrections we proceed
as follows. For a given $y$ bin, we extract $\xi_1$ from the parton
level prediction for the two-jet rate, and $\xi_2$ from the same
prediction for the three-jet rate. The shifts $\delta\xi_1$ and
$\delta\xi_2$ are then extracted in the same fashion from the hadron
level predictions. The extracted values and corresponding interpolated
functions (splines) are shown in Fig.~\ref{fig:mchadr:all} in
App.~\ref {app:had} for selected centre-of-mass energies.
The size of the derived hadronization corrections can be read off
Fig.~\ref{fig:mchadrr:all}. As expected, we see that hadronization
corrections increase at small values of the two- and three-jet rates
and that the corrections become less important at higher energies.

\section{Fit procedure}
To find the optimal value of $\as$, we used the 
\prog{MINUIT2}~\cite{minuit} package to minimize
$\chi^2=\sum_{\rm data\ sets}\chi^2(\as)_{\rm data\ set},$
where $\chi^2(\as)$ was calculated for each data set as
\begin{equation}
\chi^2(\as)=\vec{r}\,V^{-1}\,\vec{r}^T,\qquad \vec{r}\equiv(\vec{D}-\vec{P}(\as)),
\label{chi2}
\end{equation}
where $\vec{D}$ stands for the vector of data points, $\vec{P}(\as)$
is the vector of theoretical predictions, and $V$ is the covariance
matrix for the experimental data $\vec{D}$.

We choose the fit range as follows. In order to assure that the
implementation of the hadronization corrections is unitary, i.e.\ the
constraint of eq.~\eqref{eq:unitarity} is satisfied, we set the upper bound
of the fit range below the kinematical limit for four jet production,
$\log_{10}(y) = \log_{10}(1/6)\simeq -0.8$. We therefore choose
$\log_{10}(y) = -1$ as an upper bound. Moreover, we adapt the lower
bound to the centre-of-mass energy in order to take into account that
hadronization corrections become more important at lower
energies. Accordingly, we fix the lower bound $\log_{10}{y_{\rm
min}(Q)}$ of the fit range as $\log_{10}{y_{\rm
min}(Q)}=\log_{10}{y_{\rm min}}(M_{Z})+{\cal L}$ with ${\cal L}
= \log_{10}(M_Z^2/Q^2)$.
Different values used for $\log_{10}(y_{\rm min}(M_{Z}))$ for $R_2$ and
$R_3$ are indicated in the first columns in
Tabs.~\ref{tab:result:full},~\ref{tab:result:rtwothree}
and~\ref{tab:result:rthree}.

\subsection{Fit of the coupling with the two-jet rate $R_2$}
To obtain the most precise results, we first concentrate on fits that
include the two-jet rate $R_2$ solely.  The results of the fits,
together with the used fit ranges, are given in
Tab.~\ref{tab:result:full}.  We show the results obtained via a fit of
$R_2$ both at N$^3$LO and N$^3$LO+NNLL, with different hadronization
models.  The corresponding values of $\chi^2$ divided by the number of
degrees of freedom ($\ndof$) in the global fit is also reported.  The
values $\chi^2/\ndof$ in the global fit as well as the values for
every data set (not shown) are all of order unity, which can be viewed
as a support for our correlation model. These values can be compared
to the ones obtained in similar analyses. For instance, in
Ref.~\cite{OPAL:2011aa}, the $\chi^2/\ndof$ for fits with statistical
uncertainties only varies for different observables between $0.5$ and
$60$.

From the results given in the table one notices that the effect of the
resummation is to move the fitted $\alpha_s(M_Z)$ to slightly lower
values. The quality of the fit, with or without resummation, is very
similar. The benefit of including the resummation will become evident
when perturbative uncertainties of the results are discussed.

\renewcommand{\arraystretch}{1.0}
\begin{table}[htbp]\centering
\addtolength{\tabcolsep}{-3pt}
\begin{tabular}{|c||c|c|}\hline
Fit ranges, $\log_{10}(y)$& N$^3$LO & N$^3$LO+NNLL \\ Hadronization &
$\chi^{2}/ndof$ & $\chi^{2}/ndof$ \\\hline
\JRTtabularresultNNLOandNNLOREStwo
\end{tabular}
\caption{
Fit of $\alpha_s(M_Z)$ from experimental data for $R_2$ obtained using
N$^3$LO and N$^3$LO+NNLL predictions, three different hadronization
models and four different choices of the fit range, as given in the
brackets, with ${\cal L} = \log_{10}(M_Z^2/Q^2)$.
The reported uncertainty is the fit 
uncertainty as given by \prog{MINUIT2}.
}
\label{tab:result:full}
\end{table}

As our best fit we quote the result obtained from the fits of the $R_2$
observable with the $H^{L}$ hadronization
model in the fit range \JRTfitrangethreeRtwo, that reads
\begin{equation}
\as(M_{Z})=
\JRTresultHNNLORtwo\,,
\label{eq:alphasfinal}
\end{equation}
where the quoted uncertainties are coming from  \prog{MINUIT2} $(exp.)$, variation of  
 renormalization scale $(ren.)$, variation of resummation scale $(res.)$ and choice of 
hadronization model $(hadr.)$.
The estimation of these uncertainties is described in the
following subsection.

\begin{figure}[htbp]\centering
\includegraphics[width=\FIGWIDTH,height=\FIGHEIGHT]{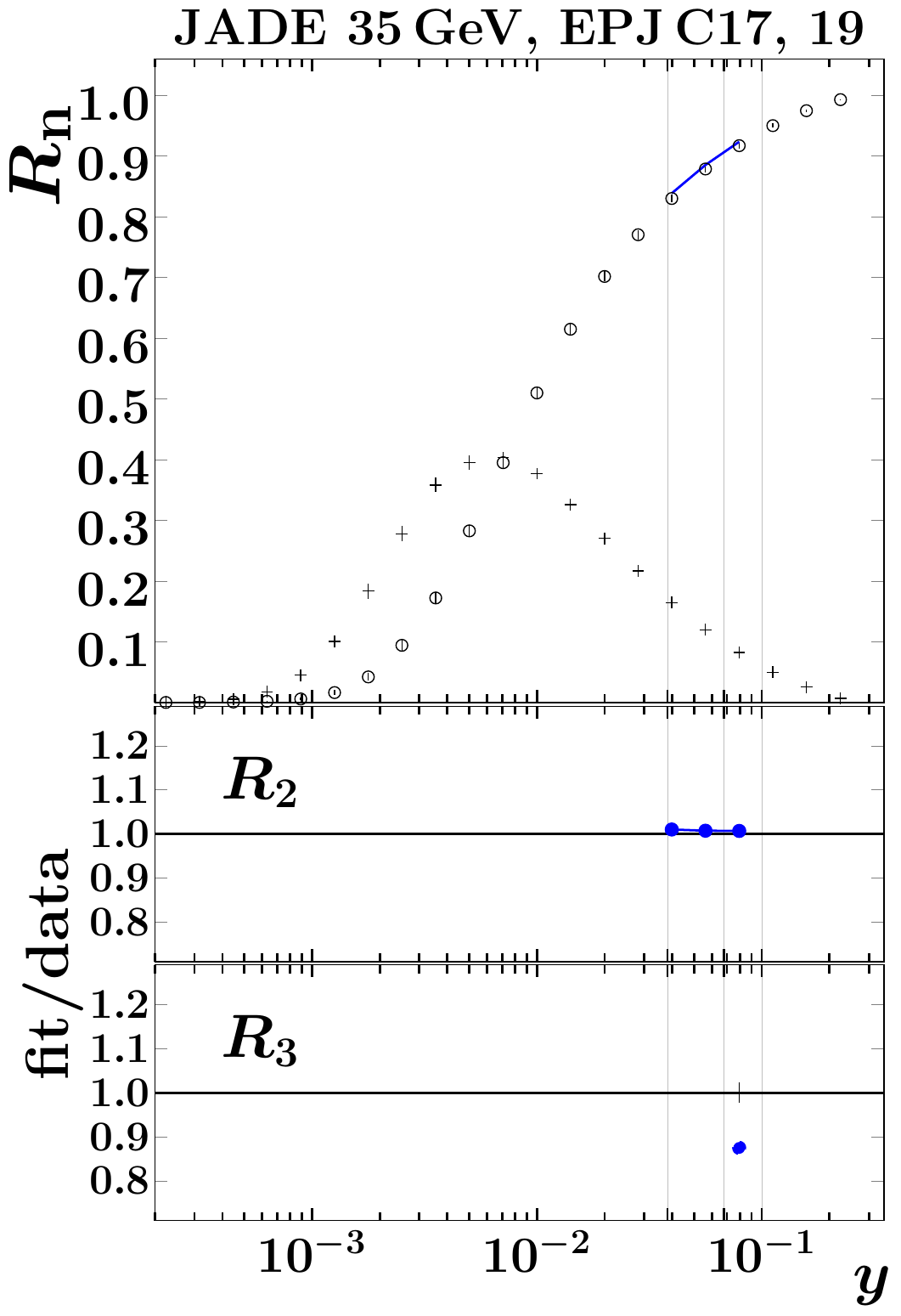}\includegraphics[width=\FIGWIDTH,height=\FIGHEIGHT]{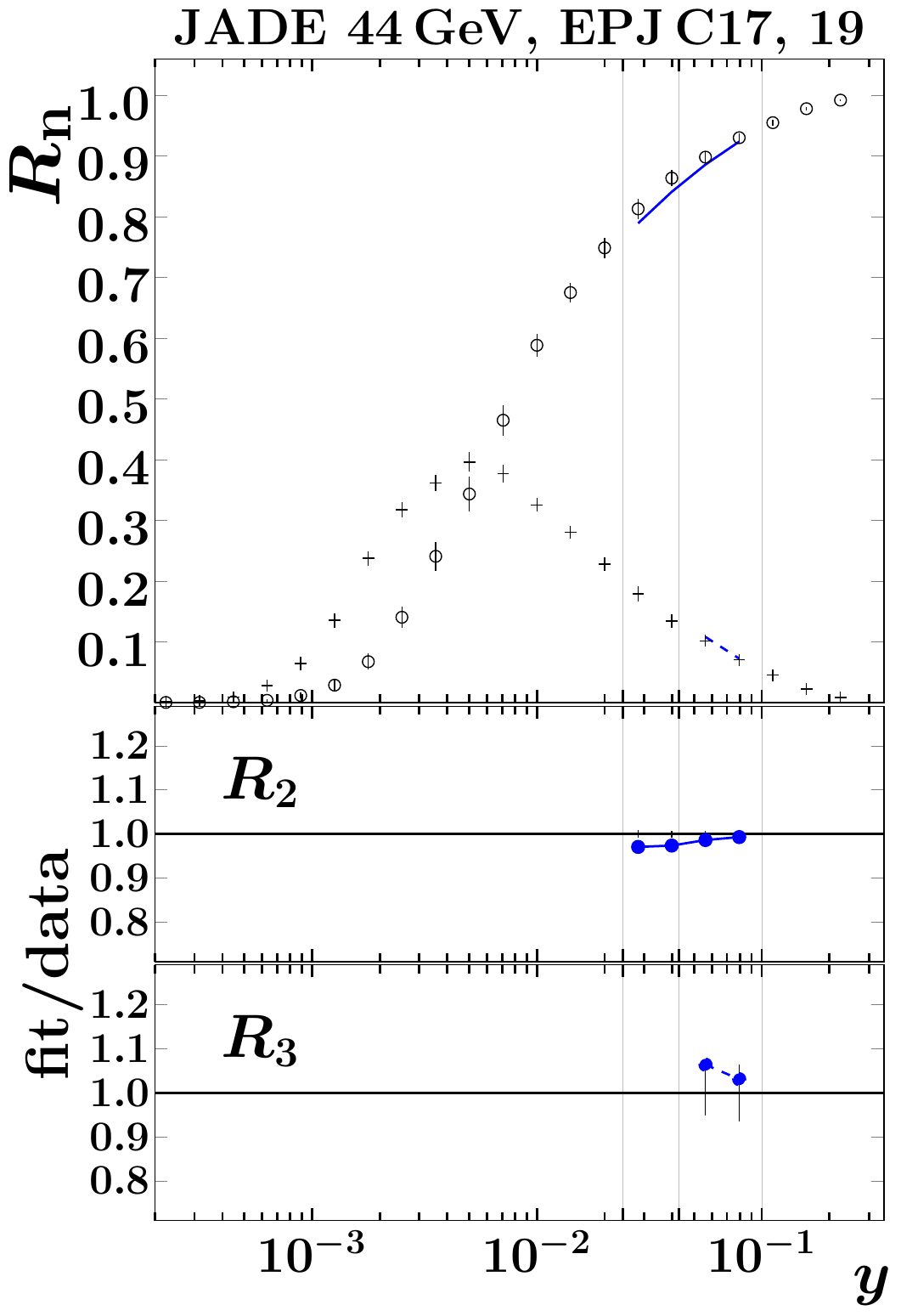}\epjcbreak{}\includegraphics[width=\FIGWIDTH,height=\FIGHEIGHT]{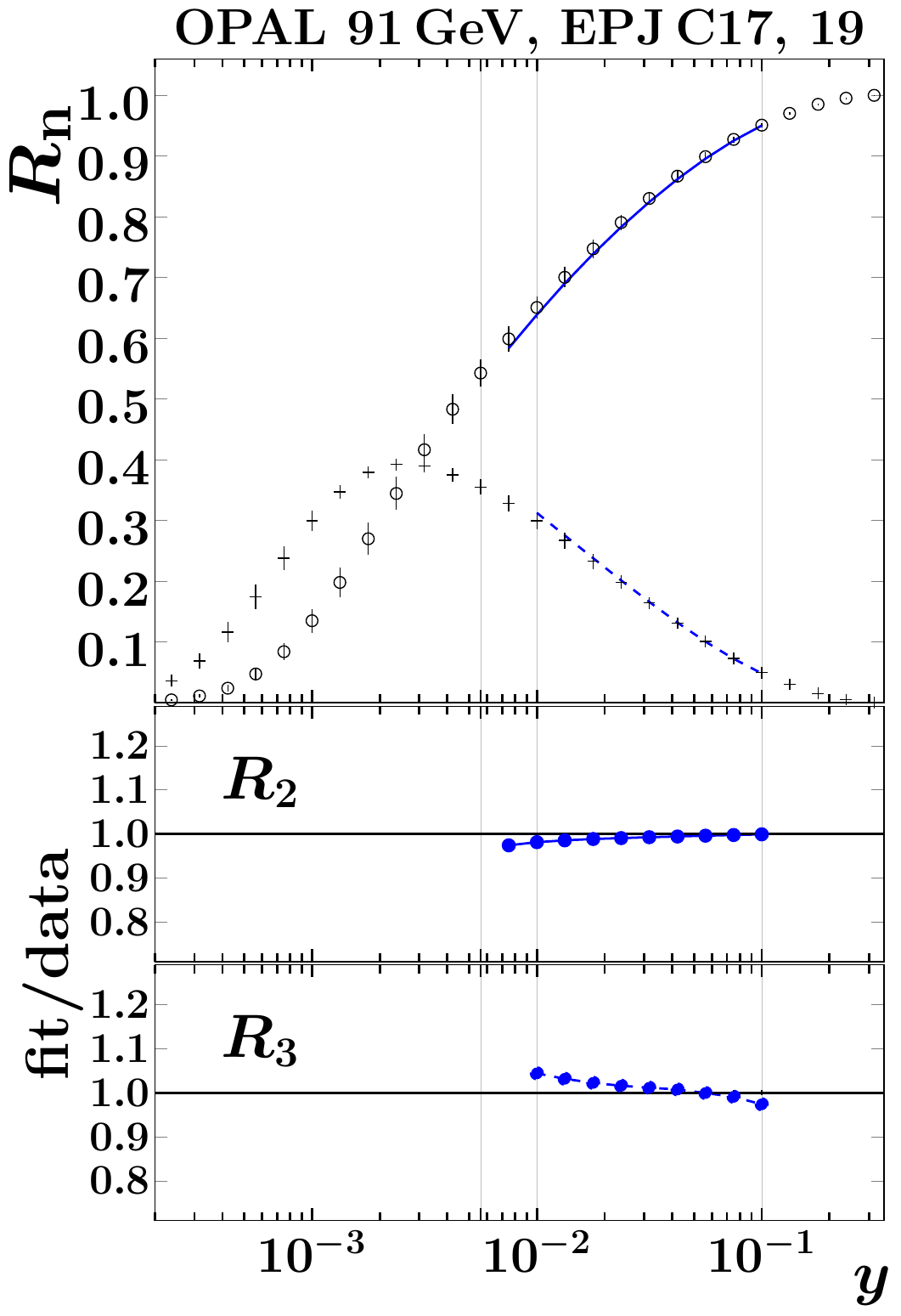}\draftbreak{}\arxivbreak{}\includegraphics[width=\FIGWIDTH,height=\FIGHEIGHT]{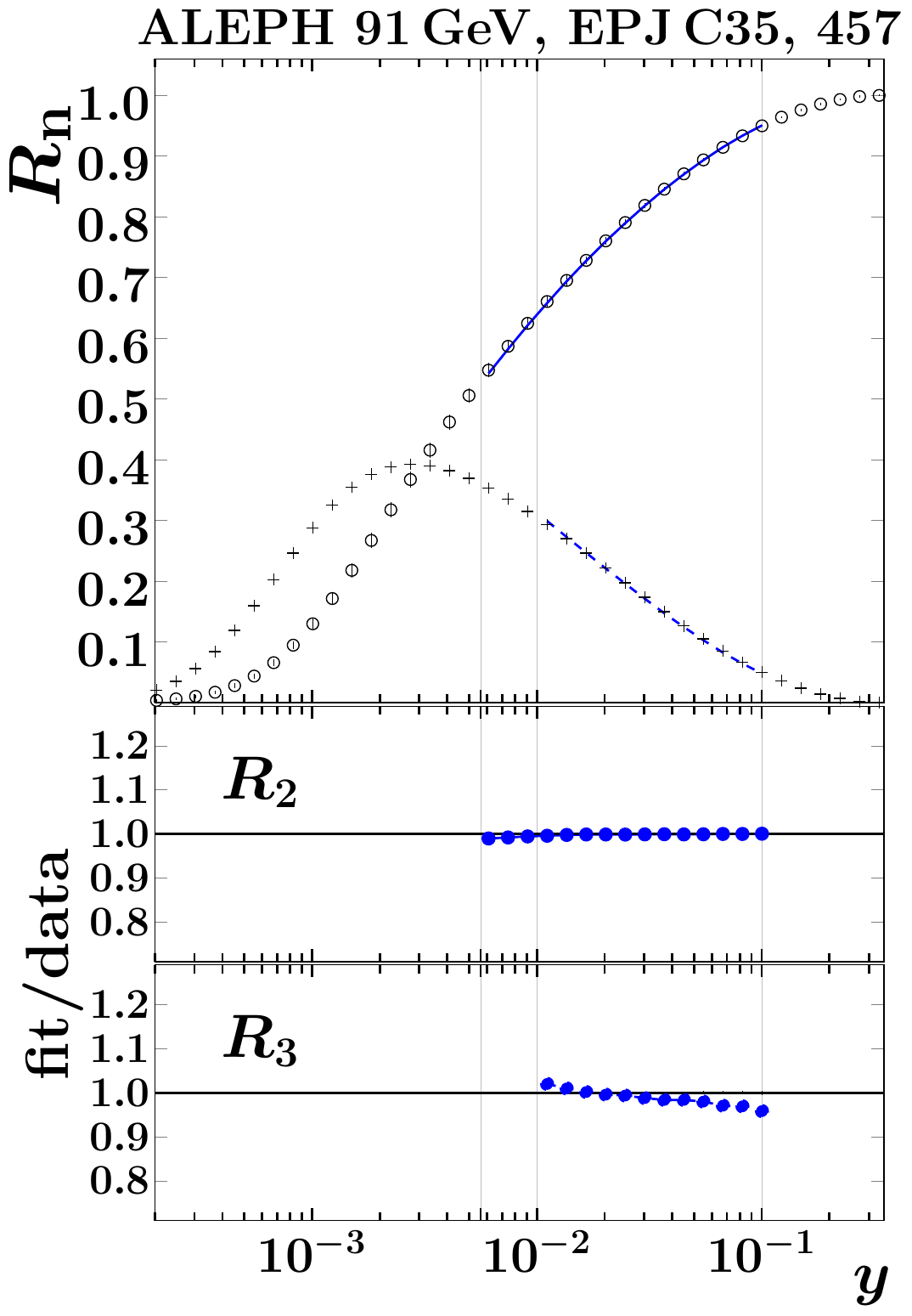}\epjcbreak{}\includegraphics[width=\FIGWIDTH,height=\FIGHEIGHT]{Figures/figresultOPAL-figure0.pdf}\includegraphics[width=\FIGWIDTH,height=\FIGHEIGHT]{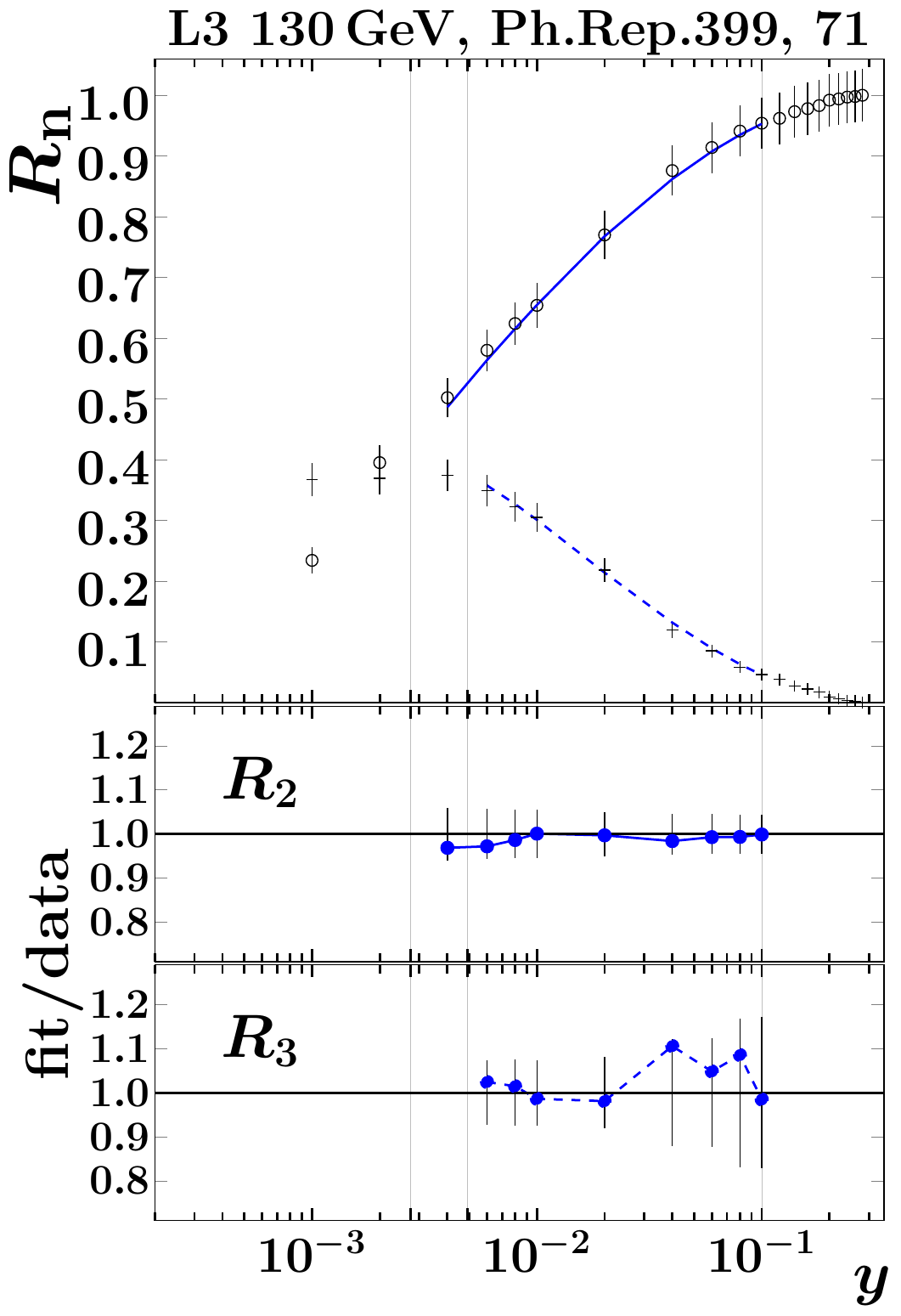}\draftbreak{}\arxivbreak{}\epjcbreak{}\includegraphics[width=\FIGWIDTH,height=\FIGHEIGHT]{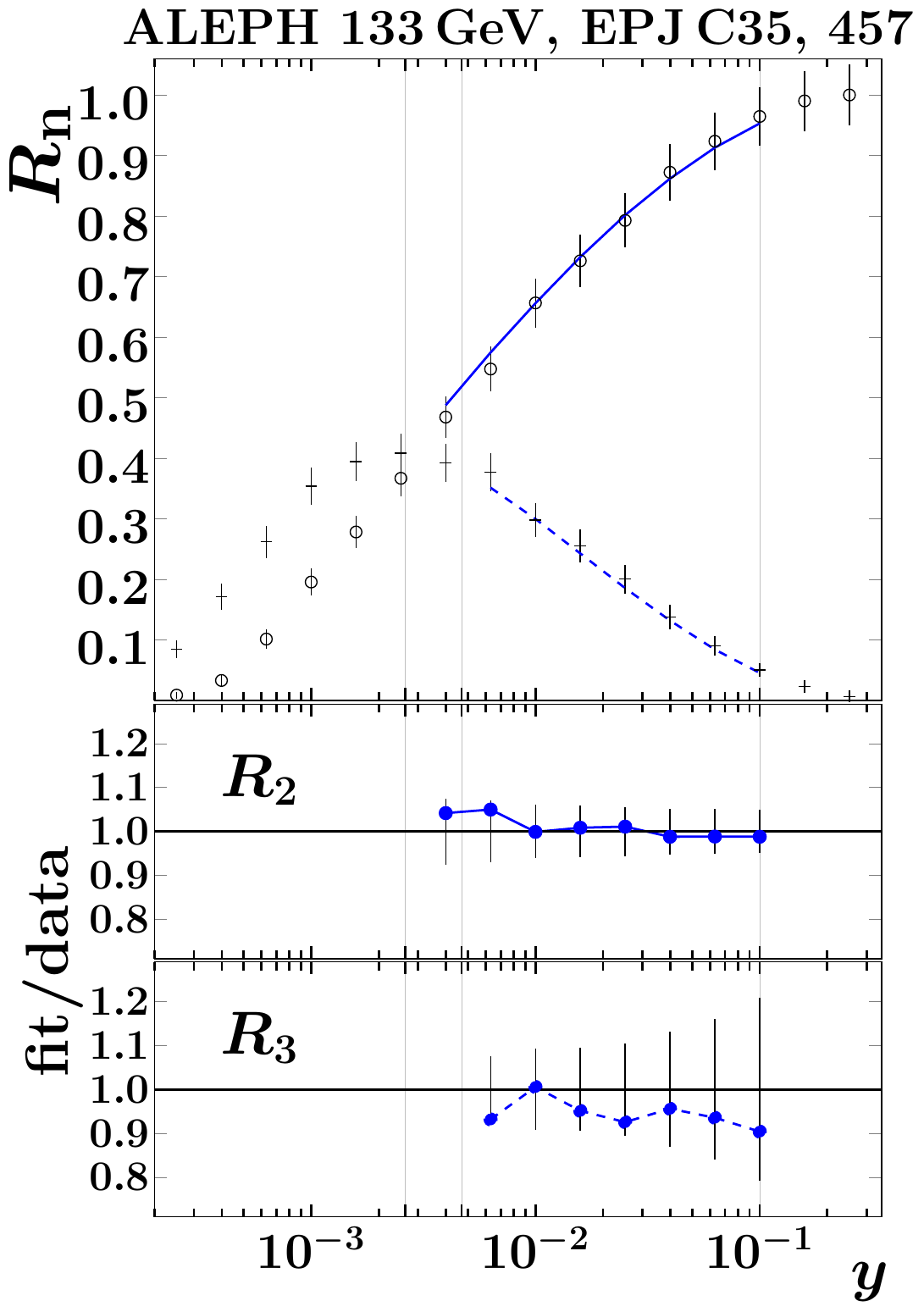}\includegraphics[width=\FIGWIDTH,height=\FIGHEIGHT]{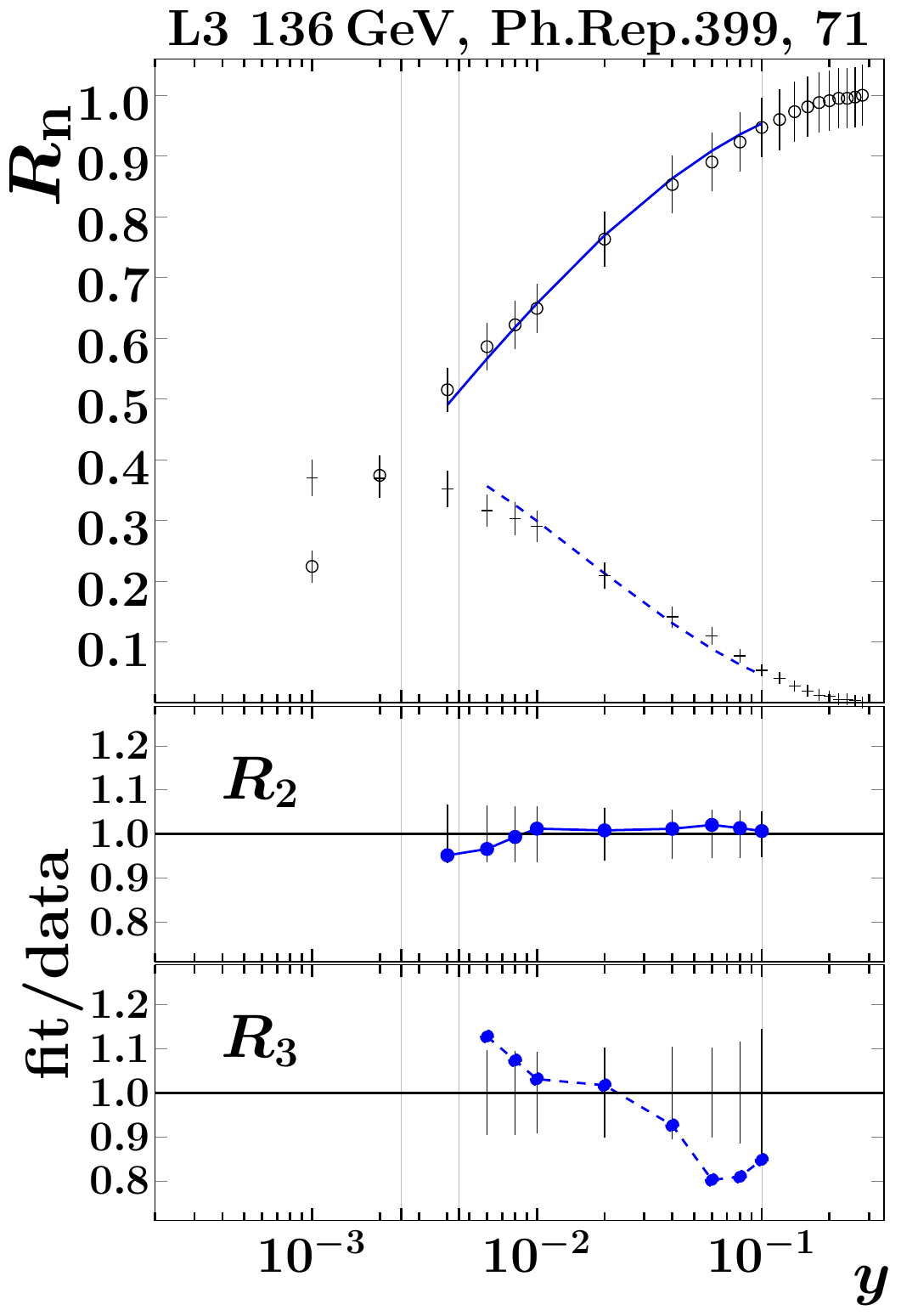}\includegraphics[width=\FIGWIDTH,height=\FIGHEIGHT]{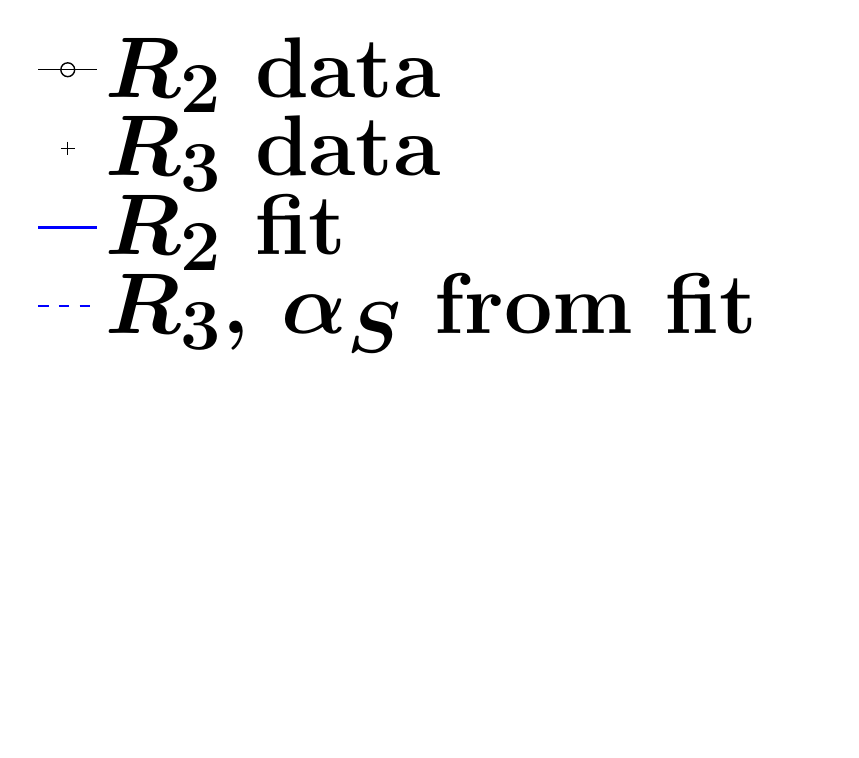}\\
\caption{Comparison of data and perturbative predictions supplemented by hadronization corrections in the $H^{L}$ model using for the strong coupling the value obtained from our global fit, eq.~\eqref{eq:alphasfinal}.} 
\label{fig:result:one}
\end{figure}

\begin{figure}[htbp]\centering
\includegraphics[width=\FIGWIDTH,height=\FIGHEIGHT]{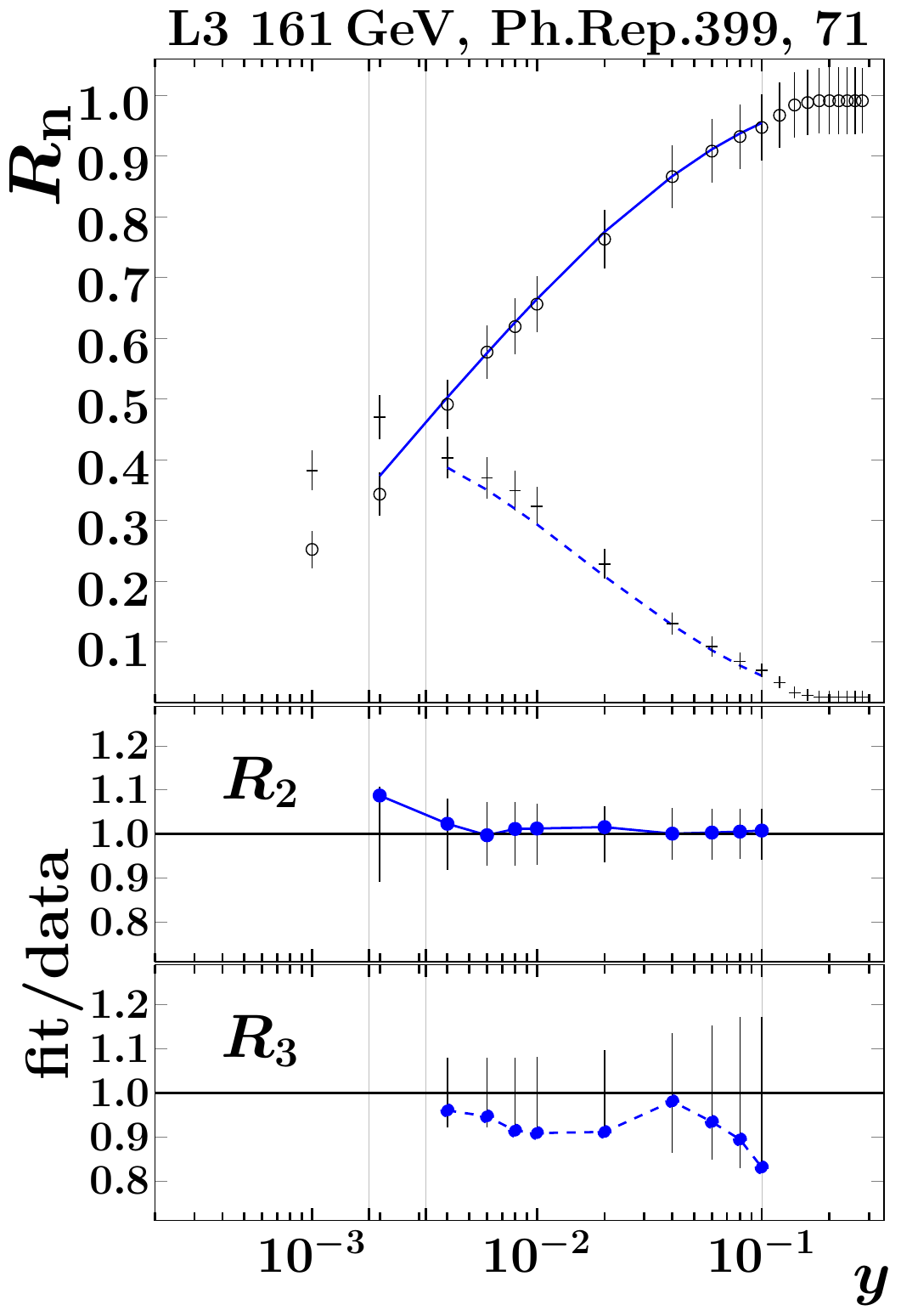}\includegraphics[width=\FIGWIDTH,height=\FIGHEIGHT]{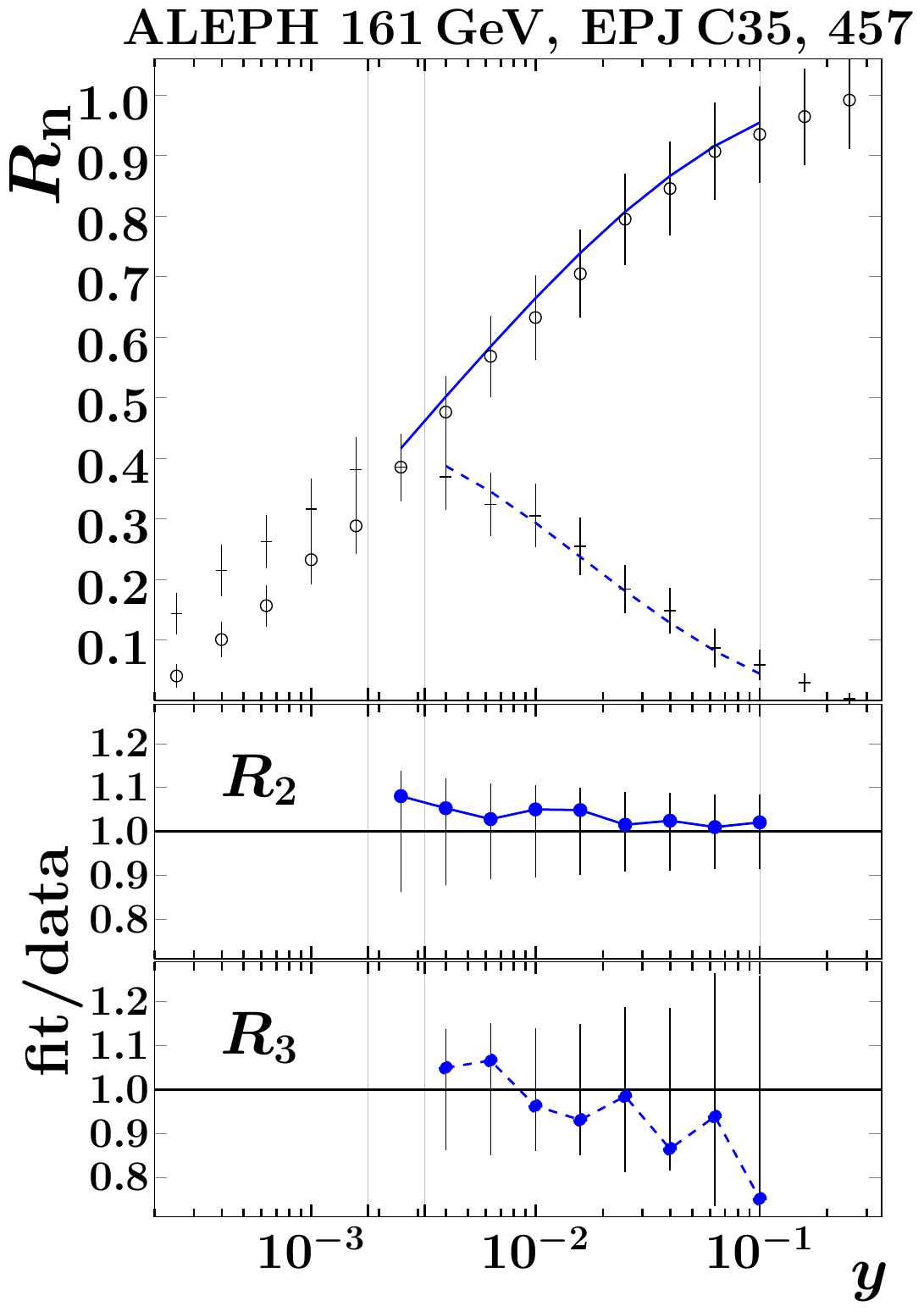}\epjcbreak{}\includegraphics[width=\FIGWIDTH,height=\FIGHEIGHT]{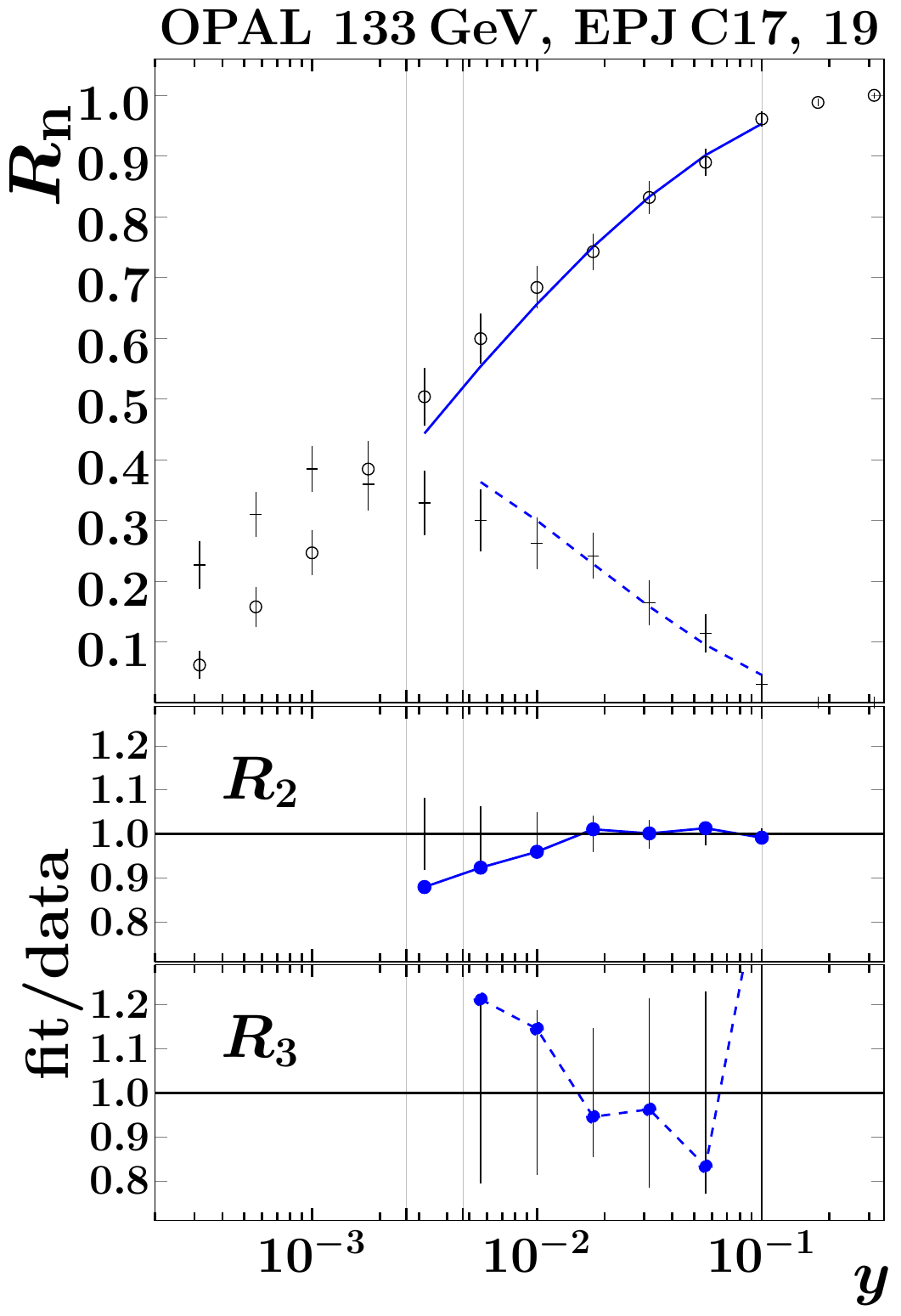}\draftbreak{}\arxivbreak{}\includegraphics[width=\FIGWIDTH,height=\FIGHEIGHT]{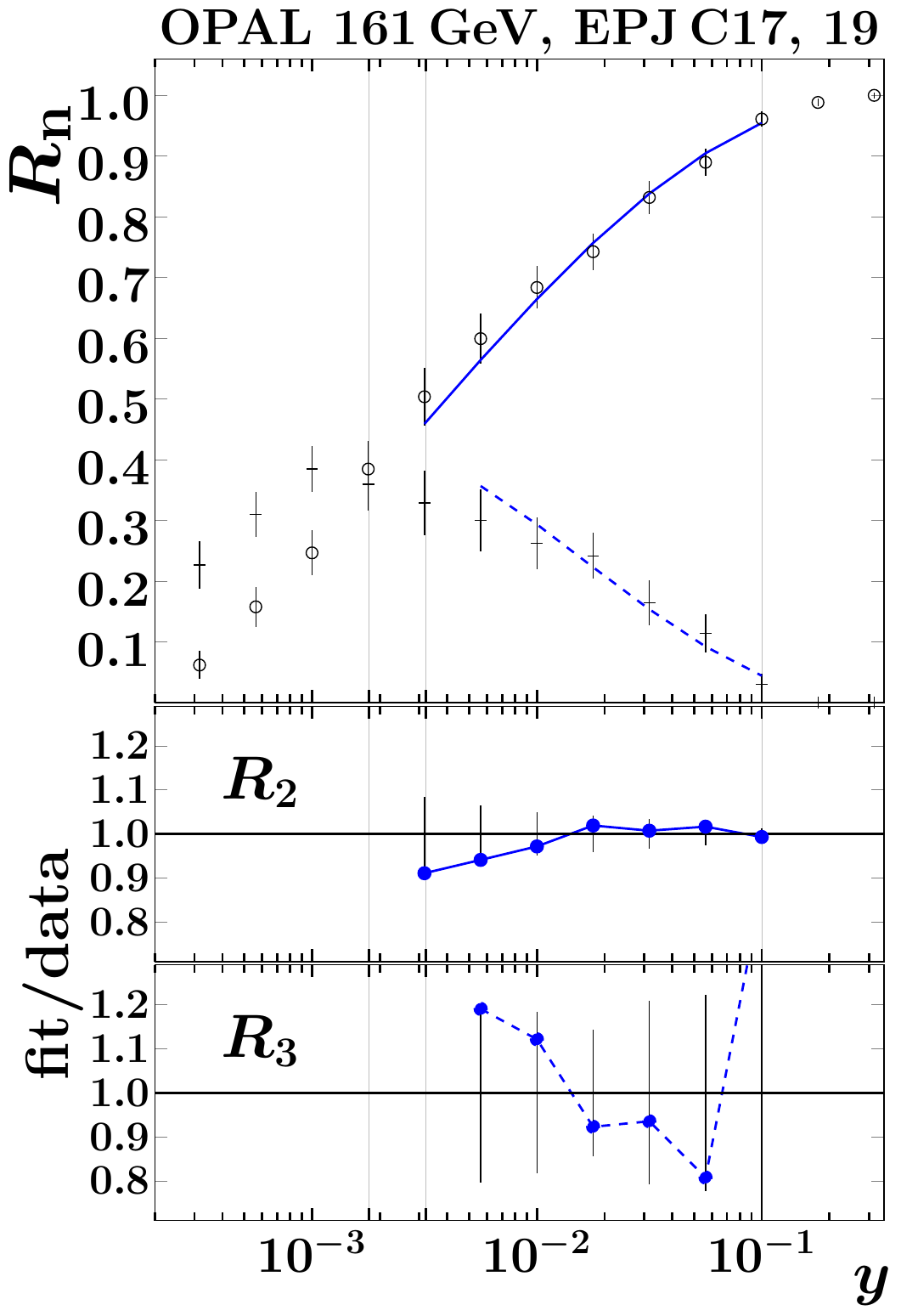}\epjcbreak{}\includegraphics[width=\FIGWIDTH,height=\FIGHEIGHT]{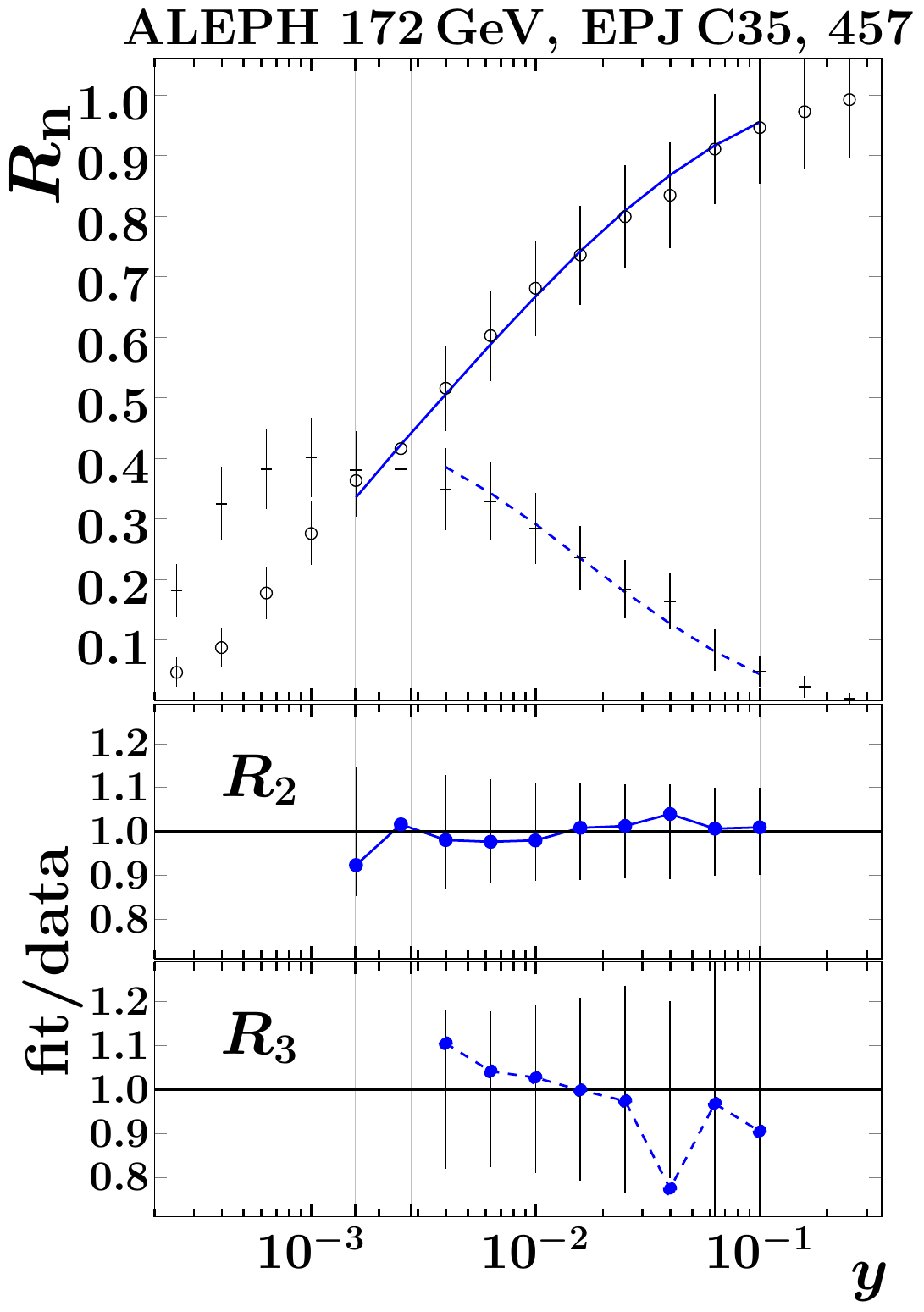}\includegraphics[width=\FIGWIDTH,height=\FIGHEIGHT]{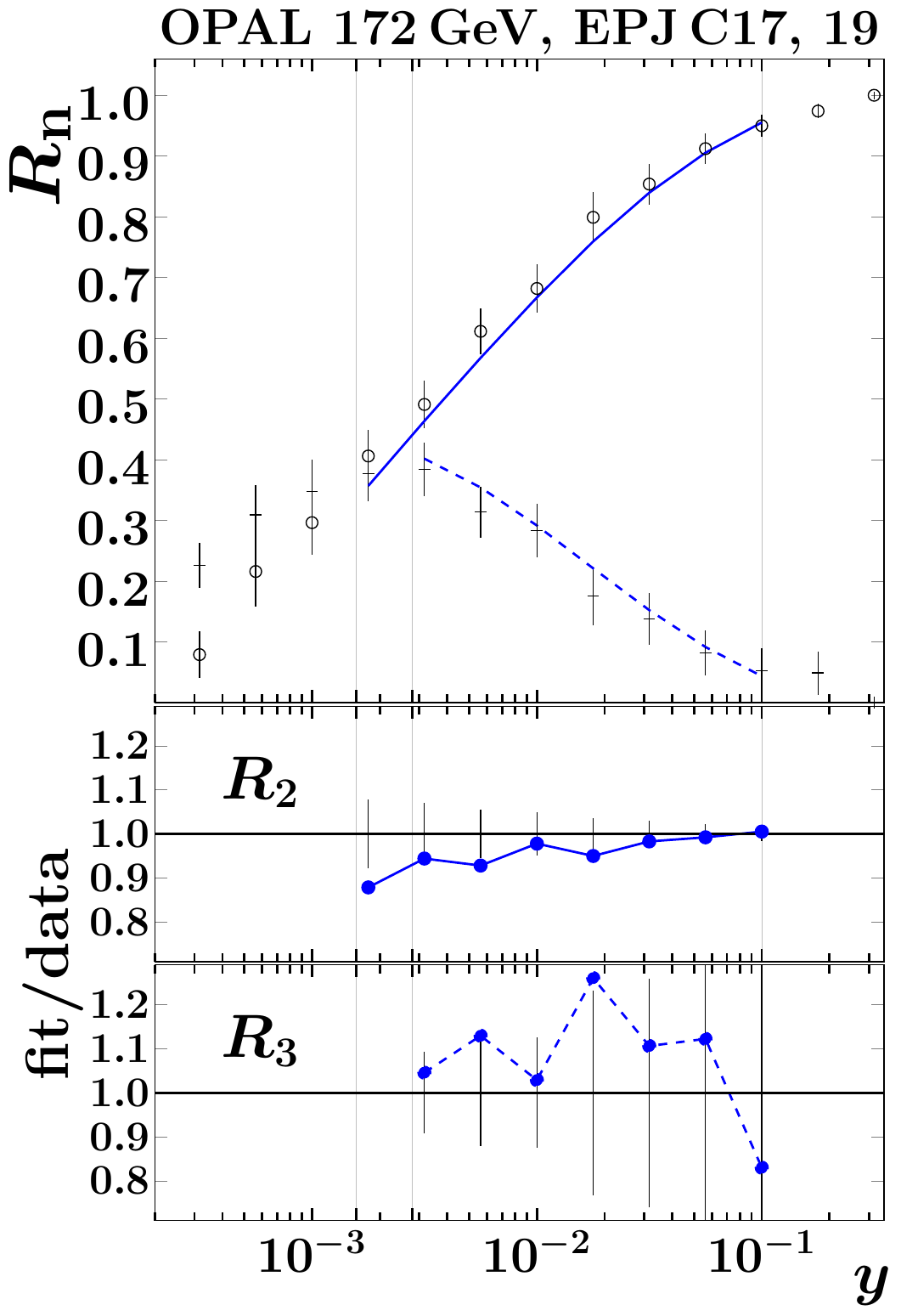}\draftbreak{}\arxivbreak{}\epjcbreak{}\includegraphics[width=\FIGWIDTH,height=\FIGHEIGHT]{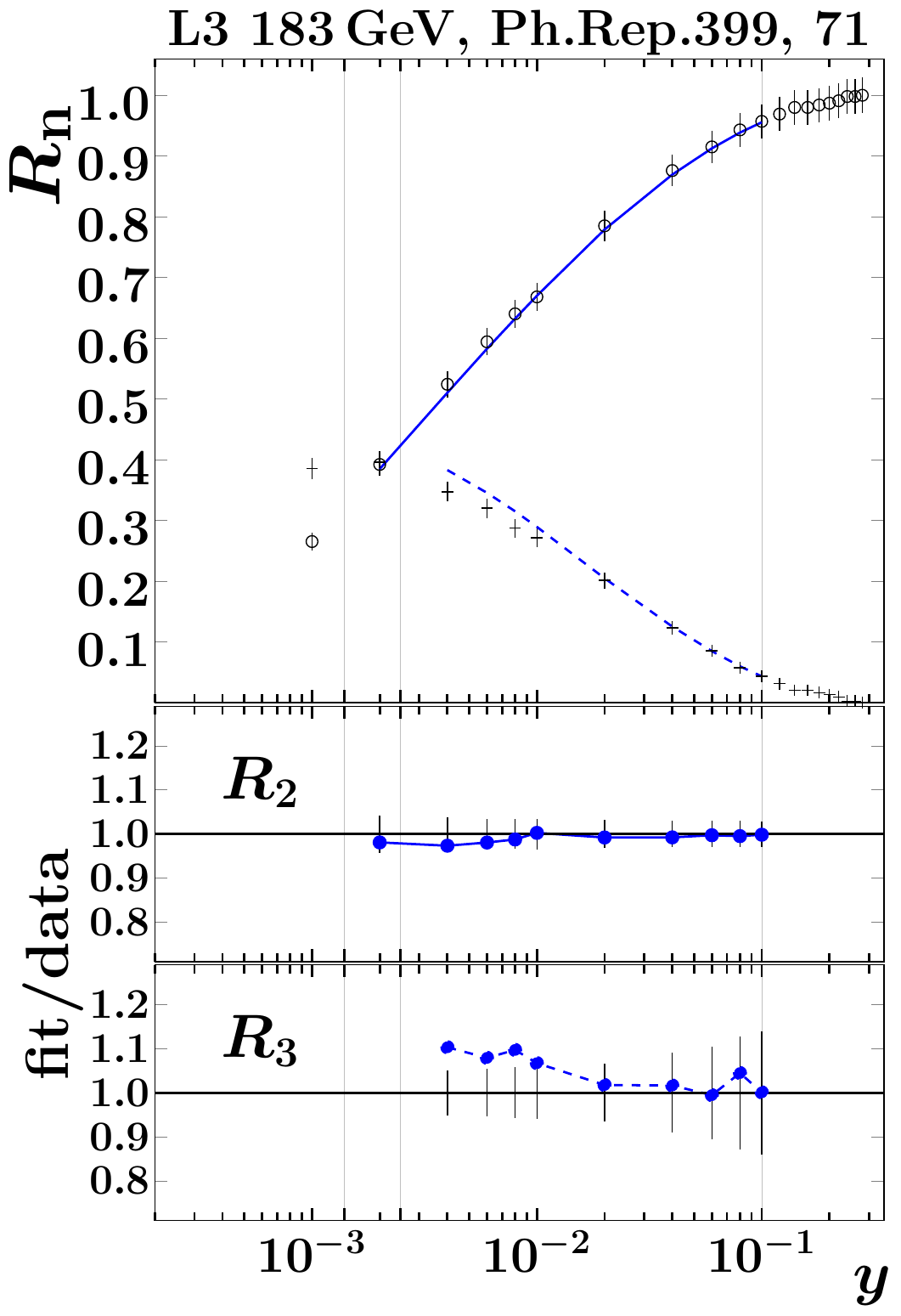}\includegraphics[width=\FIGWIDTH,height=\FIGHEIGHT]{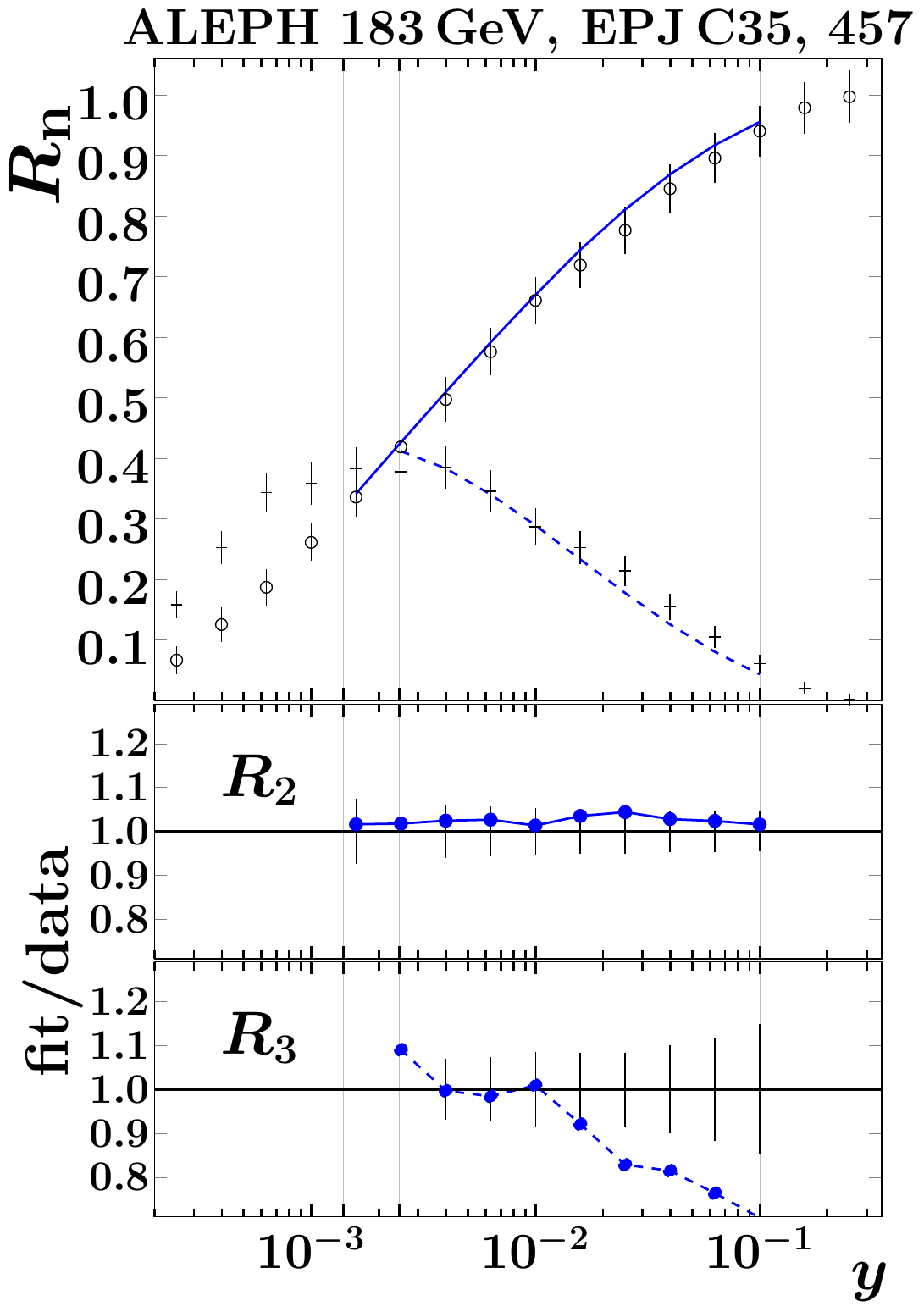}\includegraphics[width=\FIGWIDTH,height=\FIGHEIGHT]{Figures/figresultfoolegend-figure0.pdf}\\
\caption{Comparison of data and perturbative predictions supplemented by hadronization corrections in the $H^{L}$ model using for the strong coupling the value obtained from our global fit, eq.~\eqref{eq:alphasfinal}.}
\label{fig:result:two}
\end{figure}

\begin{figure}[htbp]\centering
\includegraphics[width=\FIGWIDTH,height=\FIGHEIGHT]{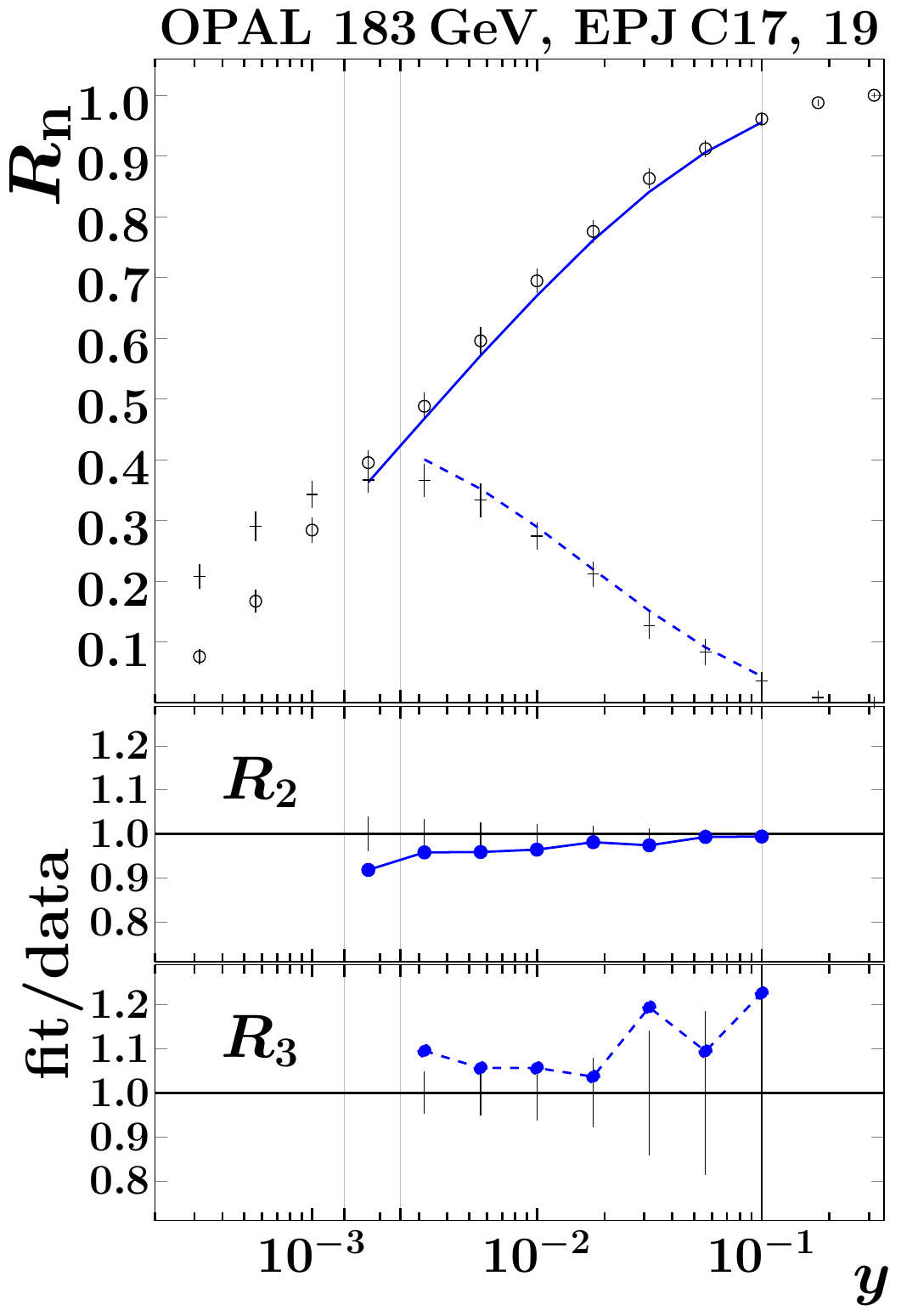}\includegraphics[width=\FIGWIDTH,height=\FIGHEIGHT]{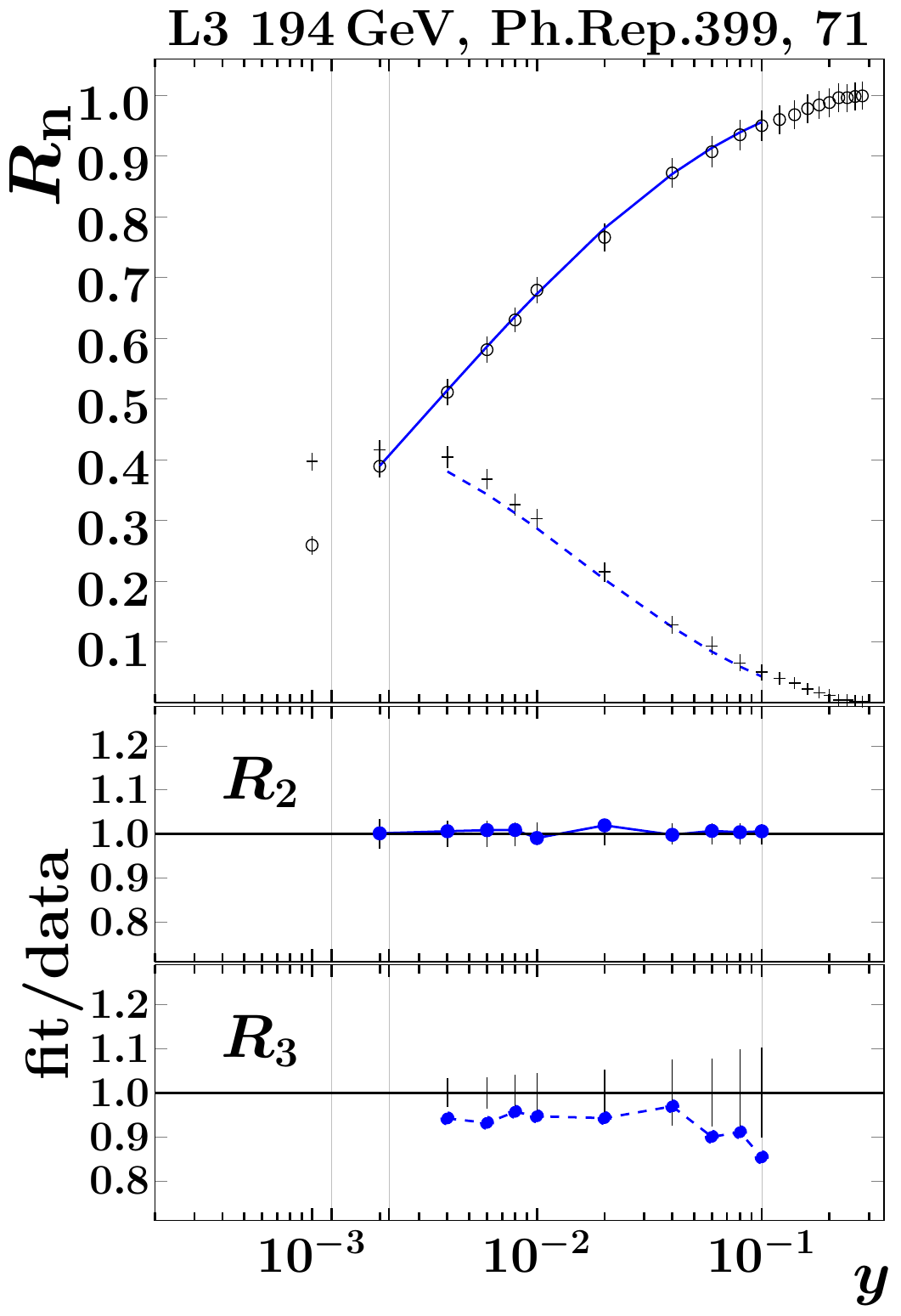}\epjcbreak{}\includegraphics[width=\FIGWIDTH,height=\FIGHEIGHT]{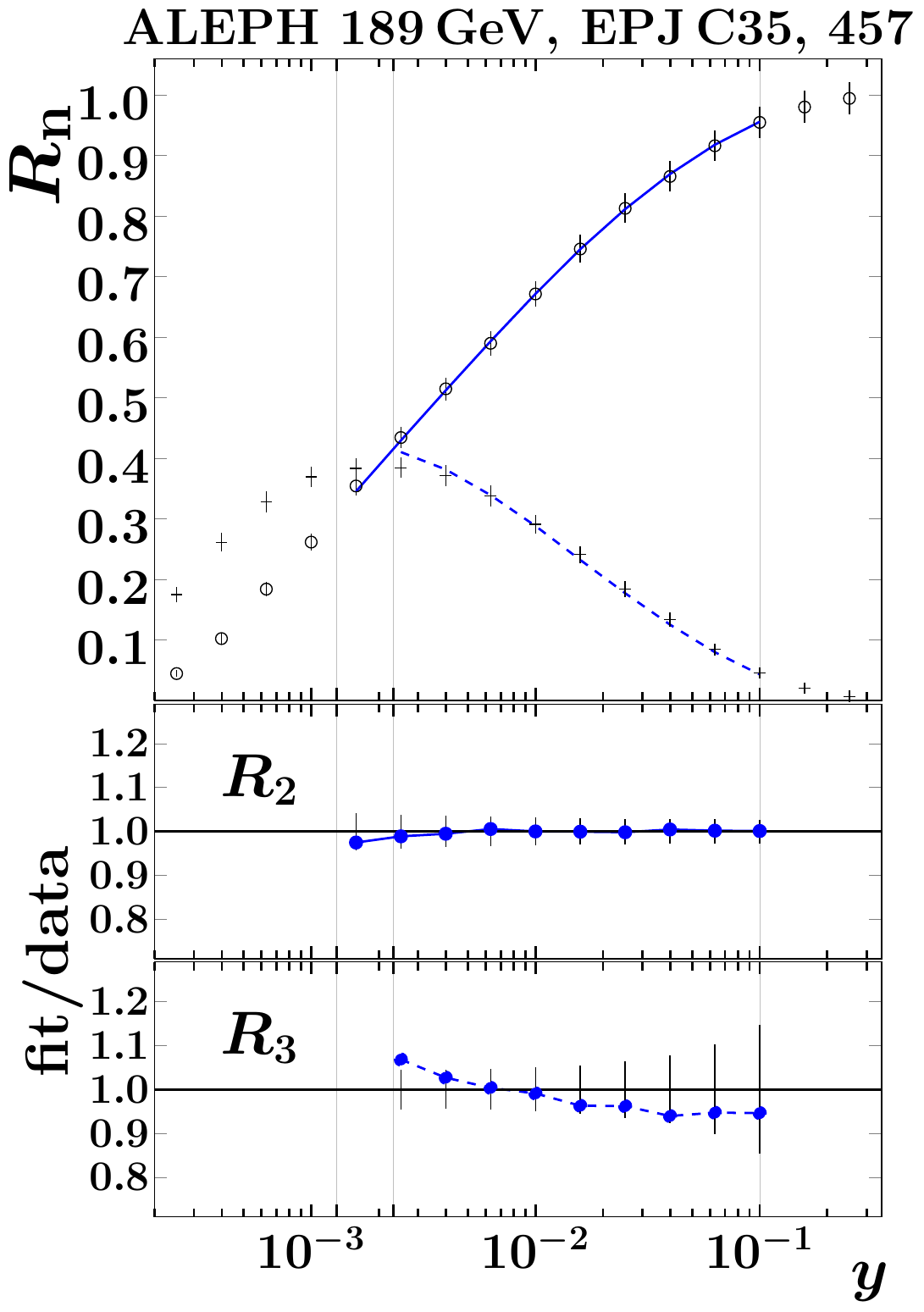}\draftbreak{}\arxivbreak{}\includegraphics[width=\FIGWIDTH,height=\FIGHEIGHT]{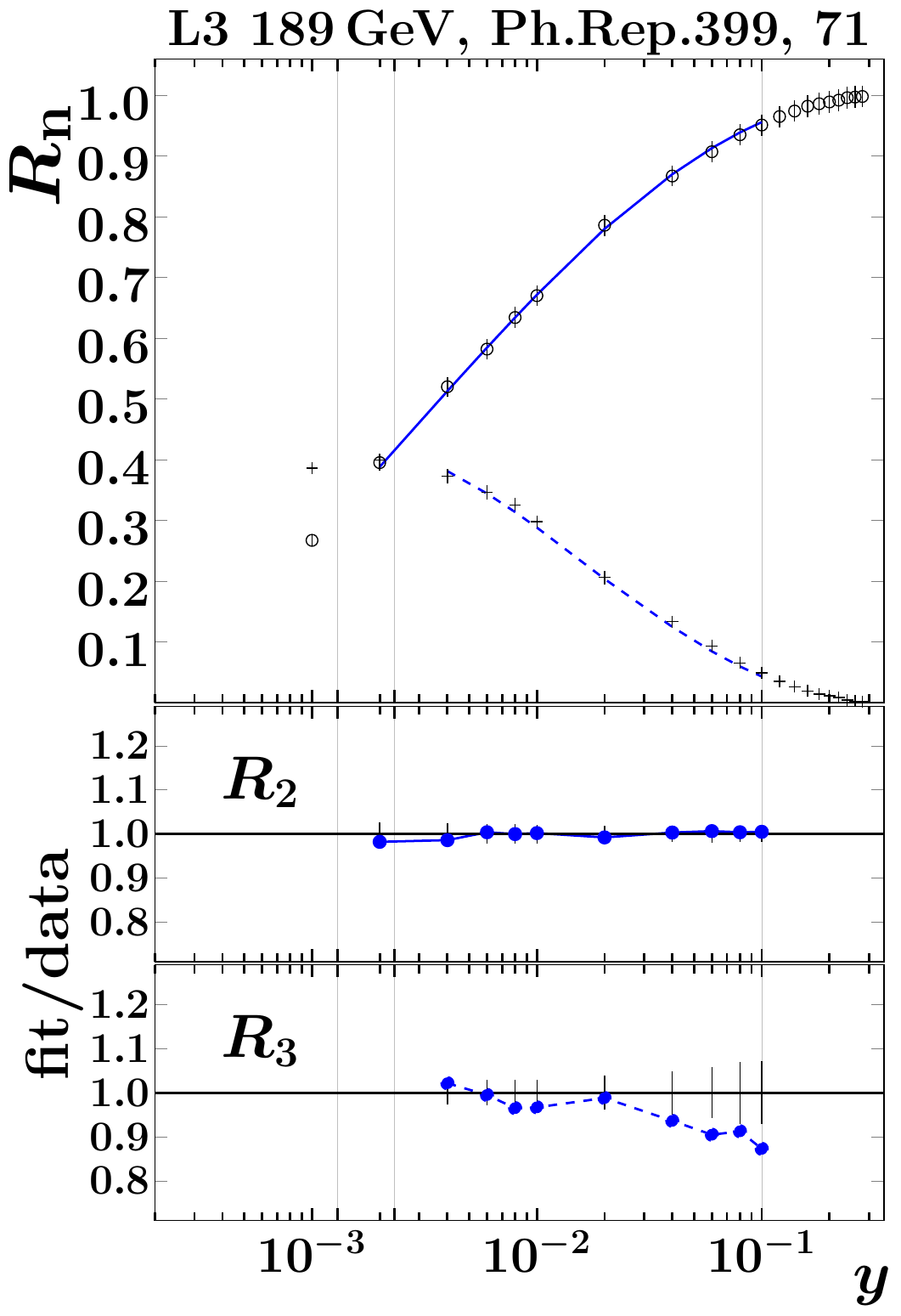}\epjcbreak{}\includegraphics[width=\FIGWIDTH,height=\FIGHEIGHT]{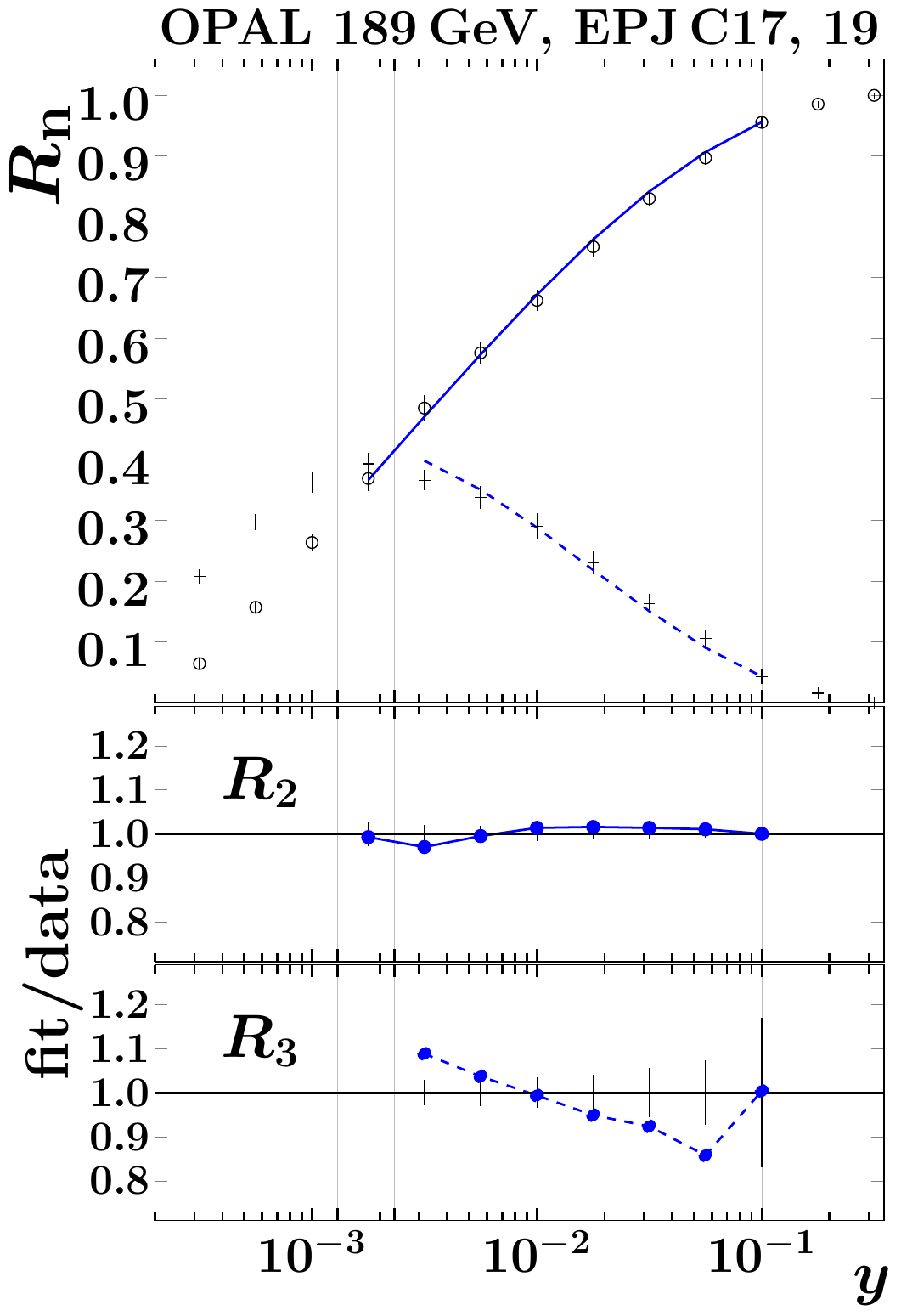}\includegraphics[width=\FIGWIDTH,height=\FIGHEIGHT]{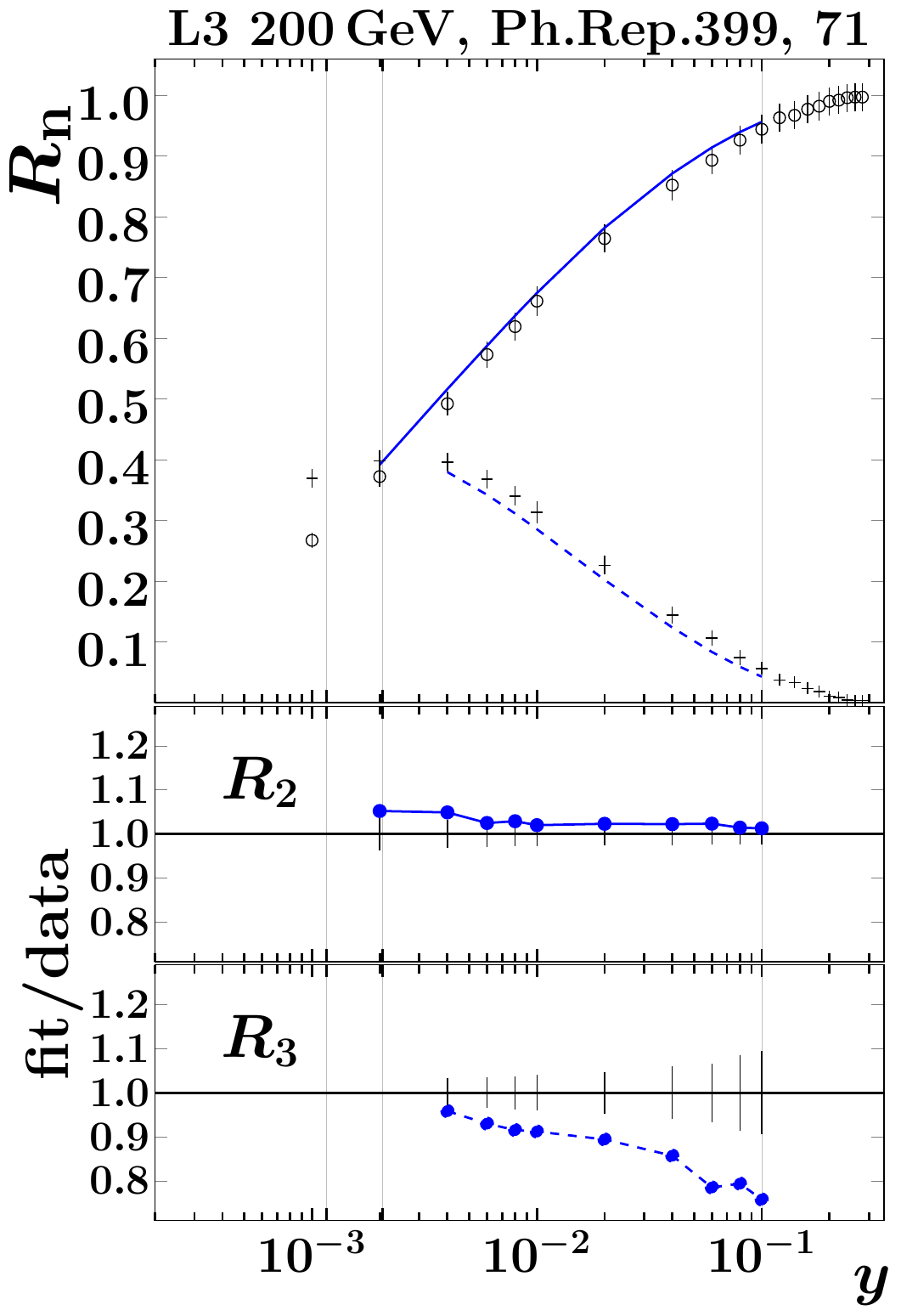}\draftbreak{}\arxivbreak{}\epjcbreak{}\includegraphics[width=\FIGWIDTH,height=\FIGHEIGHT]{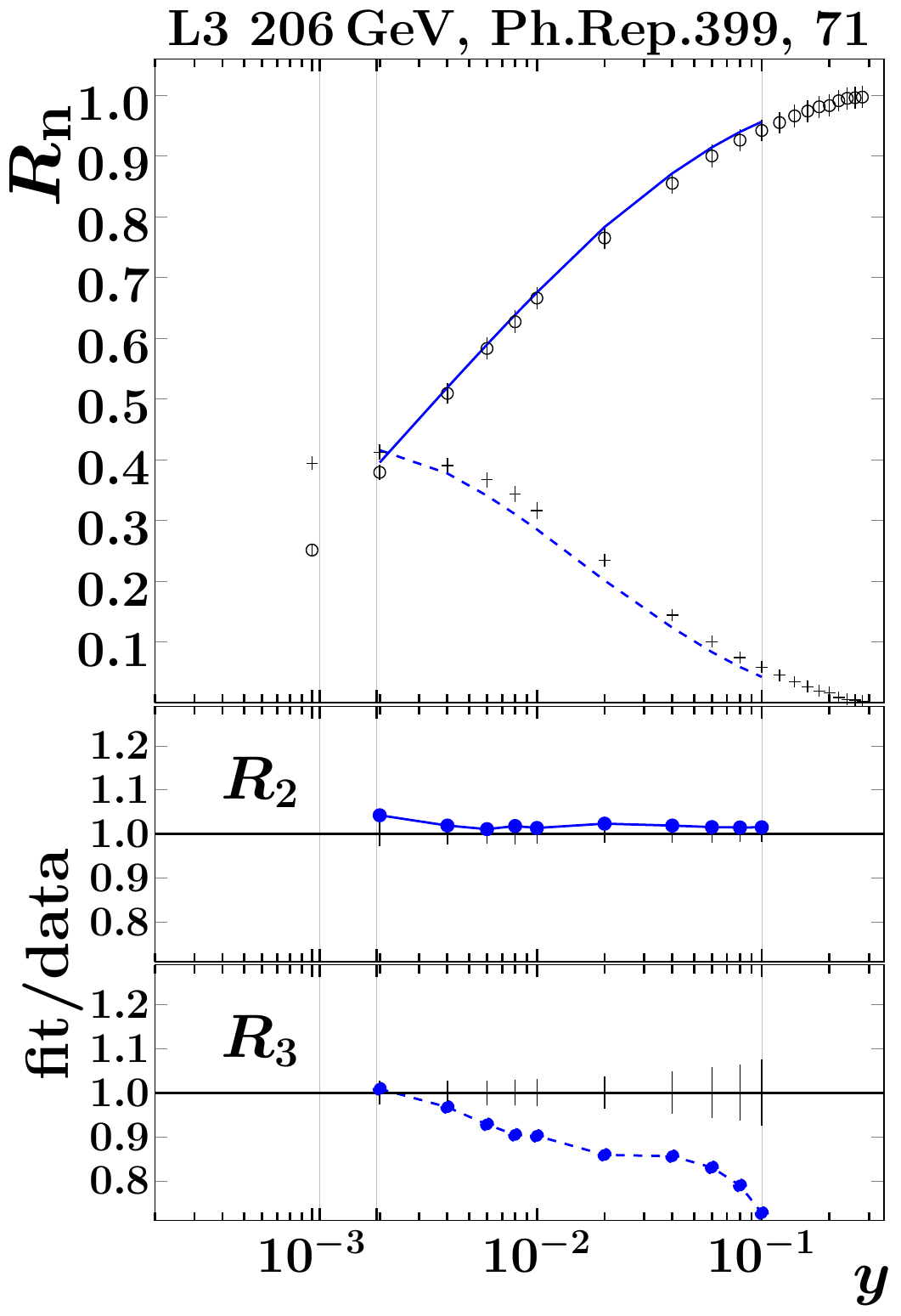}\includegraphics[width=\FIGWIDTH,height=\FIGHEIGHT]{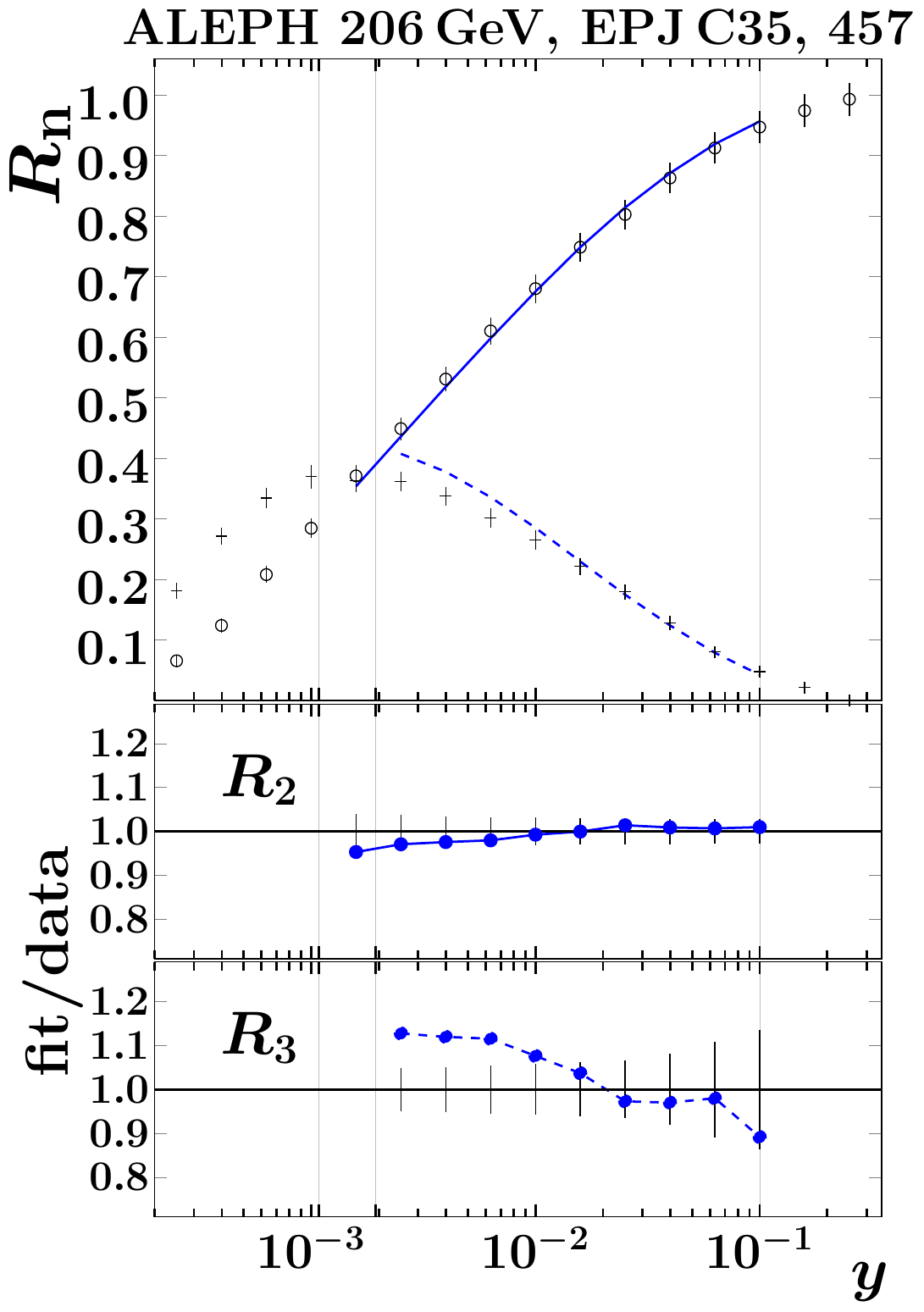}\includegraphics[width=\FIGWIDTH,height=\FIGHEIGHT]{Figures/figresultfoolegend-figure0.pdf}\\
\caption{Comparison of data and perturbative predictions supplemented by hadronization corrections in the $H^{L}$ model using for the strong coupling the value obtained from our global fit, eq.~\eqref{eq:alphasfinal}.} 
\label{fig:result:three}
\end{figure}

Finally, we show in Figs.~\ref{fig:result:one},~\ref{fig:result:two}
and
\ref{fig:result:three} the comparison of data at different energies with theory predictions using $\alpha_s(M_Z)$ obtained from our global fit, eq.~\eqref{eq:alphasfinal}.

\FloatBarrier

\subsection{Estimation of uncertainties}
\label{sec:unc}
\begin{figure}\centering
\includegraphics[width=\FIGWIDTHTWO,height=\FIGHEIGHTTWO]{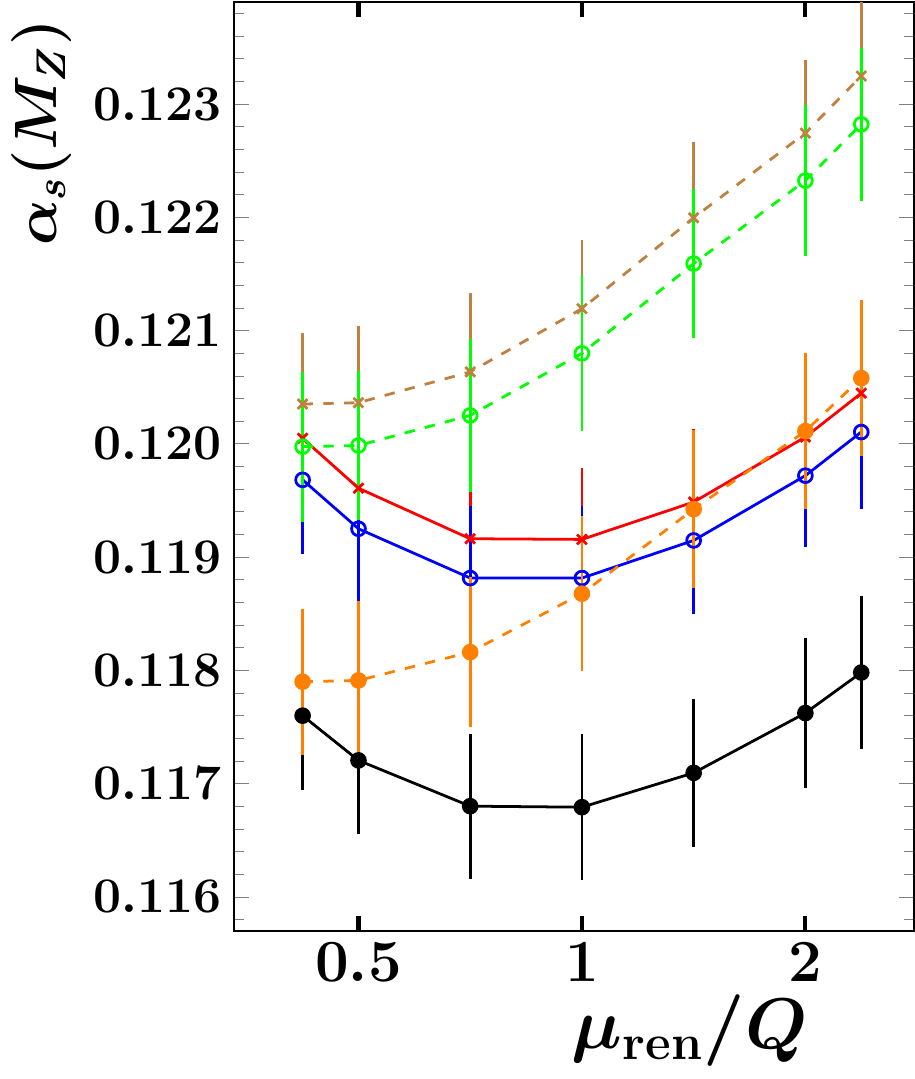}\includegraphics[width=\FIGWIDTHTWO,height=\FIGHEIGHTTWO]{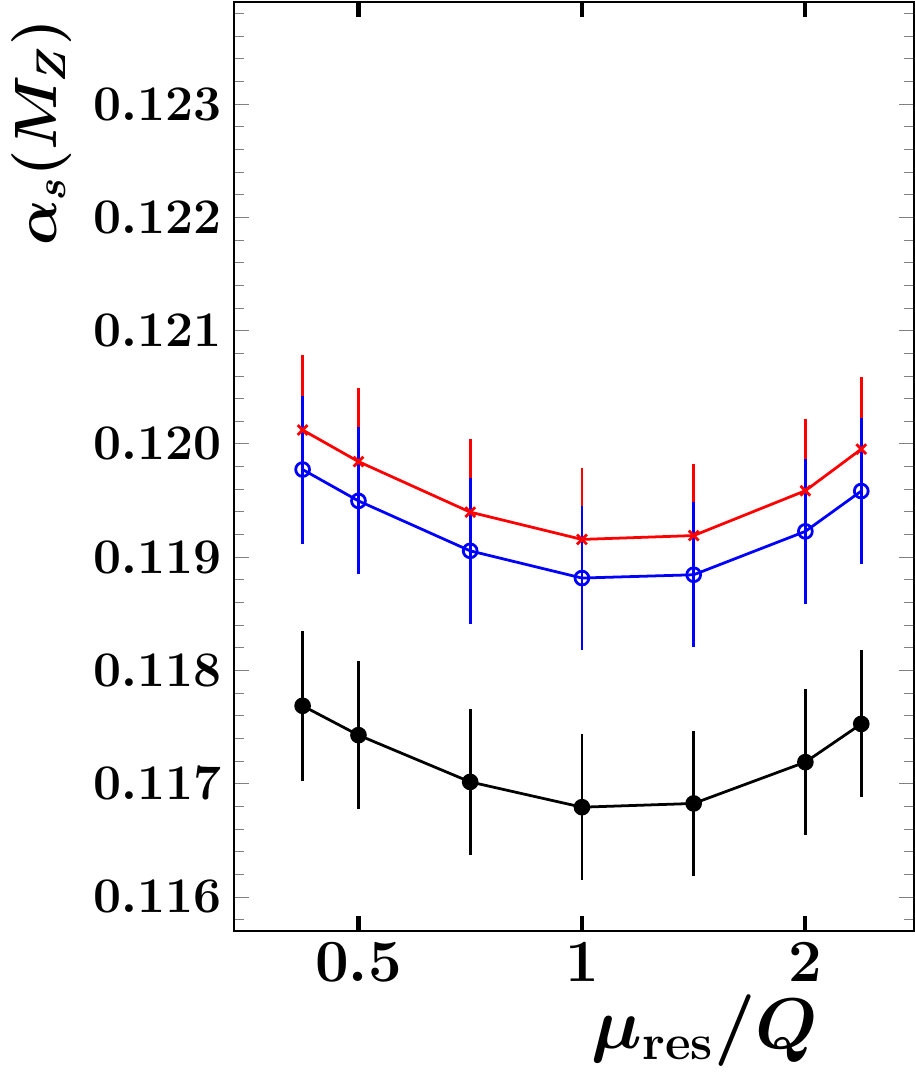}\includegraphics[width=\FIGWIDTHTWO,height=\FIGHEIGHTTWO]{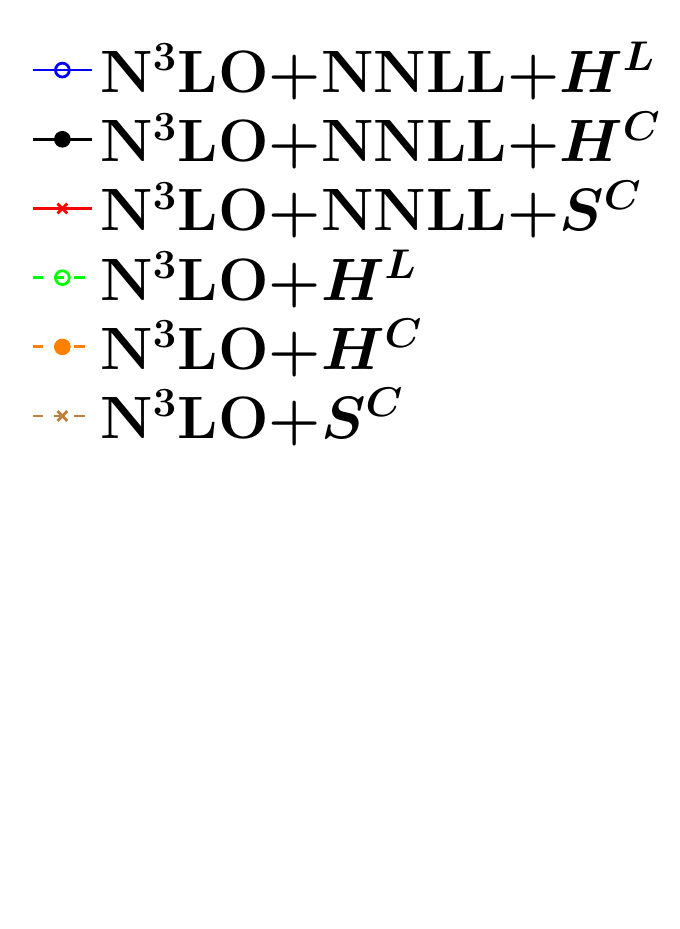}\\
\caption{Dependence of ($R_{2}$) fit results on the renormalization and
  resummation scales. The fit range for $S^{C}$, $H^{C}$ and $H^{L}$
  setups is \JRTfitrangethreeRtwo with 
  ${\cal L} = \log_{10}(M_Z^2/Q^2)$.  }
\label{fig:result:dependence}
\end{figure}

The systematic uncertainties in $\alpha_s$ are determined following
the procedure of~\cite{Jones:2003yv}. To estimate the size of
higher-order terms in the perturbative prediction, we vary the
renormalization scale in the range $\mu_{{\rm ren}}=Q/2$ and
$\mu_{{\rm ren}}=2Q$. Moreover, while keeping $\mu_{{\rm ren}}=Q$ fixed, we
vary the resummation scale in the range $\mu_{{\rm res}}=Q/2$ and
$\mu_{{\rm res}}=2Q$. The effects of the individual variations are
displayed in Fig.~\ref{fig:result:dependence}, where different
hadronization models are compared. We notice that, when resummation is
included, a much reduced dependence on the renormalization scale is
observed.

The bias due to the selection of the hadronization model is studied by
means of the difference between the $H^{L}$ and $H^{C}$ setups, see
Fig.~\ref{fig:result:dependence}.  In particular, considering the
results from the Lund string and cluster hadronization models,
the desired systematic uncertainty is obtained as half of the
difference between the $\alpha_S(M_Z)$ results obtained in nominal
fits with $H^{L}$ and $H^{C}$ setups.  The obtained numerical value is
close to $0.001$ (i.e.\ slightly below 1\%). This can be briefly
compared to previous estimations obtained with Monte Carlo event
generator models in similar analyses. Namely, the values
$0.001$~\cite{Dissertori:2007xa,OPAL:2011aa} and
$0.0005$~\cite{Dissertori:2009ik} obtained previously allow us to
validate our estimation.  
We stress that we have not performed any tuning of the adopted
hadronization models to the data in order to artificially reduce the
related uncertainties. This leads to a more conservative, and thus more
robust estimate of hadronization uncertainties.

The final uncertainty is obtained by combining each of the above
uncertainties in quadrature.

\section{Validation of the procedure and further fits}

In this section we perform some consistency checks to validate the
fitting procedure used above. Moreover, we present an extraction of $
\alpha_s(M_Z)$ from a simultaneous fit of $R_2$ at N$^3$LO+NNLL and
$R_3$ at NNLO.
\subsection{Fit consistency tests}
We have performed a number of consistency tests, as outlined in the following:

\begin{enumerate}
\item We repeat the nominal fits in different ranges of $\sqrt{s}$
  separately, instead of simultaneously. The results of these fits are
  shown in Fig.~\ref{fig:result:dependenceStwo}. We do not observe any
  significant dependence of the results on the centre-of-mass energy.

\begin{figure}[htbp]\centering
\includegraphics[width=\FIGWIDTHTHREE,height=\FIGHEIGHTTWO]{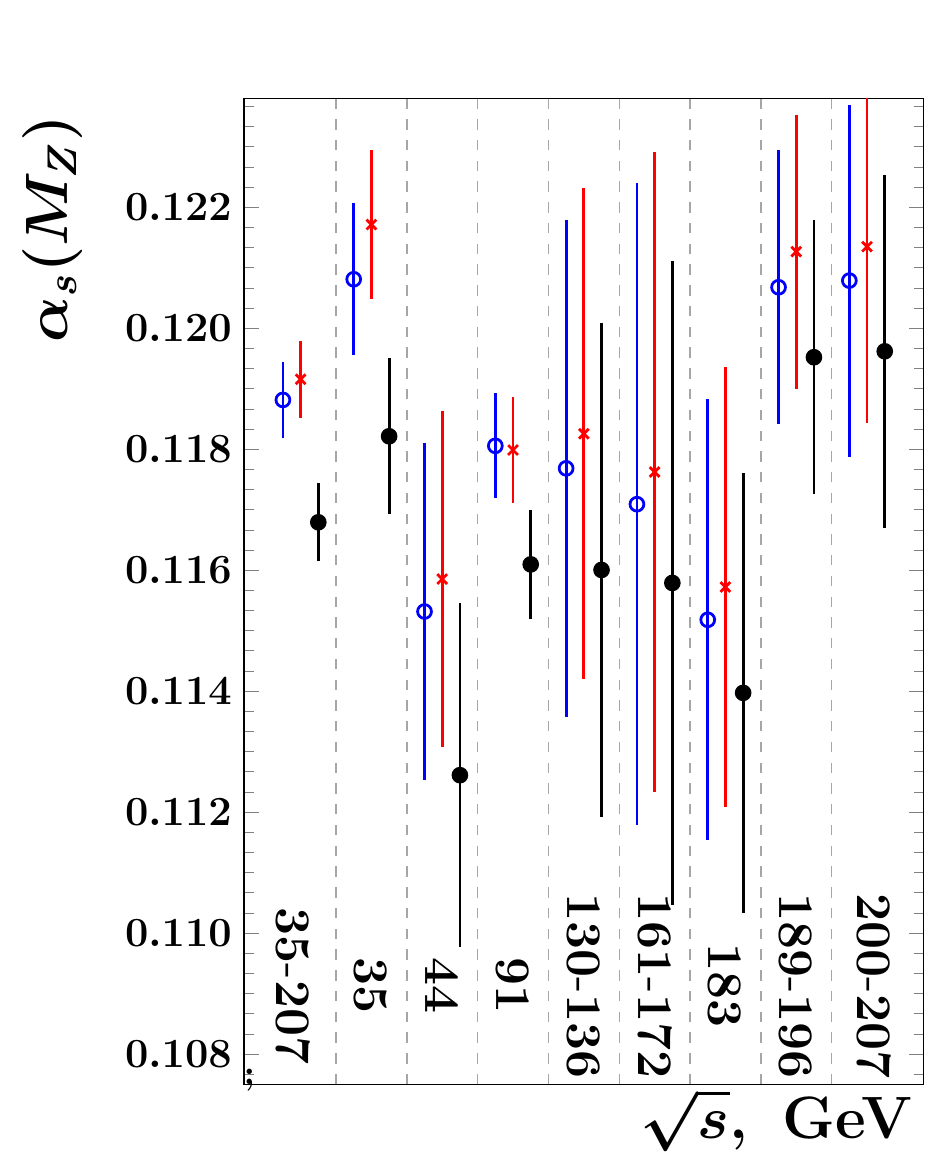}\includegraphics[width=\FIGWIDTHTWO,height=\FIGHEIGHTTWO]{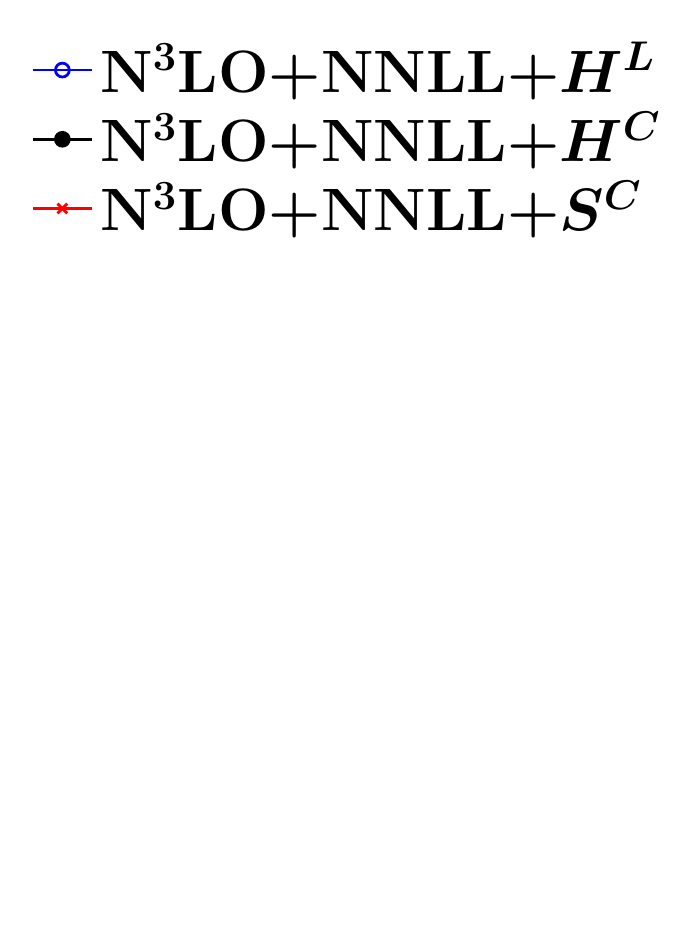}\\
\caption{Dependence of ($R_{2}$) fit results on the $\sqrt{s}$ of  measurement.	
The fit range  
is 
\JRTfitrangethreeRtwo. Only statistical uncertainties are shown.
}
\label{fig:result:dependenceStwo}
\end{figure}

\item We repeat the nominal fit for $R_2$ by implementing the
  hadronization corrections on a bin-by-bin basis as
  $R_{2,{\rm hadron}}=R_{2,{\rm parton}}f_2(y)$, where $f_2(y)$ is
  derived from the MC generated samples. Within this scheme, we find

$\as(M_{Z})=\JRTresultHNNLOMULTRtwo.$

\noindent
This value is close to the reference result, but the hadronization
uncertainty estimated from this setup is more sensitive to changes in
the fit range.

\item
To test the reliability of the correlation model used for the systematic
 uncertainties in the reference fit, we use the OPAL data and systematic shifts (uncertainties) from Ref.~\cite{Verbytskyi201813}. 
With this data  we perform  the fits  using the $\chi^{2}$ definition from  
eq.~\eqref{chi2} and the definition that explicitly includes set of $N$ 
nuisance  parameters $B=\{b_{1},b_{2}\dots b_{N}\}$ and vectors of systematic 
shifts (uncertainties) $\vec{S_{1}},\vec{S_{2}}\dots \vec{S_{N}}$\footnote{Such a treatment of systematic shifts (uncertainties) is widely used  in QCD analyses, see Ref.~\cite{Alekhin:2014irh} as an example.}:
\begin{equation}
\chi^2(\as,B)=(\vec{r}-\sum_{i=1}^{N}b_{i}\vec{S_{i}})V^{-1}(\vec{r}-\sum_{i=1}^{N}b_{i}\vec{S_{i}})^{T}+\sum_{i=1}^{N}b_{i}^{2}.
\label{eq:nu}
\end{equation}
The  result is 

\noindent
$\as(M_{Z})=\JRTresultHNNLOOPALRtwo $ for the definition in eq.~\eqref{chi2}

\noindent
and 

\noindent
$\as(M_{Z})=\JRTresultHNNLOOPALSHIFTSRtwo$  for the definition in eq.~\eqref{eq:nu}.

\noindent
These results are in fair agreement, and the corresponding values of
$\chi^{2}$ are close, which demonstrates the relatively good performance
of the selected correlation model.

\item We repeat the  reference fit without the data sets from JADE,
  for which the explicit verification of correlation model using Ref.~\cite{Verbytskyi201813} was not possible.
  The obtained result 
$\as(M_{Z})=\JRTresultHNNLONOJADERtwo$  agrees well with the reference value 
$\as(M_{Z})=\JRTresultHNNLORtwo$.

\item To estimate the error due to the correlation model, we vary the value of
  the parameter $\rho$  in a wide range.  No significant
  change of fit results was observed, see
  Fig.~\ref{fig:result:dependencerho}.
\begin{figure}\centering
\includegraphics[width=\FIGWIDTHTWO,height=\FIGHEIGHTTWO]{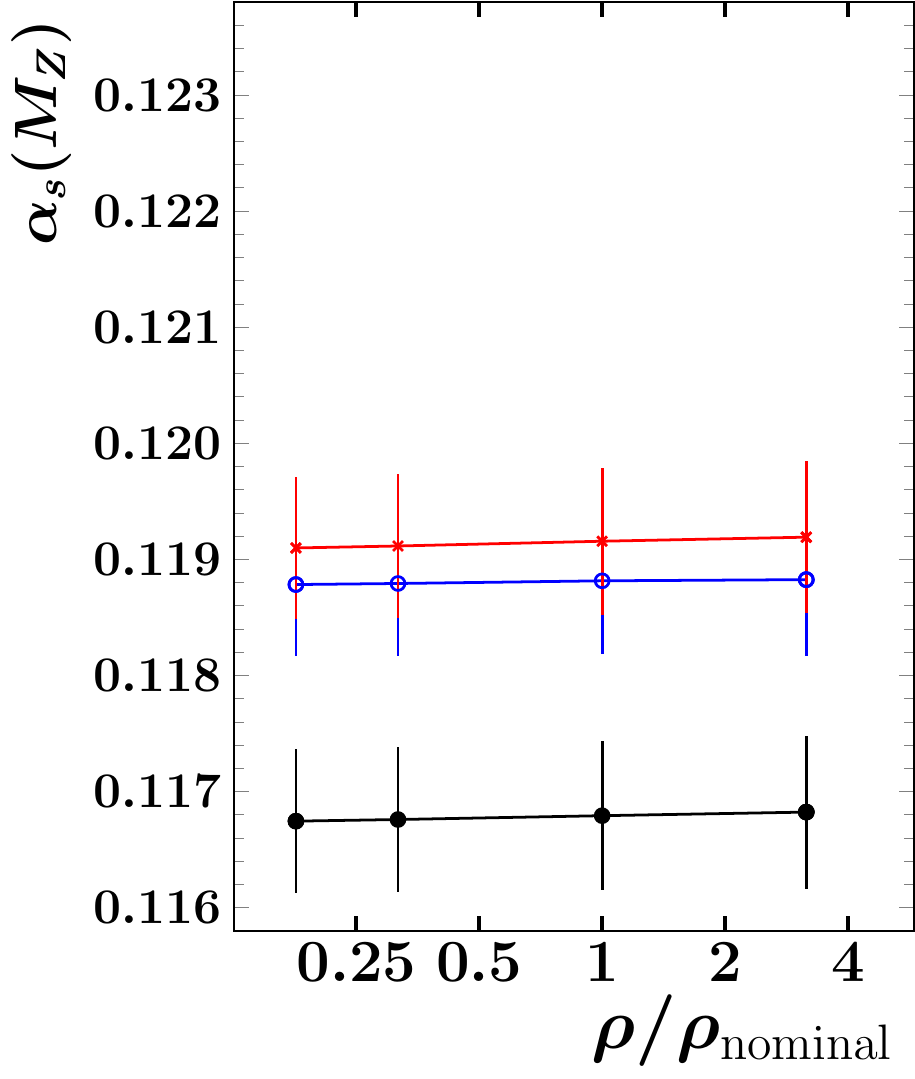}\includegraphics[width=\FIGWIDTHTWO,height=\FIGHEIGHTTWO]{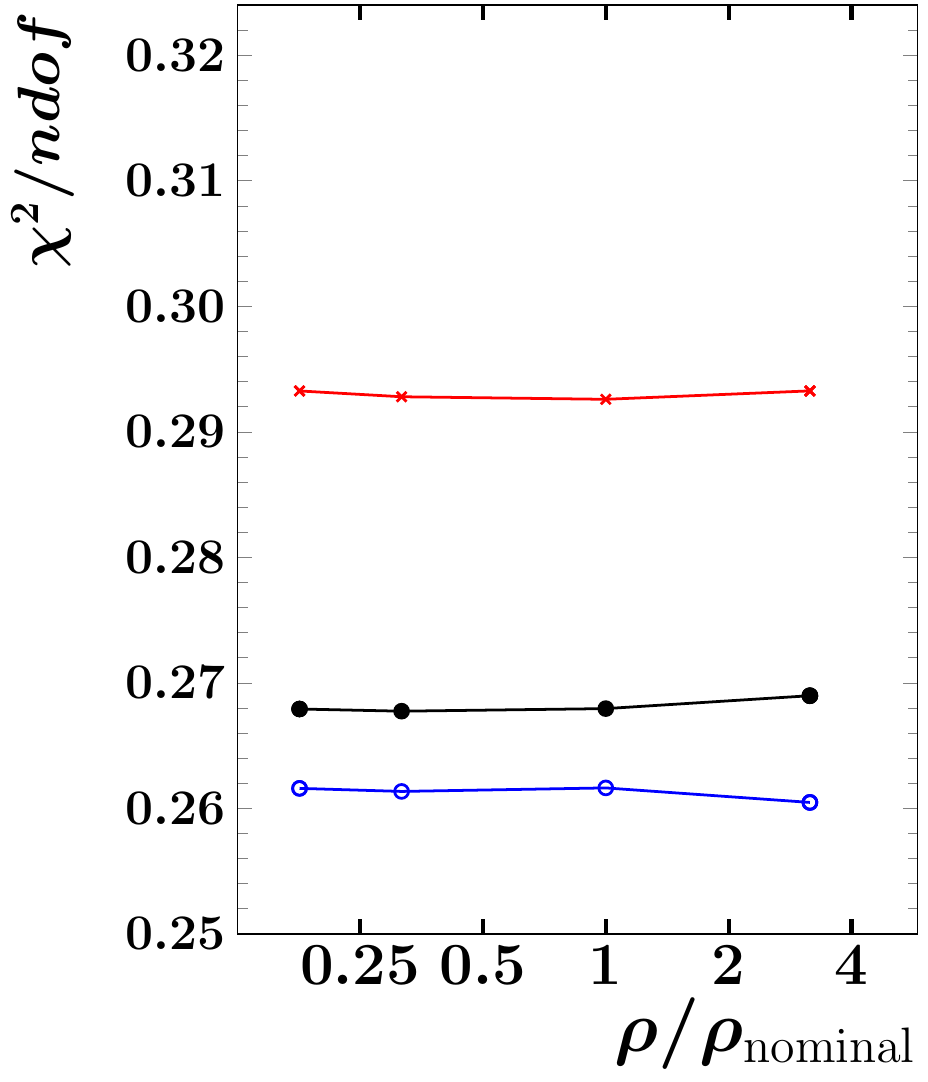}\includegraphics[width=\FIGWIDTHTWO,height=\FIGHEIGHTTWO]{Figures/figdependencescalefoohalf-figure0.pdf}\\
\caption{Dependence of the ($R_2$) fit results  on the parameter $\rho$ used in the correlation model.
The fit range is fixed to  \JRTfitrangethreeRtwo.
}
\label{fig:result:dependencerho}
\end{figure}

\item We perform variations of the renormalization and resummation
  scales simultaneously, as done in Ref.~\cite{Schieck:2012mp}.  The
  results are reported in Fig.~\ref{fig:result:dependenceboth} and
  show smaller uncertainty than in the case of independent variations.
\begin{figure}\centering
\includegraphics[width=\FIGWIDTHTWO,height=\FIGHEIGHTTWO]{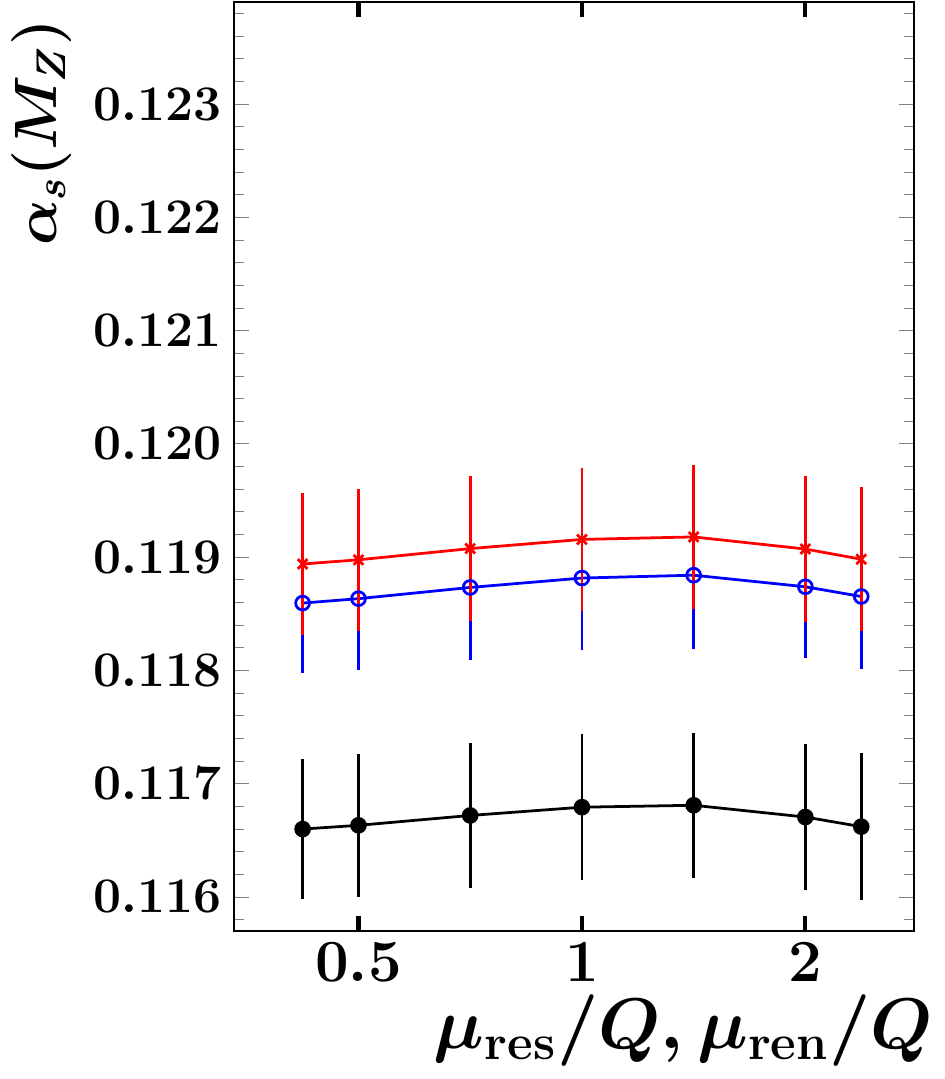}\includegraphics[width=\FIGWIDTHTWO,height=\FIGHEIGHTTWO]{Figures/figdependencescalefoohalf-figure0.pdf}\\
\caption{Dependence of the ($R_{2}$) fit results on the simultaneous variation of 
renormalization and resummation scales. The fit range is fixed to 
 \JRTfitrangethreeRtwo.
}
\label{fig:result:dependenceboth}
\end{figure}
A similar behaviour was observed in Ref.~\cite{Kardos:2018kqj}.
\end{enumerate}
\FloatBarrier

\subsection{Simultaneous fit of the coupling with the  two- and three- jet rates $R_2$ and $R_3$}

An alternative fit was performed with $R_2$ and $R_3$ observables
simultaneously, with $R_3$ computed at NNLO.  The obtained result
using the \JRTfitrangethreeRtwoRthree fit ranges for $R_2$ and $R_3$,
respectively, and the $H^{L}$ setup is

\noindent
$\as(M_{Z})=\JRTresultHNNLORtwoRthree.$

Unlike the results of the reference fits, the obtained result is
sensitive to the selected fit range, see
Tab.~\ref{tab:result:rtwothree} in App.~\ref{app:special}. Taking this
effect into account would result in another uncertainty of order
$0.001$ that is not included in the uncertainties given above.

As a final cross-check we perform a fit for a single point of ALEPH $R_3$
data at $y=0.02$ without resummation.  The obtained result (with
statistical uncertainties only),
$\as(M_{Z})=\JRTresultHNNLOALEPHRthree$, can be compared to the
results from Ref.~\cite{Dissertori:2009qa}, 
$\as(M_{Z})=\DissertoriNNLOALEPHRthree$. The results agree well
despite the differences in the implementation of the hadronization
corrections.

\subsection{Discussion}
\label{sec:disc}
The value obtained from the analysis relying on N$^3$LO+NNLL
predictions for $R_2$ is in agreement with the world average as of
2017~\cite{Bethke:2017uli}, however it is visibly lower than the
results from measurements performed for other $e^{+}e^{-}$ observables
using NNLO perturbative QCD predictions and MC hadronization
models~\cite{Bethke:2017uli}.  The estimated uncertainties are on the
other hand approximately of the same sizes.

\section{Summary}                      
\label{sec:summary}
The main result of this paper is a first fit of the strong coupling
for the two-jet rate that relies on very accurate (N$^3$LO+NNLL)
predictions. Our main result reads 
\begin{equation*}
\as(M_{Z})=
\JRTresultHNNLORtwo.
\end{equation*}
The uncertainty on $\alpha_s$ induced by scale variations is now
considerably smaller than that related to  hadronization modelling. This is not the
case if the fit is performed using only N$^3$LO
predictions, see Fig.~\ref{fig:result:dependence}.
Furthermore the experimental uncertainty is now comparable to the
perturbative one.

After combining the uncertainties in quadrature we obtain 
\begin{equation*}
\as(M_{Z})=\JRTresultHNNLORtwocomb,
\end{equation*}
where the largest estimation bias comes from the used hadronization model.

Our results agrees  with the last PDG average
\begin{equation*}
\as(M_Z)_{\rm PDG 2018}  = 0.1181 \pm 0.0011
\end{equation*}
with an uncertainty that is of the same size.

We have also performed a combined fit of the two- and three-jet rate,
taking for the first time the correlation between these observables
into account. The results of the two fits are fully compatible.
However, the fit including $R_3$ shows a stronger dependence on the
fit range than the results of our reference fit based on $R_2$ only.
An accurate resummation for the $R_3$ observable could potentially
reduce the sensitivity to the fit range selection and lead to an even
more precise determination of $\as(M_Z)$.

\section*{Acknowledgements}
\label{sec:acknowledgements}
We are grateful to Simon Pl\"{a}tzer and Johannes Bellm for fruitful
discussions about the calculation of NLO predictions with {\tt
Herwig7.1.4} and to Carlo Oleari for providing us the
\prog{Zbb4} code. 

The work of A.B. is supported by the Science Technology and Facilities Council (STFC) under grant number ST/P000819/1.
A.K. acknowledges financial support from the Premium Postdoctoral
Fellowship program of the Hungarian Academy of Sciences. A.K. is also
grateful to D\'avid Zsoldos for his help with processing the raw data of predictions at fixed order. 
This work was supported by grant K 125105 of the National
Research, Development and Innovation Fund in Hungary.  The research of
P.M. was supported by the European Commission through the Marie
Sk\l{}odowska Curie Individual Fellowship, contract number 702610.
The work of G.Z was supported in part by ERC Consolidator Grant HICCUP
(No. 614577).

\FloatBarrier

\appendix

\newpage 

\section{Perturbative ingredients}
\label{app:PTcoeffs}

We report in Tab.~\ref{tab:predictions} the numerical results for the
fixed-order coefficients introduced in eq.~\eqref{eq:ABC}.

\renewcommand{\arraystretch}{1.0}
\begin{table}[!htbp]\centering\small
\addtolength{\tabcolsep}{-3pt}
\begin{tabular}{|c|c|c|c|c|c|c|}\hline
 $y$      & $A_3$ & $A_4$ & $A_5$& $B_3$& $B_4$& $C_3$  \\
\hline\hline
\JRTpredictions\hline
\end{tabular}
\caption{
The perturbative coefficients entering eq.~\eqref{eq:ABC} as
calculated by the \prog{MCCSM} program. }
\label{tab:predictions}
\end{table}

\section{MC simulations at hadron and parton levels}
\label{app:MC}

\begin{figure}[htbp]\centering
\includegraphics[width=\FIGWIDTH,height=\FIGHEIGHT]{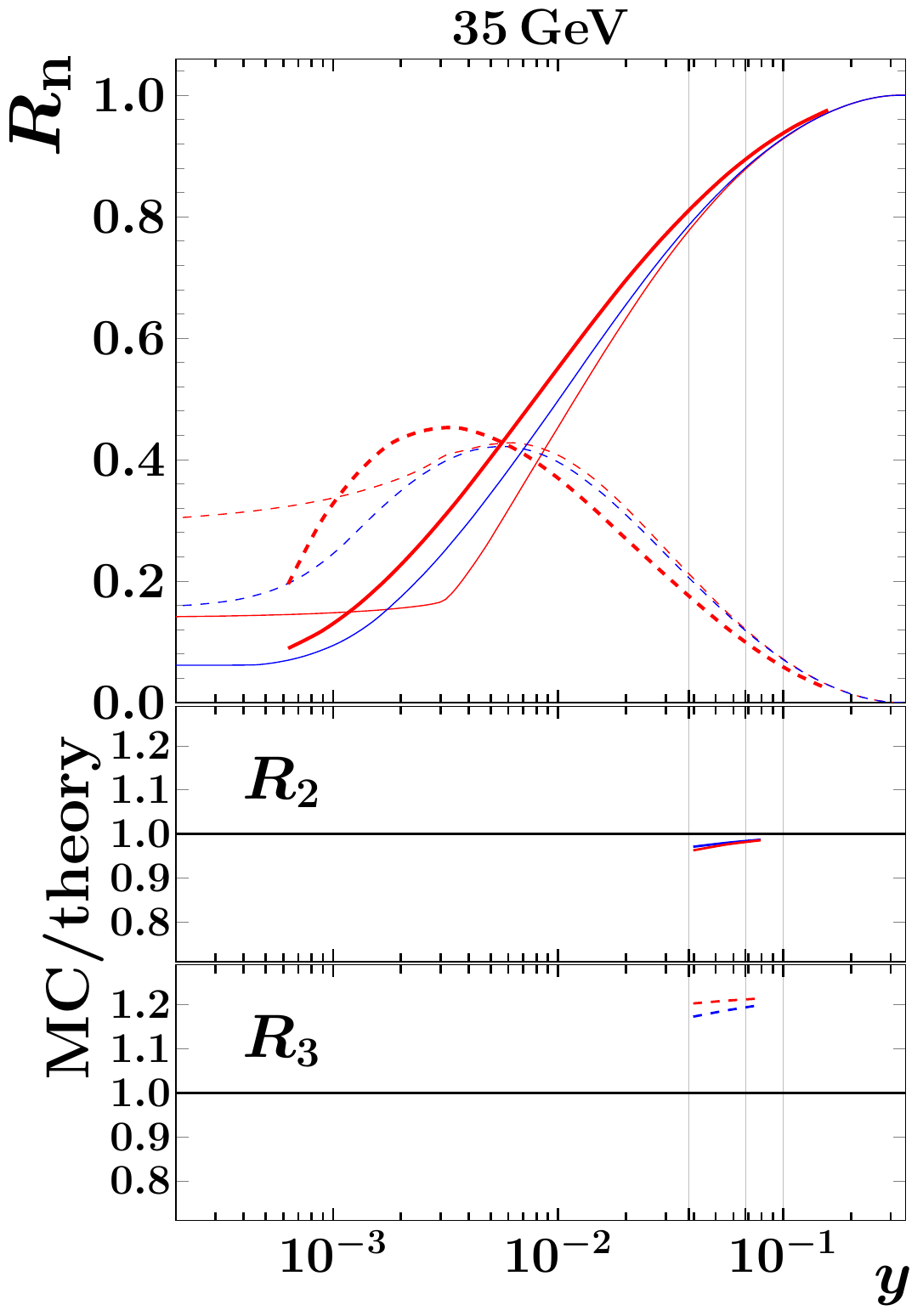}\includegraphics[width=\FIGWIDTH,height=\FIGHEIGHT]{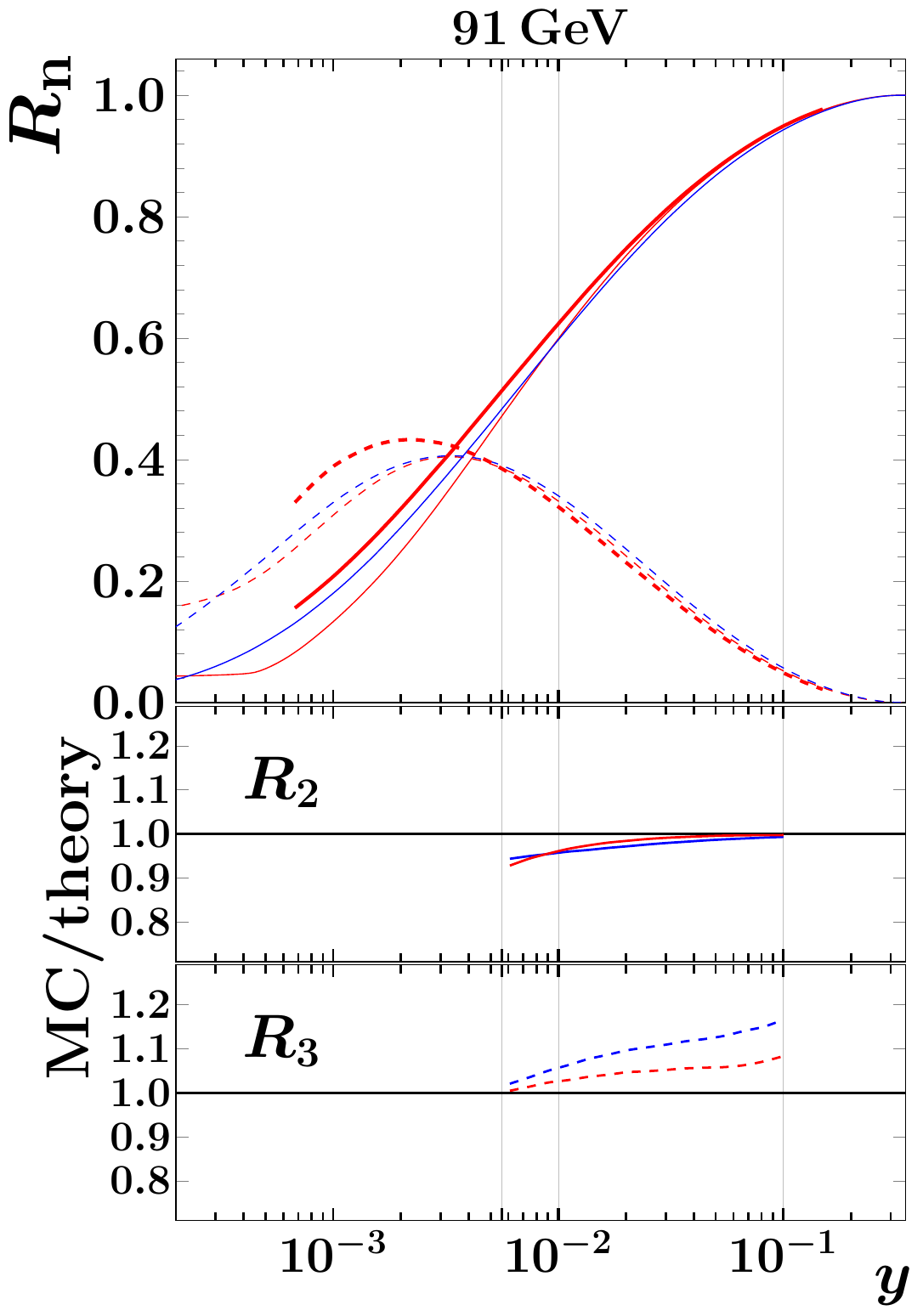}\epjcbreak{}\includegraphics[width=\FIGWIDTH,height=\FIGHEIGHT]{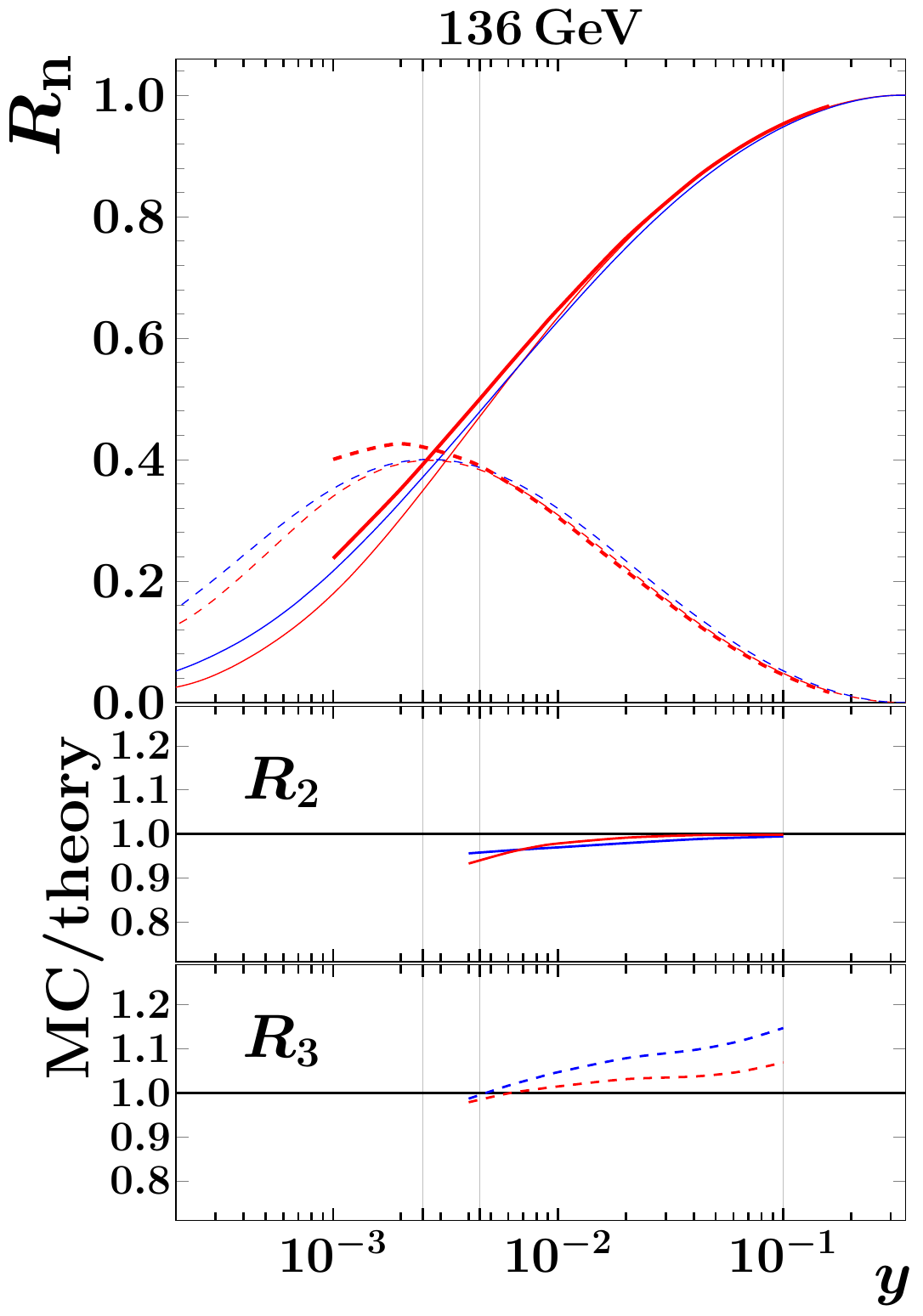}\epjcbreak{}\draftbreak{}\arxivbreak{}\includegraphics[width=\FIGWIDTH,height=\FIGHEIGHT]{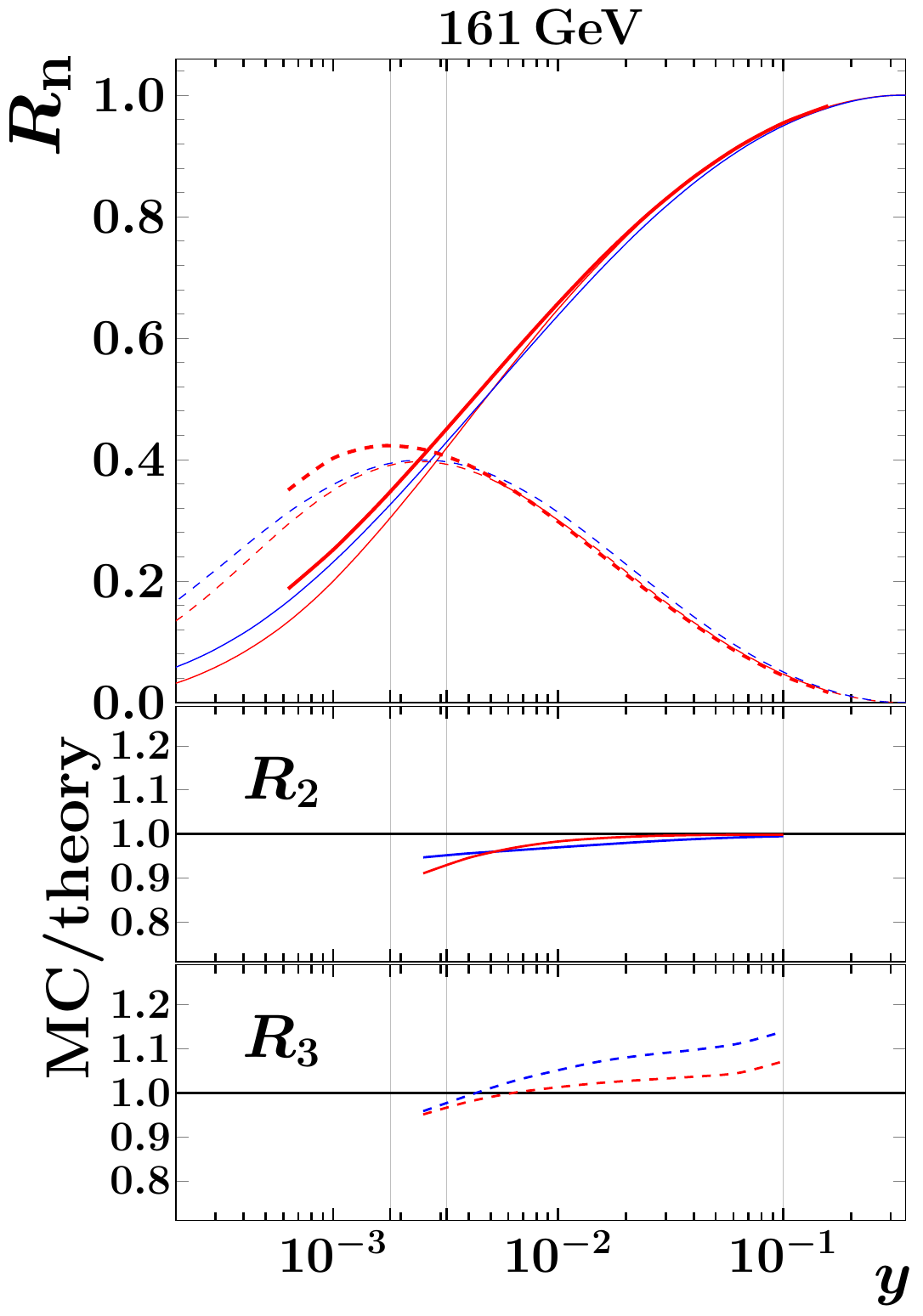}\includegraphics[width=\FIGWIDTH,height=\FIGHEIGHT]{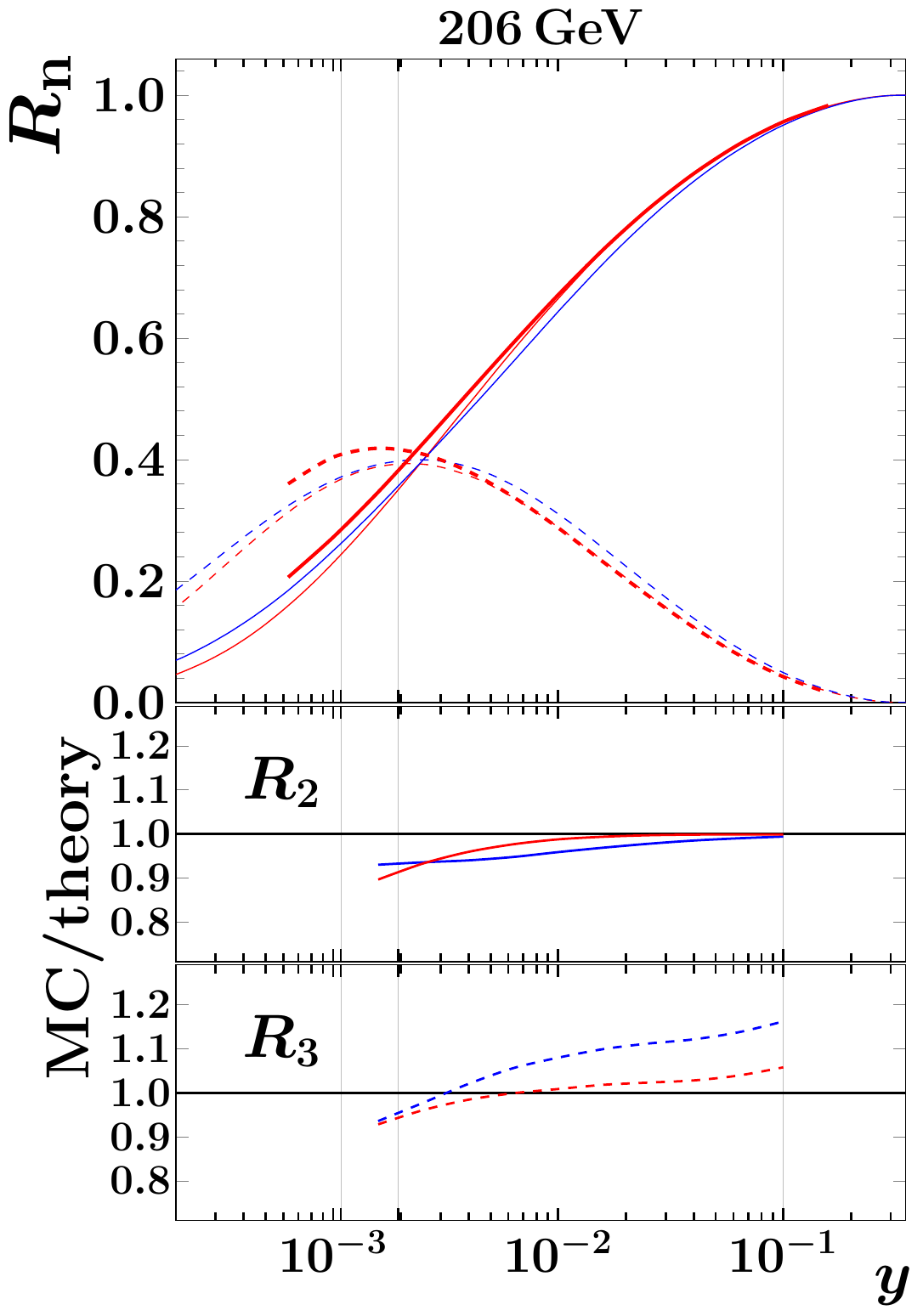}\includegraphics[width=\FIGWIDTH,height=\FIGHEIGHT]{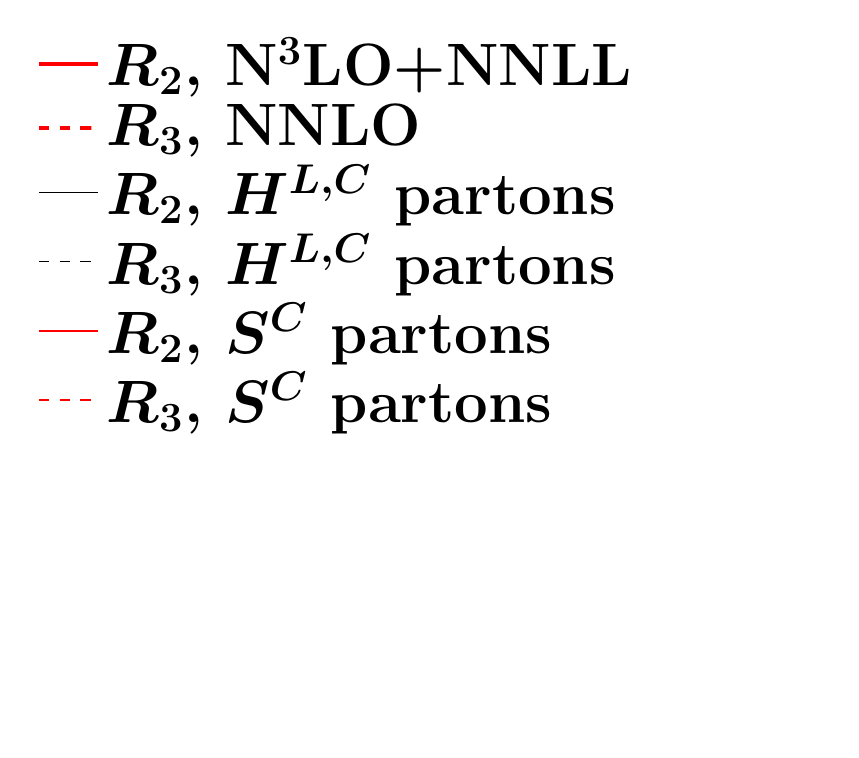}\\
\caption{Selected predictions  obtained  with $S^{C}$, $H^{C}$ and $H^{L}$ MC setups at  parton level
and theory predictions obtained with $\alpha_{S}(M_{Z})=0.1181$.}
\label{fig:partons:all}
\end{figure}

\begin{figure}[htbp]\centering
\includegraphics[width=\FIGWIDTH,height=\FIGHEIGHT]{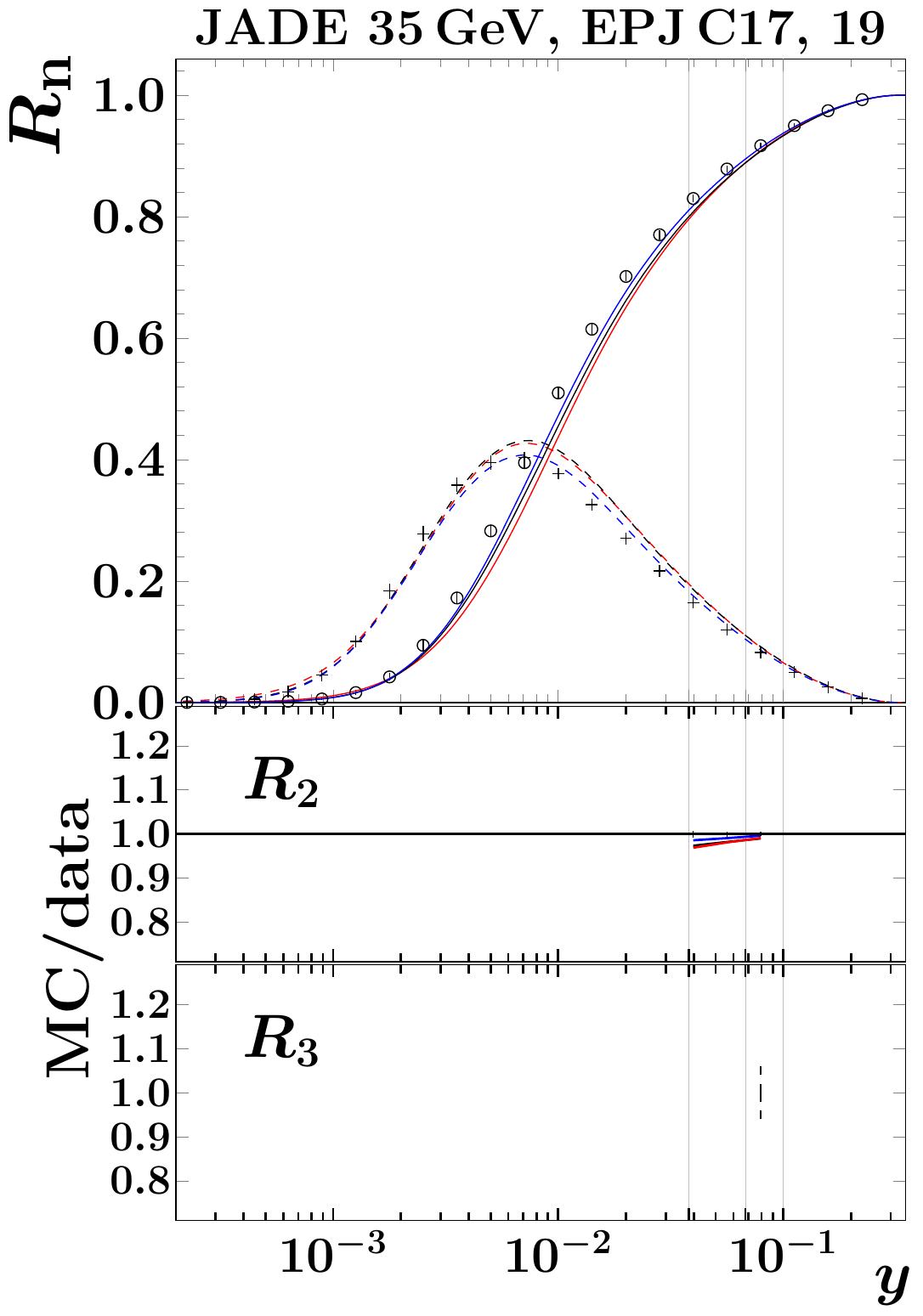}\includegraphics[width=\FIGWIDTH,height=\FIGHEIGHT]{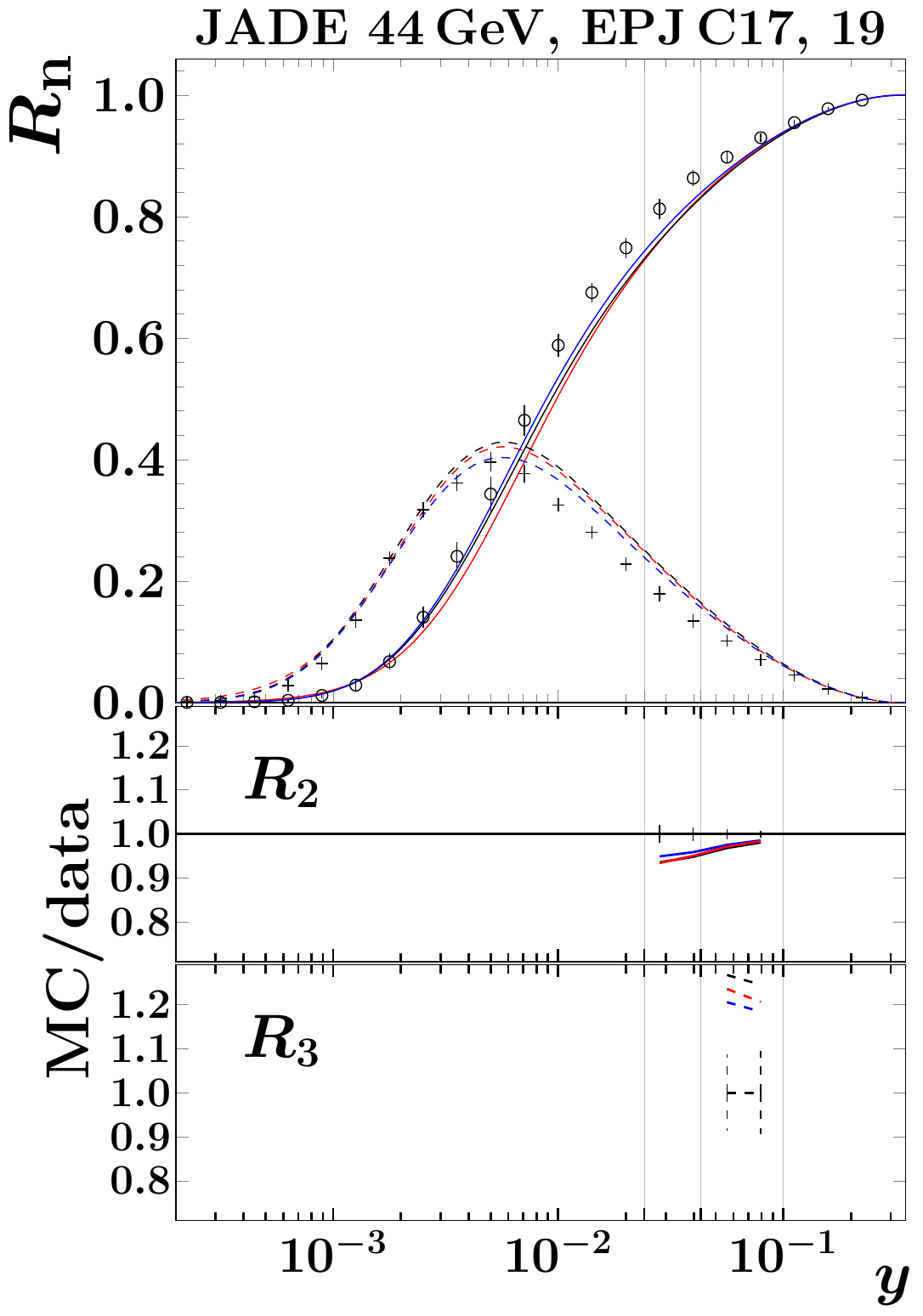}\epjcbreak{}\includegraphics[width=\FIGWIDTH,height=\FIGHEIGHT]{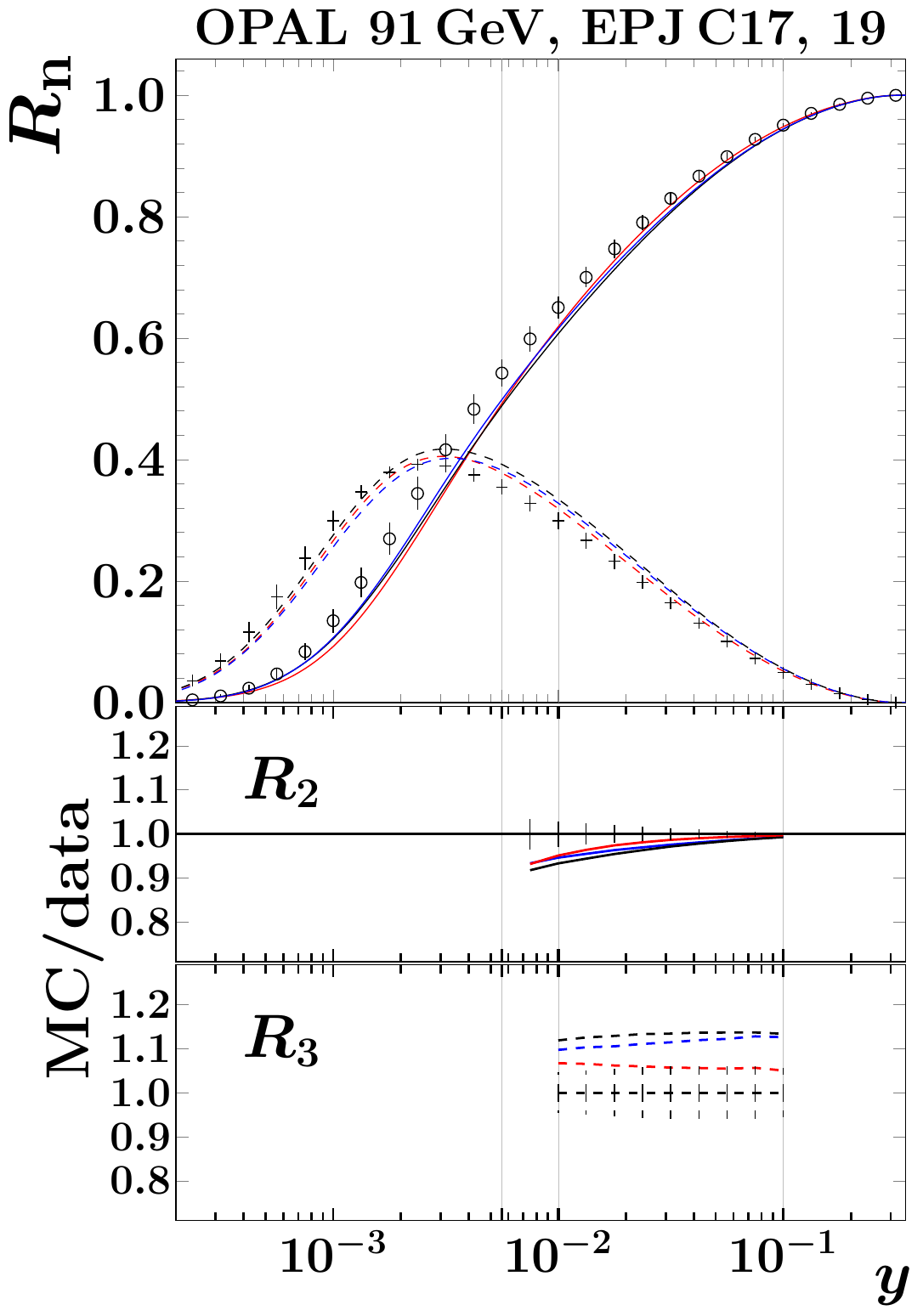}\draftbreak{}\arxivbreak{}\includegraphics[width=\FIGWIDTH,height=\FIGHEIGHT]{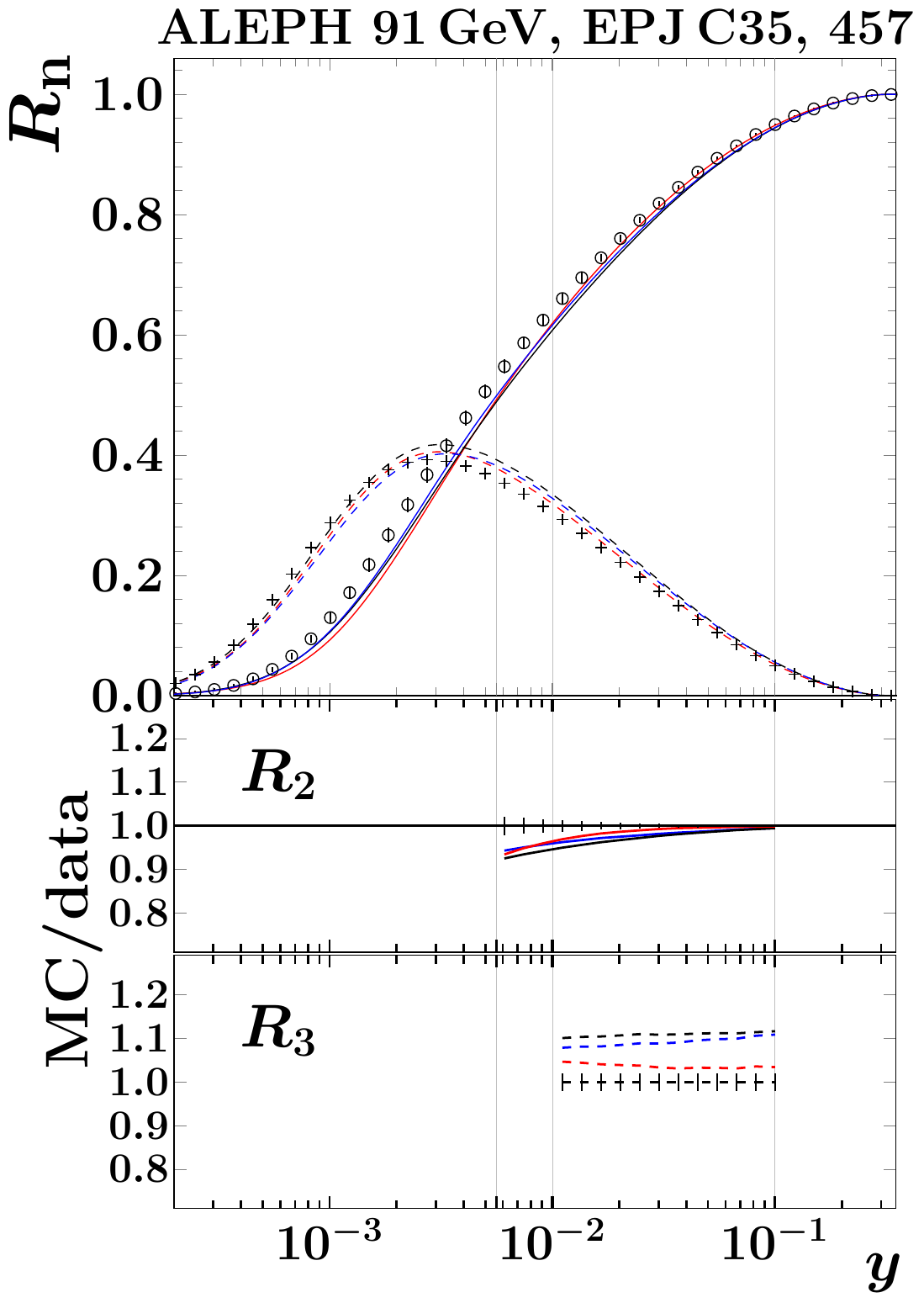}\epjcbreak{}\includegraphics[width=\FIGWIDTH,height=\FIGHEIGHT]{Figures/fighadronsOPAL-figure0.pdf}\includegraphics[width=\FIGWIDTH,height=\FIGHEIGHT]{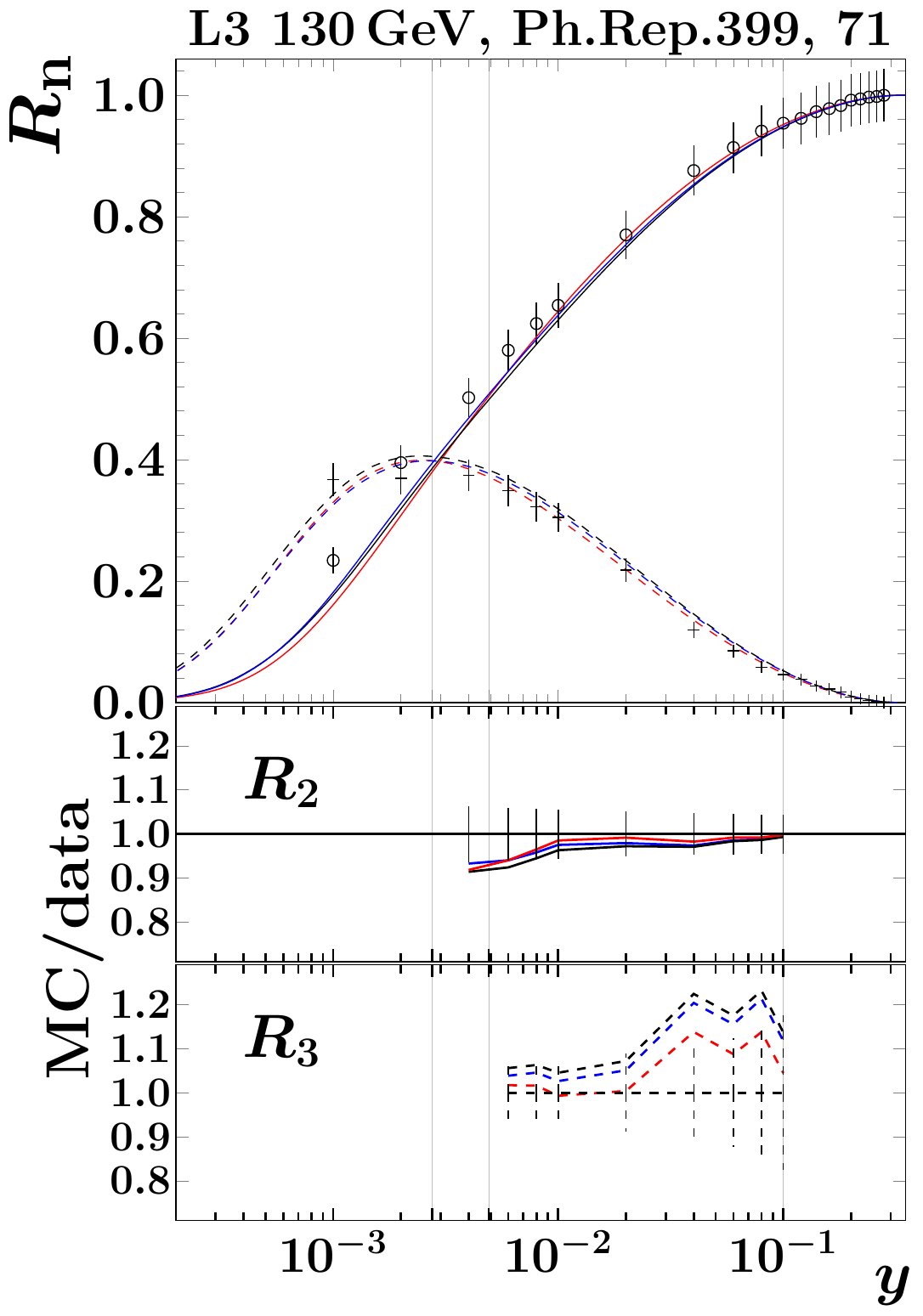}\epjcbreak{}\draftbreak{}\arxivbreak{}\includegraphics[width=\FIGWIDTH,height=\FIGHEIGHT]{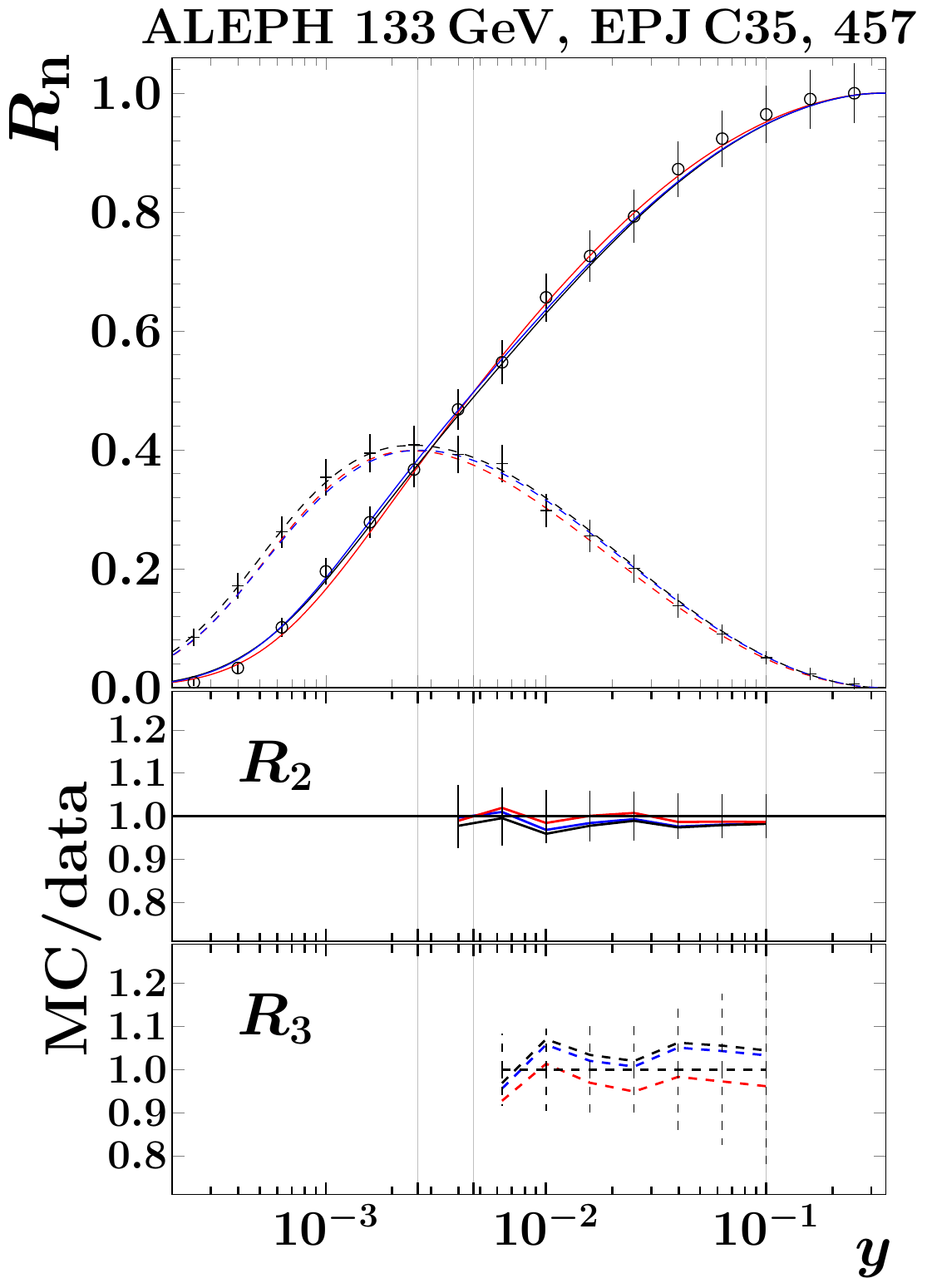}\includegraphics[width=\FIGWIDTH,height=\FIGHEIGHT]{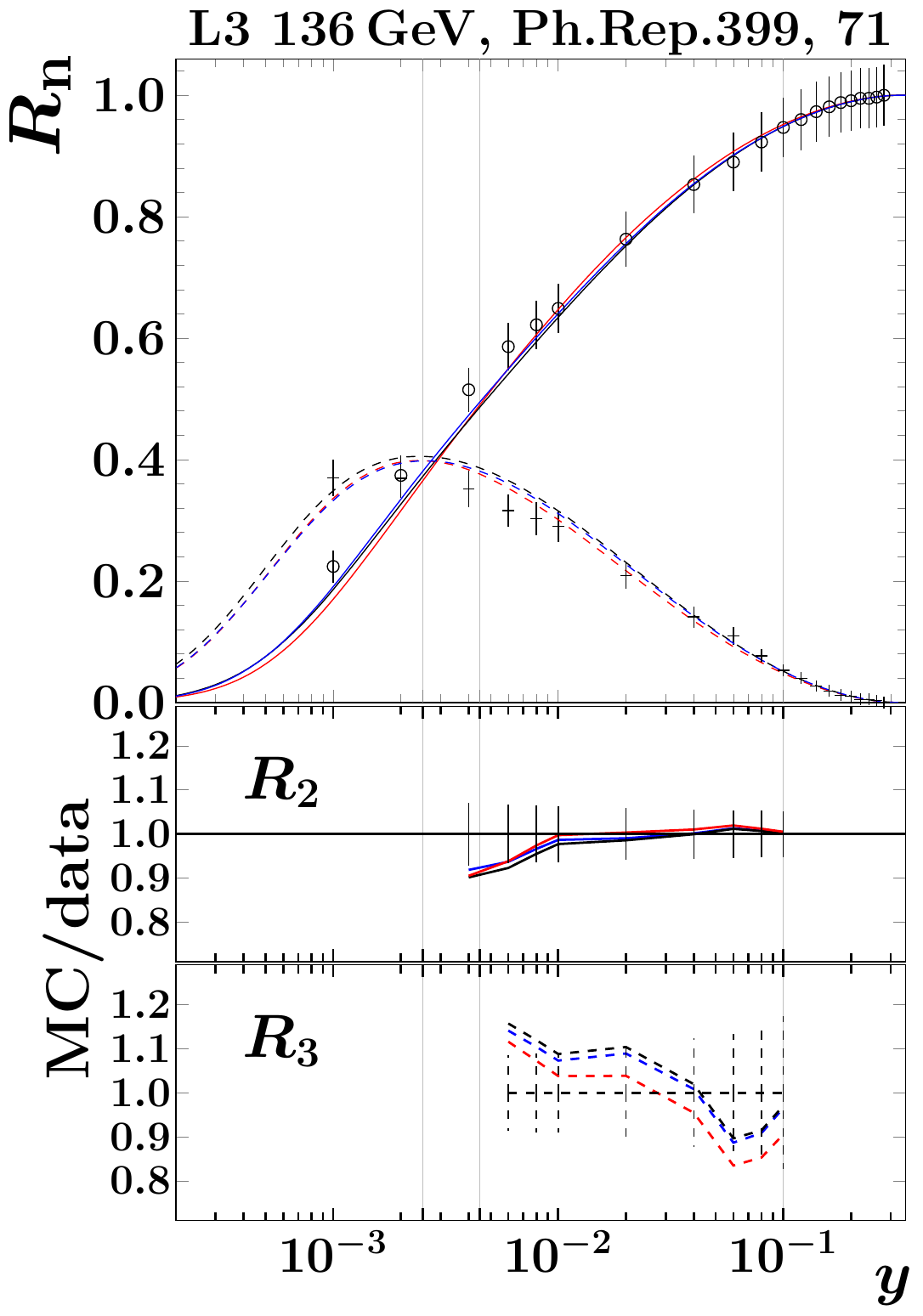}\includegraphics[width=\FIGWIDTH,height=\FIGHEIGHT]{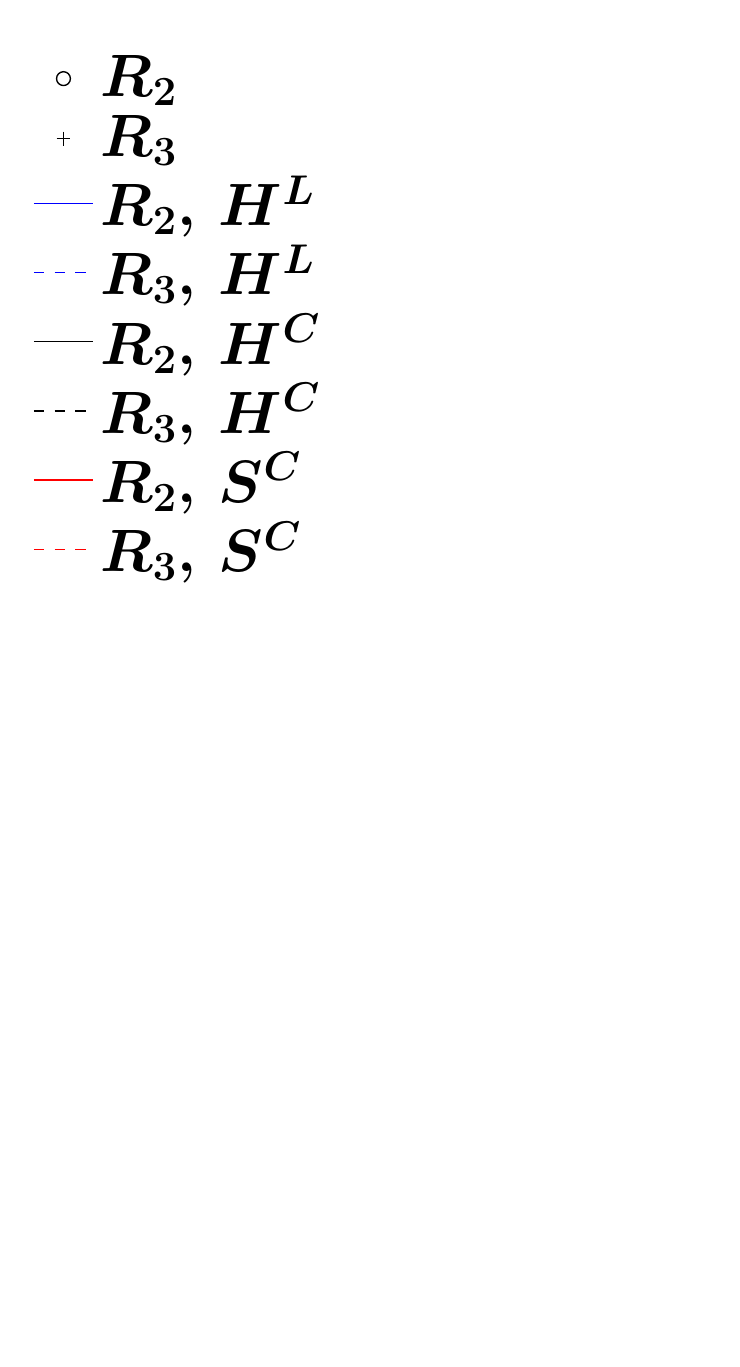}\\
\caption{Predictions  obtained  with $S^{C}$, $H^{C}$ and $H^{L}$ MC setups at  hadron level.}
\label{fig:hadrons:one}
\end{figure}

\begin{figure}[htbp]\centering
\includegraphics[width=\FIGWIDTH,height=\FIGHEIGHT]{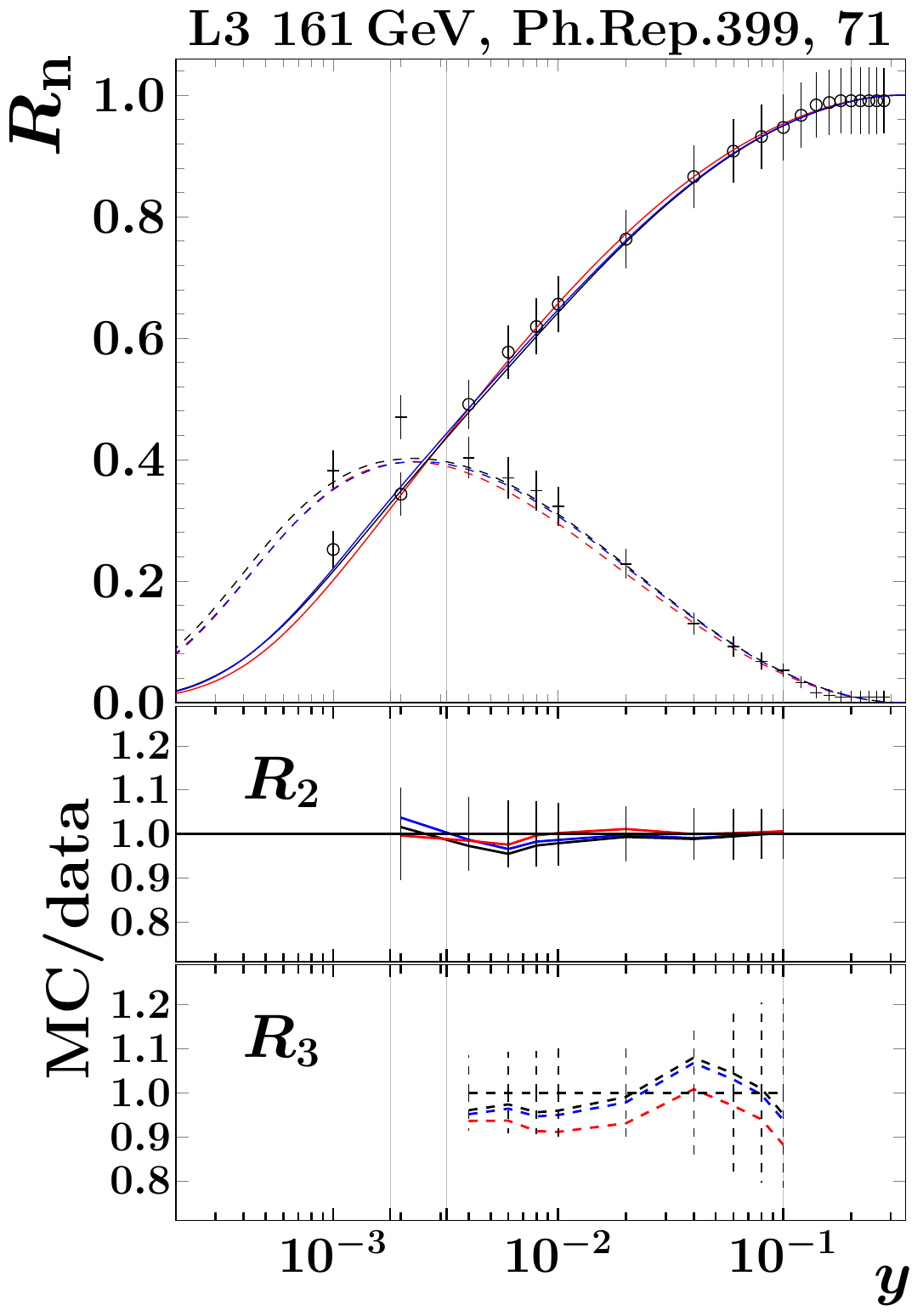}\includegraphics[width=\FIGWIDTH,height=\FIGHEIGHT]{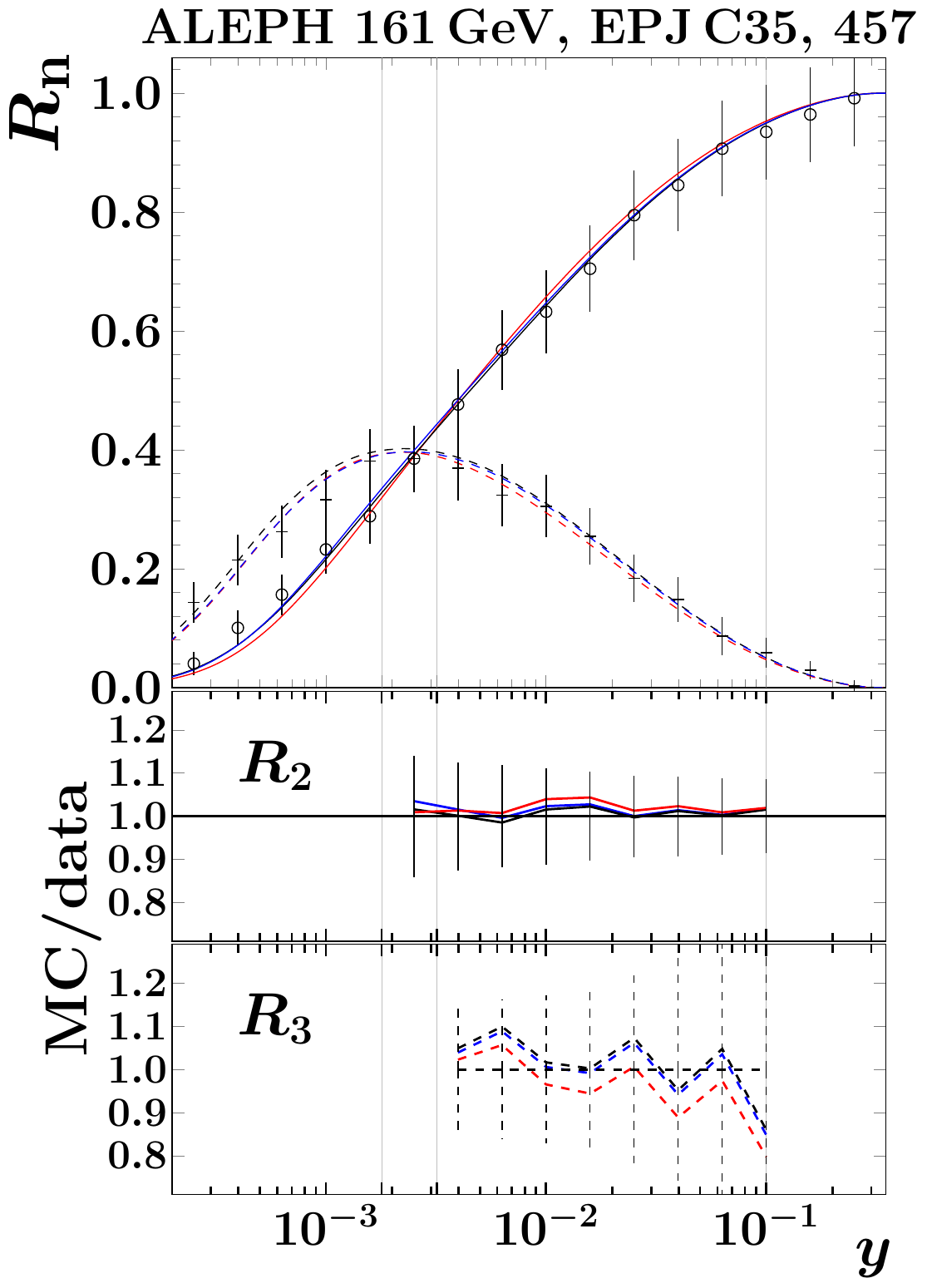}\epjcbreak{}\includegraphics[width=\FIGWIDTH,height=\FIGHEIGHT]{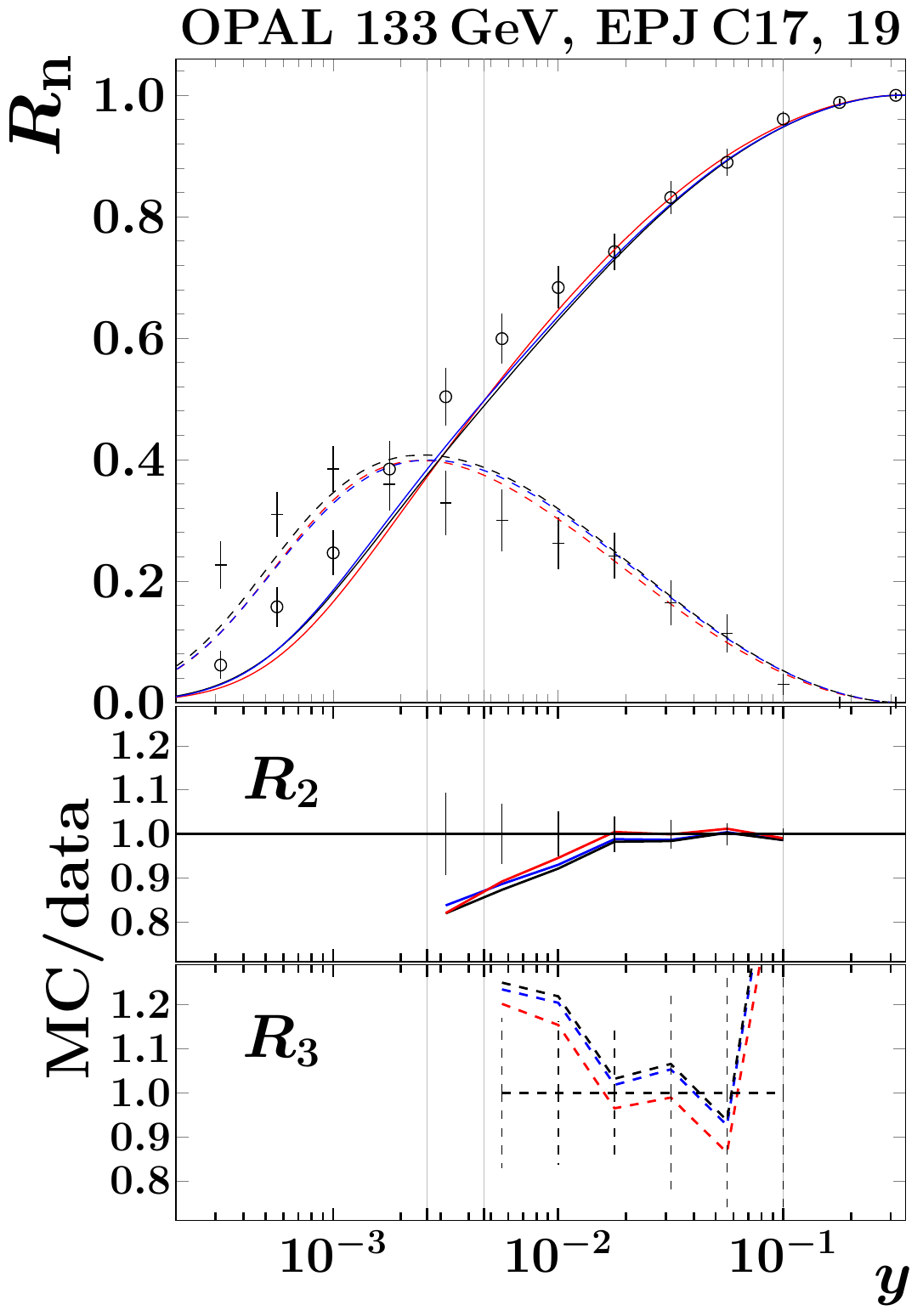}\draftbreak{}\arxivbreak{}\includegraphics[width=\FIGWIDTH,height=\FIGHEIGHT]{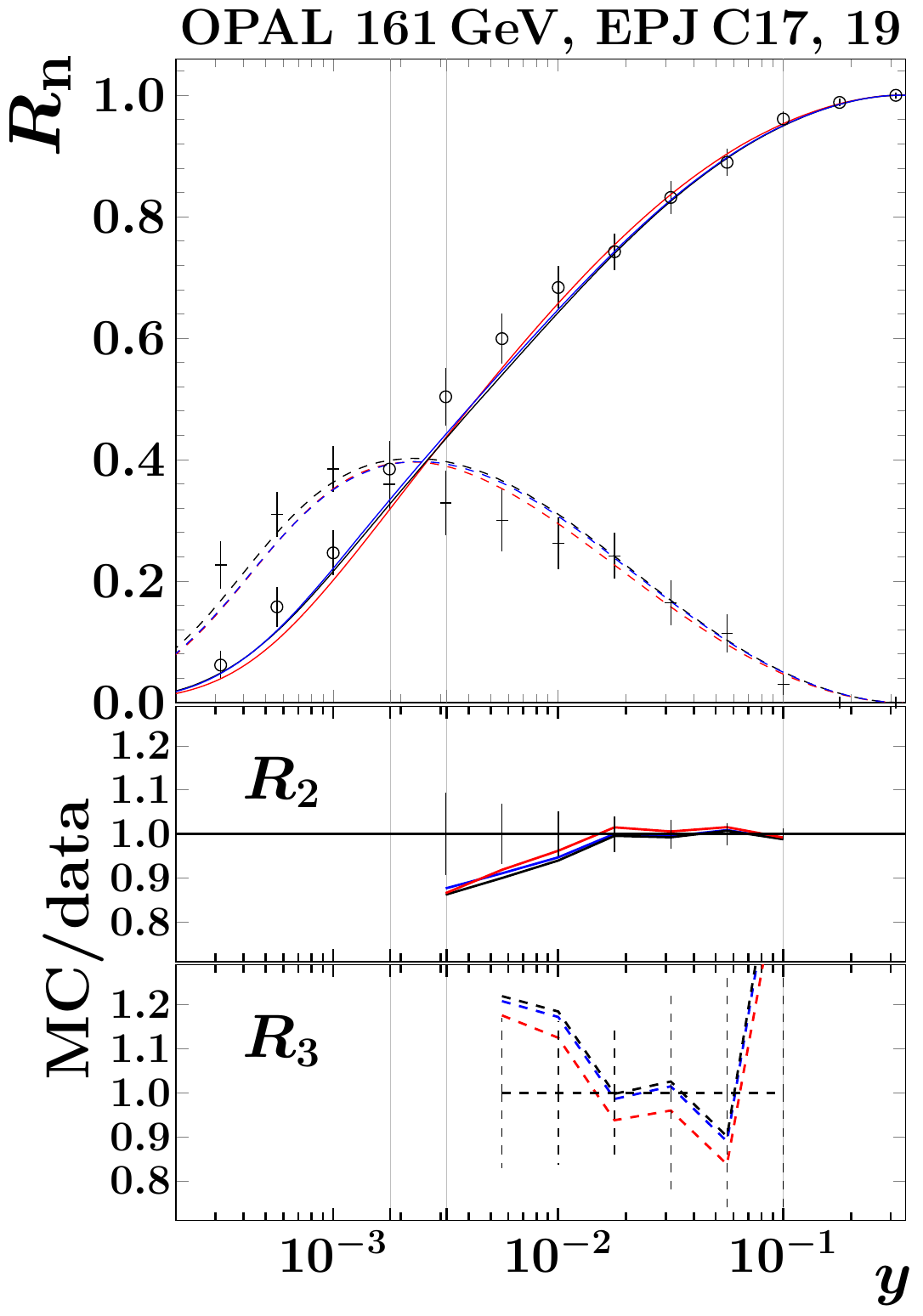}\epjcbreak{}\includegraphics[width=\FIGWIDTH,height=\FIGHEIGHT]{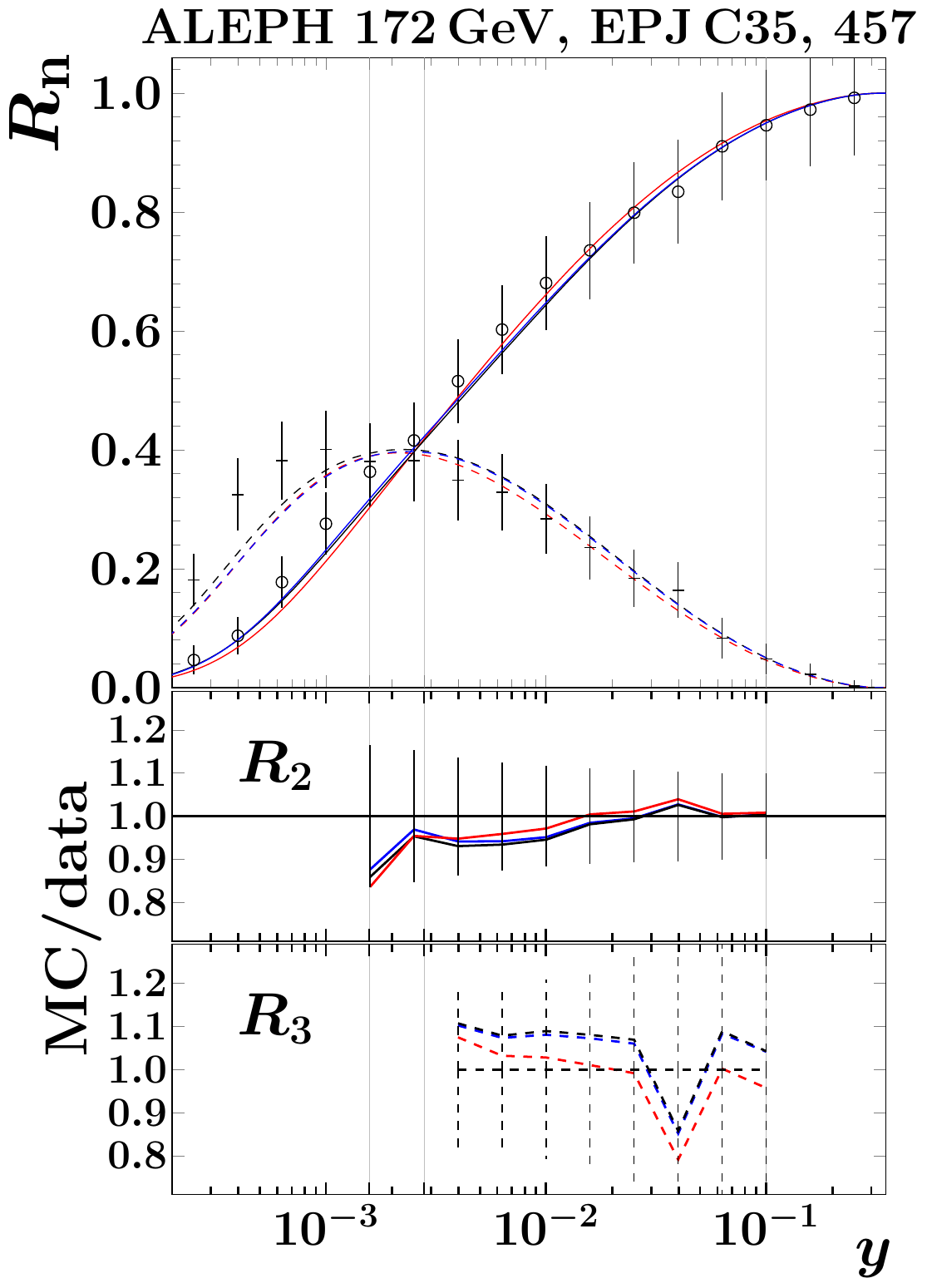}\includegraphics[width=\FIGWIDTH,height=\FIGHEIGHT]{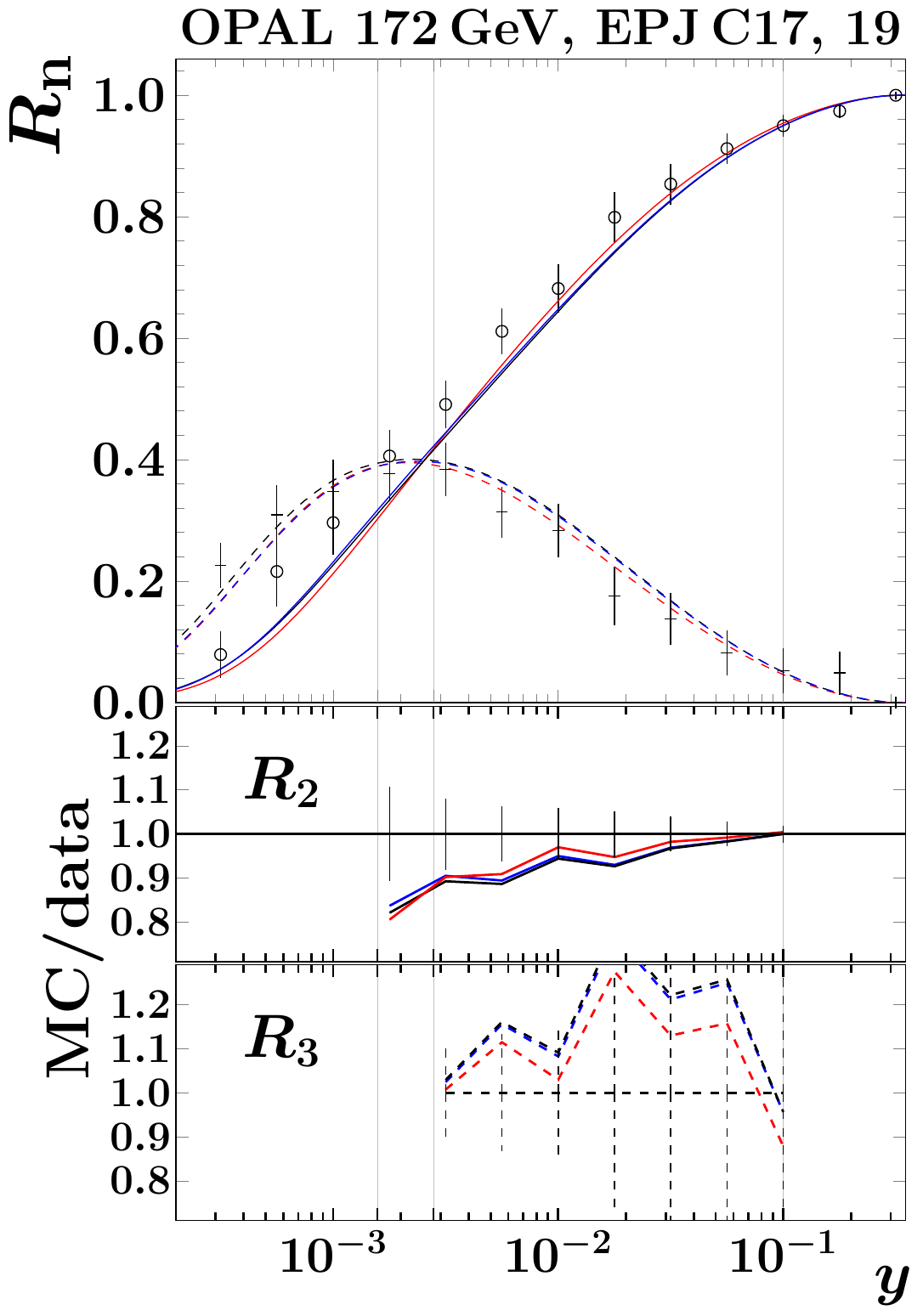}\epjcbreak{}\draftbreak{}\arxivbreak{}\includegraphics[width=\FIGWIDTH,height=\FIGHEIGHT]{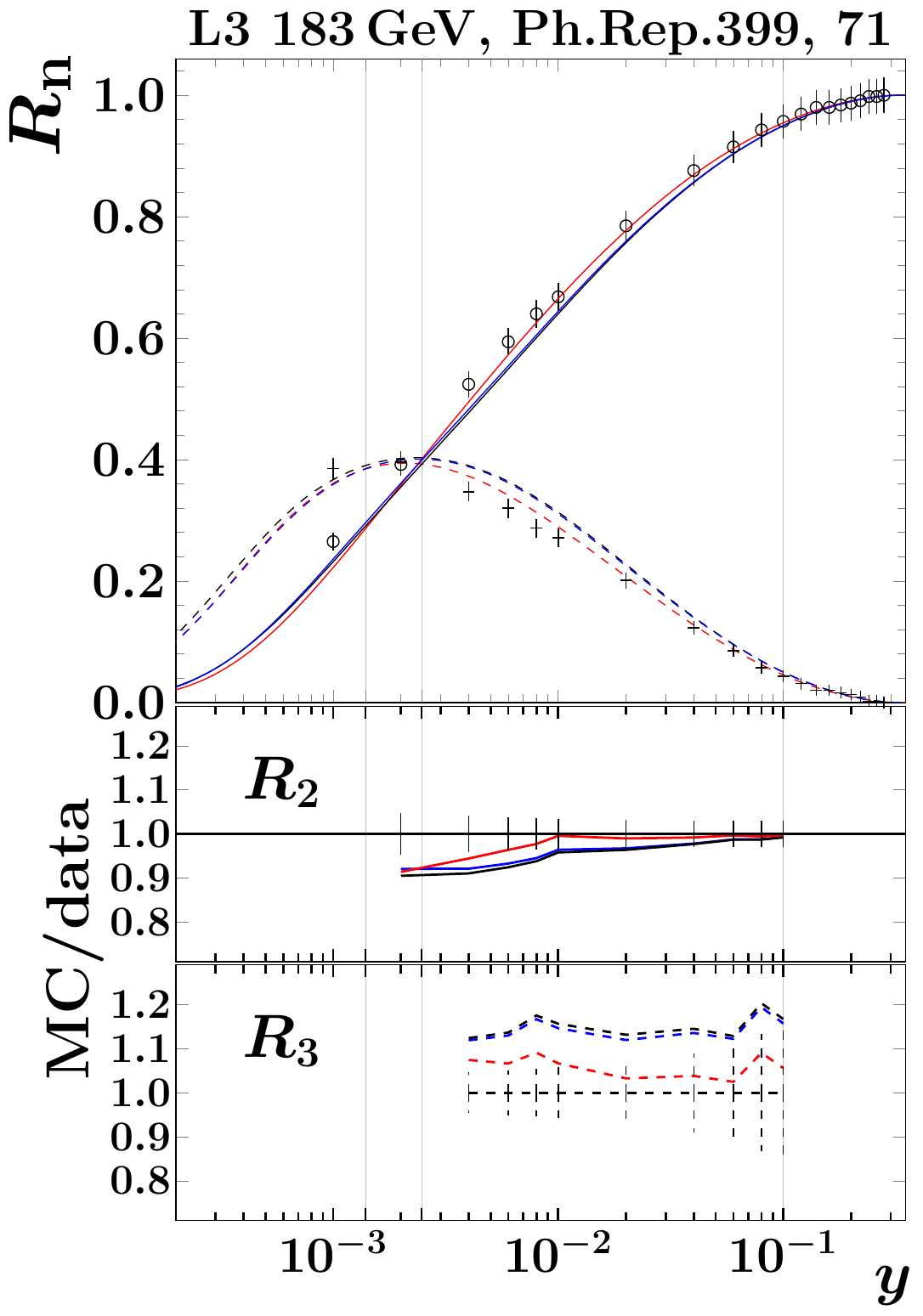}\includegraphics[width=\FIGWIDTH,height=\FIGHEIGHT]{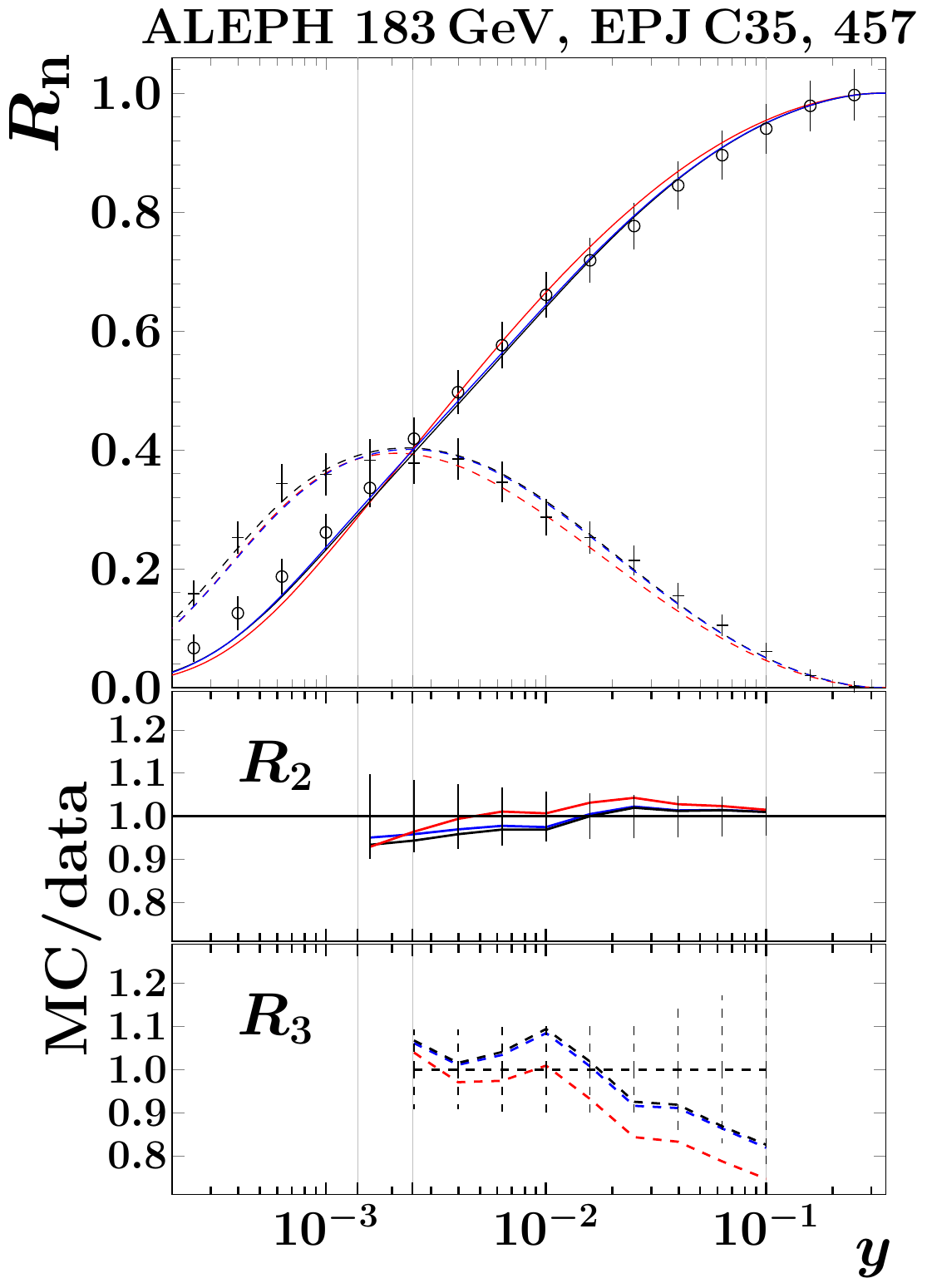}\includegraphics[width=\FIGWIDTH,height=\FIGHEIGHT]{Figures/fighadronsfoo-figure0.pdf}\\
\caption{Predictions  obtained  with $S^{C}$, $H^{C}$ and $H^{L}$ MC setups at  hadron level.}
\label{fig:hadrons:two}
\end{figure}

\begin{figure}[htbp]\centering
\includegraphics[width=\FIGWIDTH,height=\FIGHEIGHT]{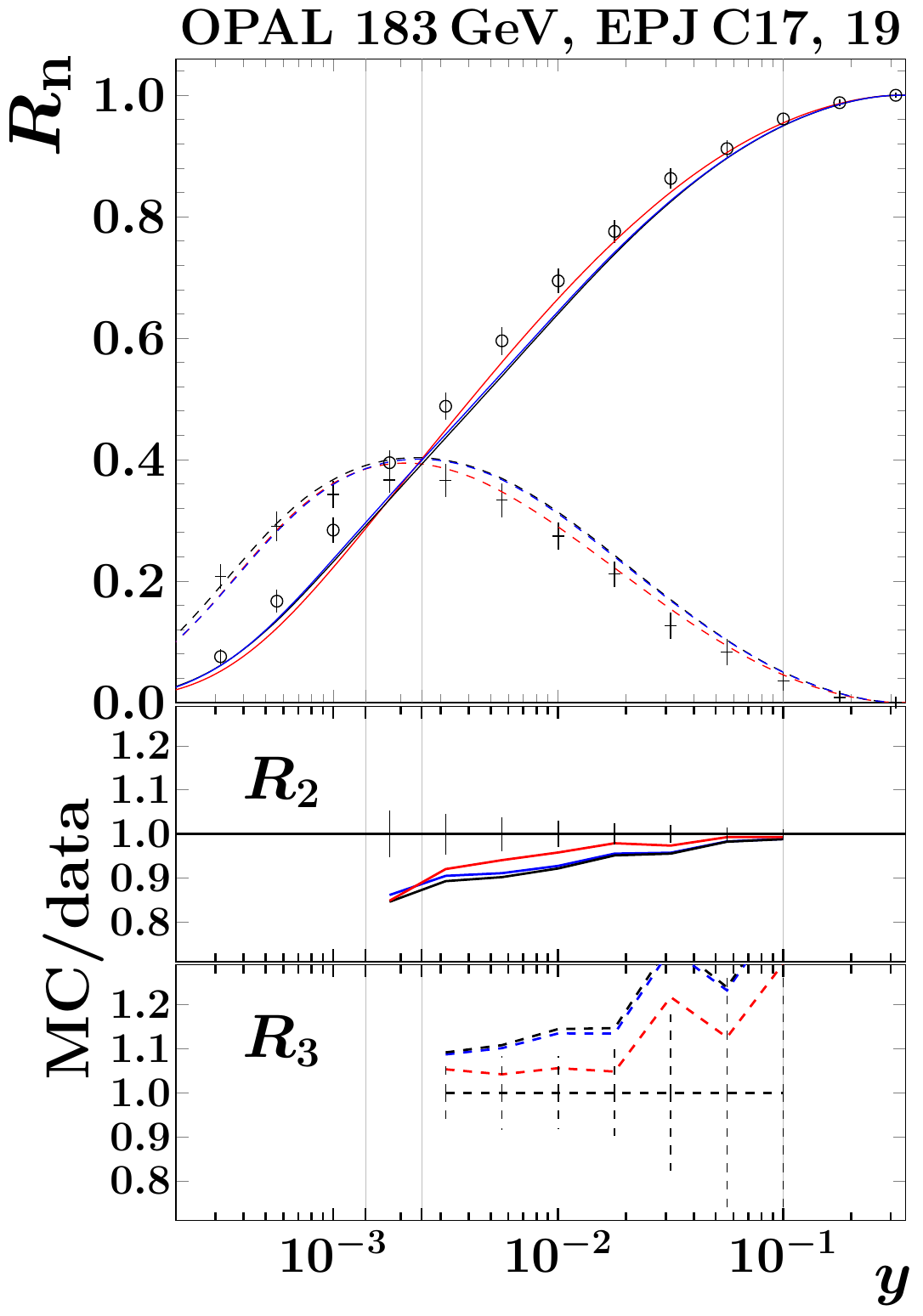}\includegraphics[width=\FIGWIDTH,height=\FIGHEIGHT]{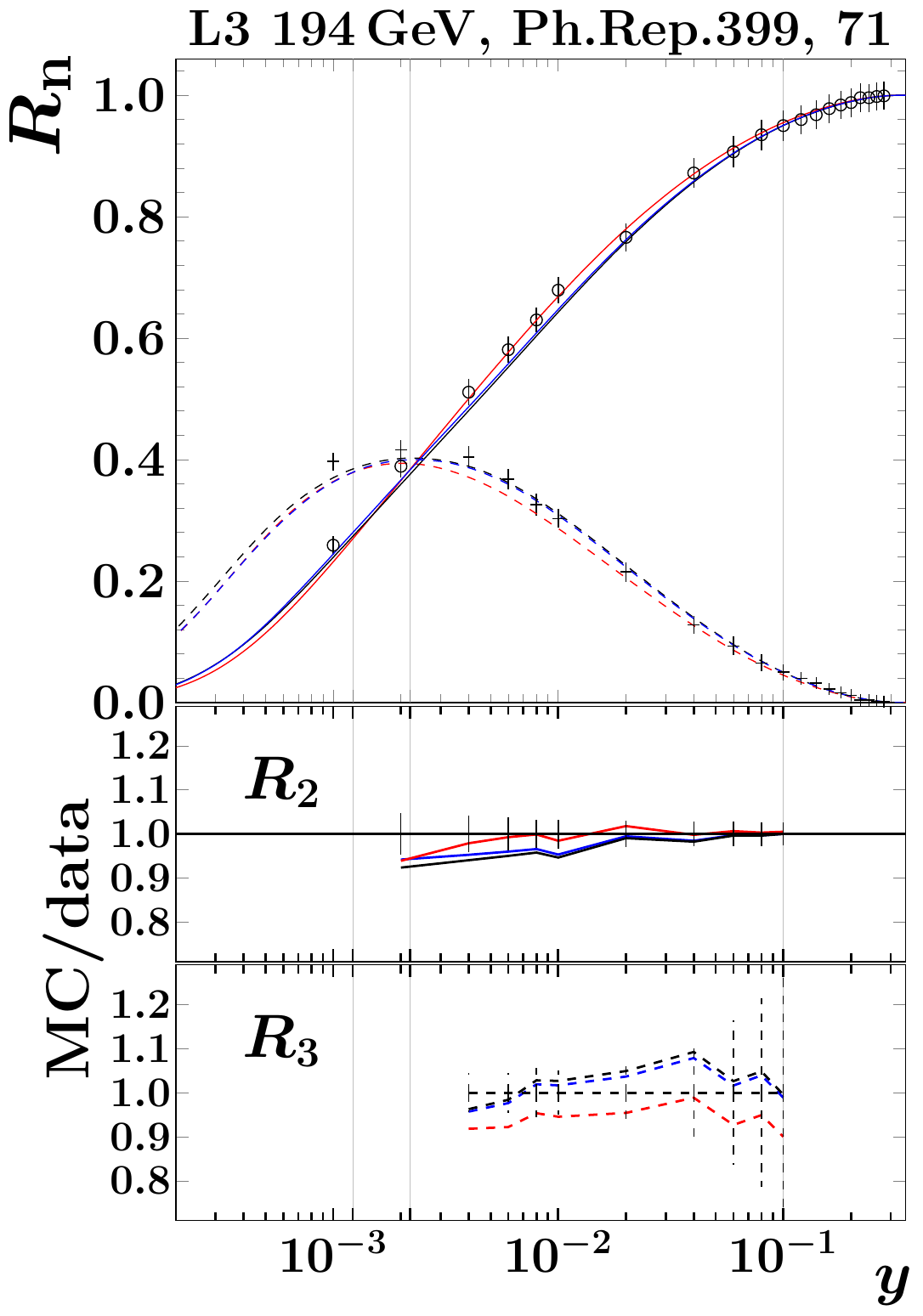}\epjcbreak{}\includegraphics[width=\FIGWIDTH,height=\FIGHEIGHT]{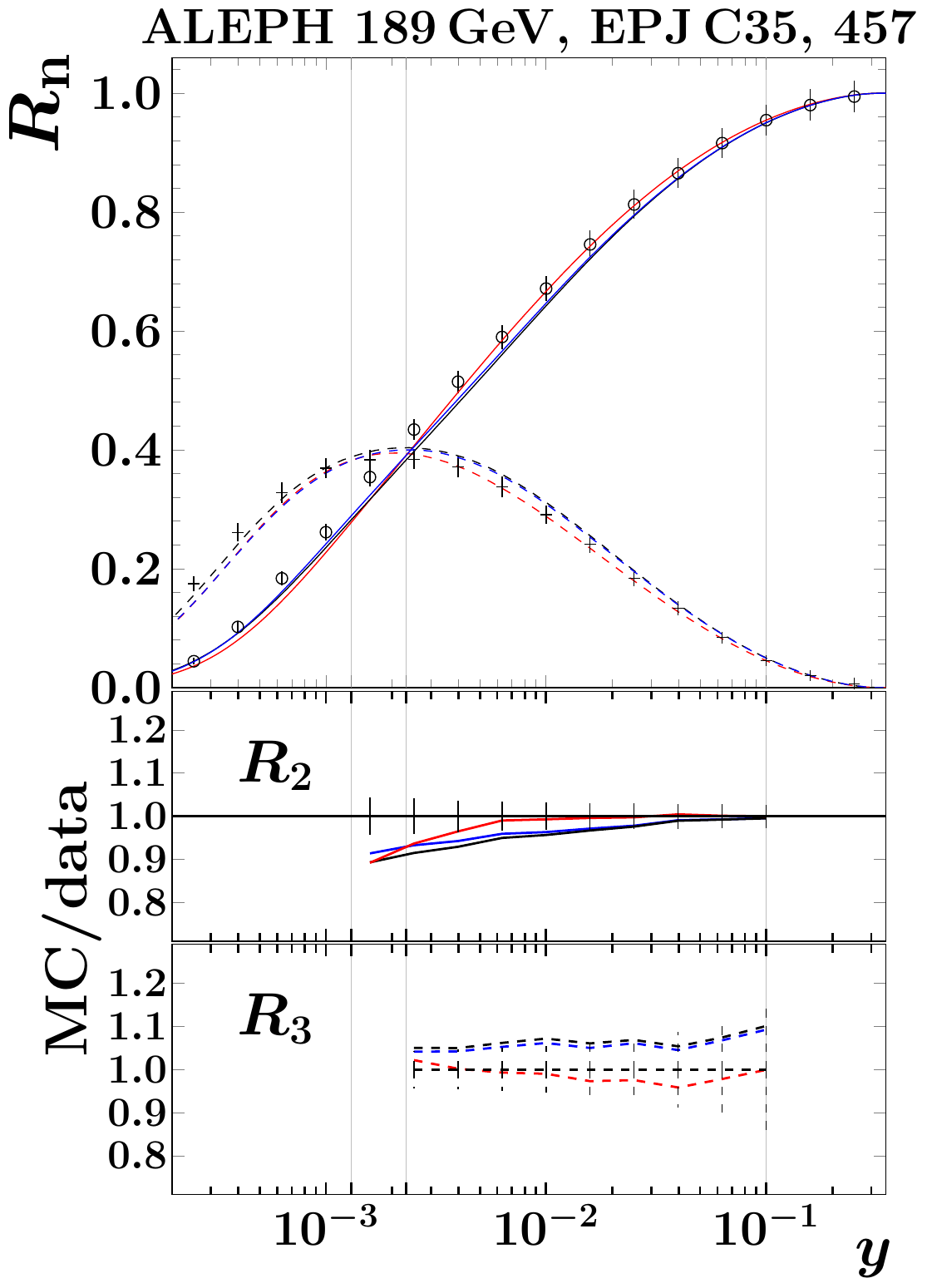}\draftbreak{}\arxivbreak{}\includegraphics[width=\FIGWIDTH,height=\FIGHEIGHT]{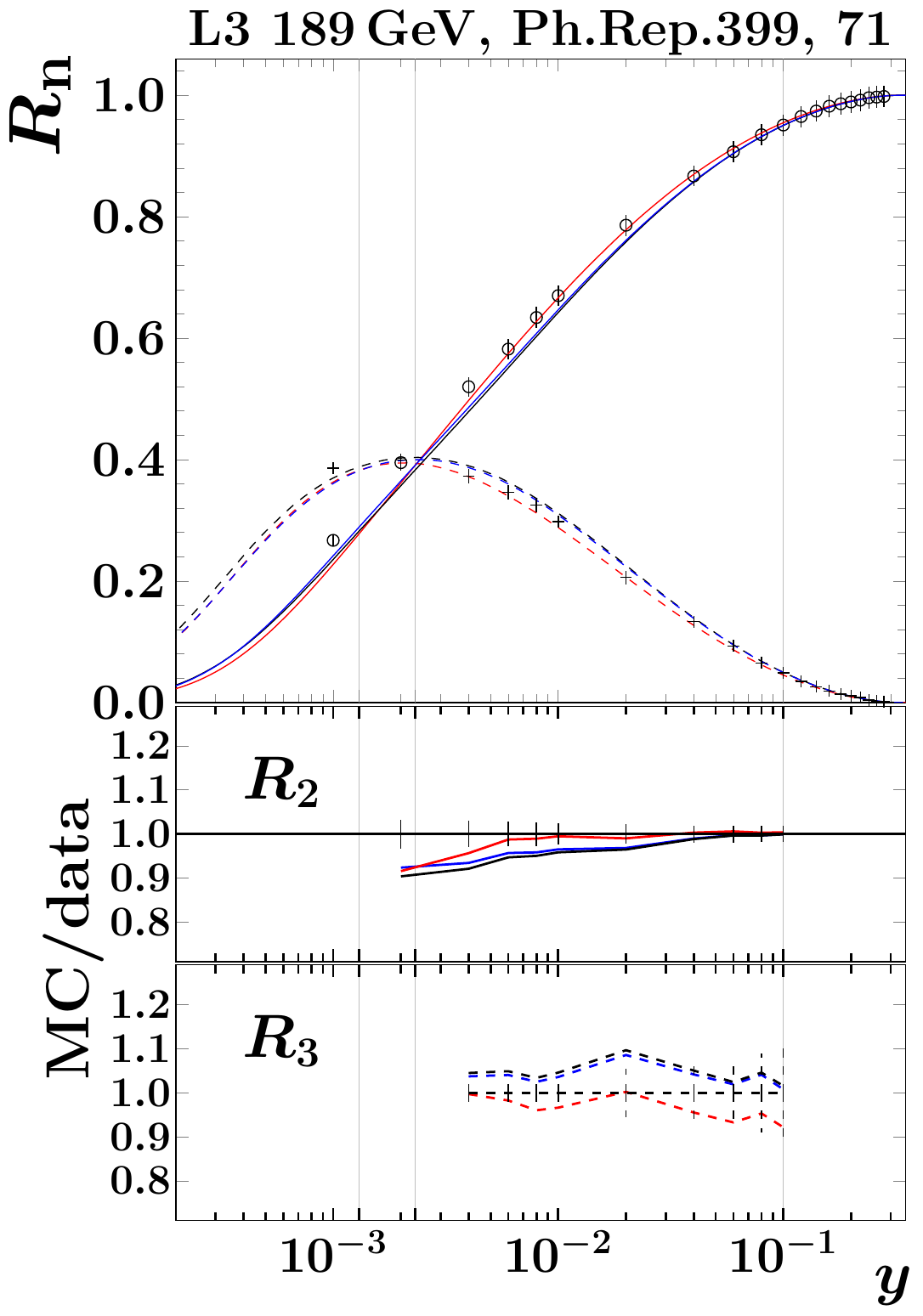}\epjcbreak{}\includegraphics[width=\FIGWIDTH,height=\FIGHEIGHT]{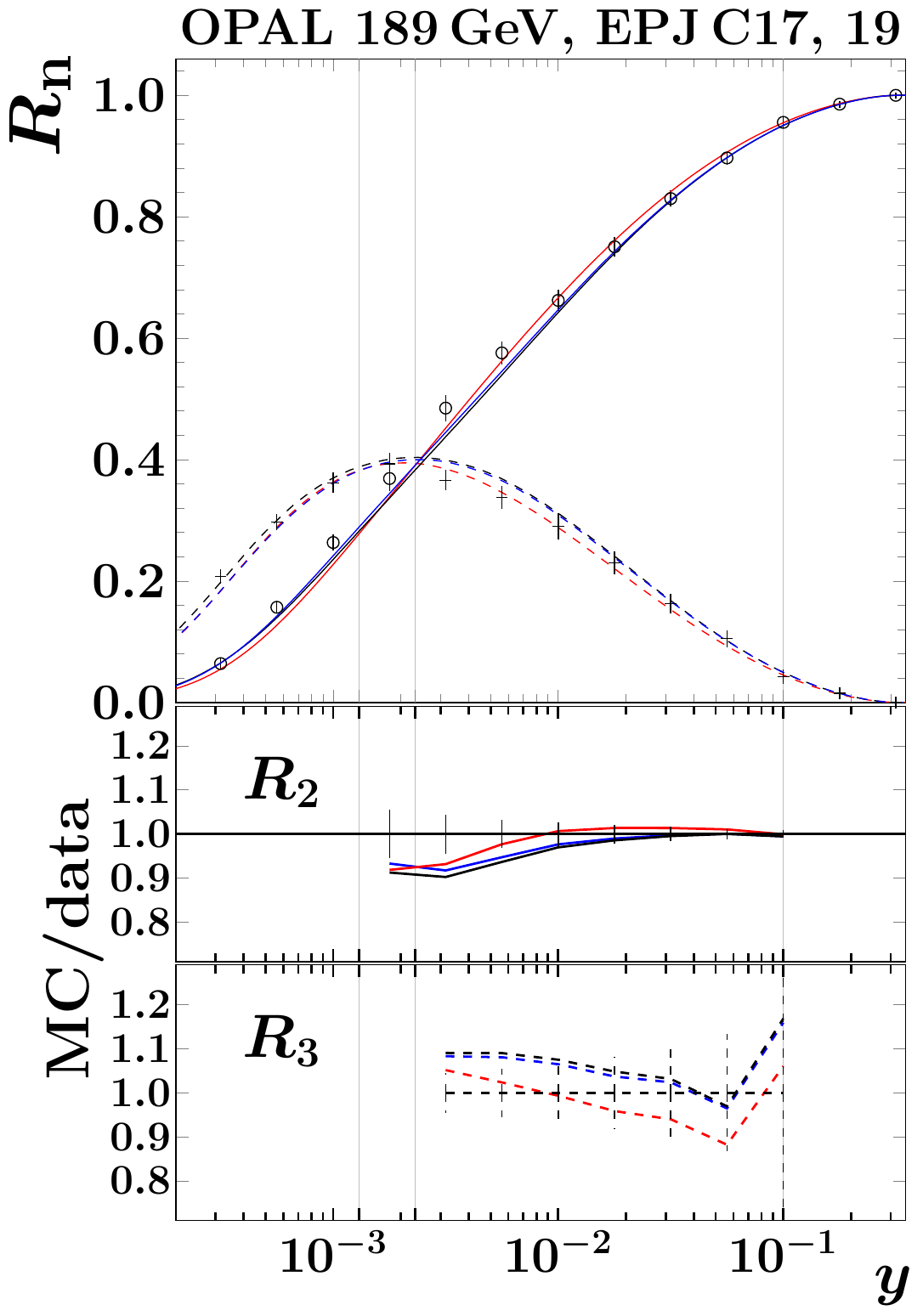}\includegraphics[width=\FIGWIDTH,height=\FIGHEIGHT]{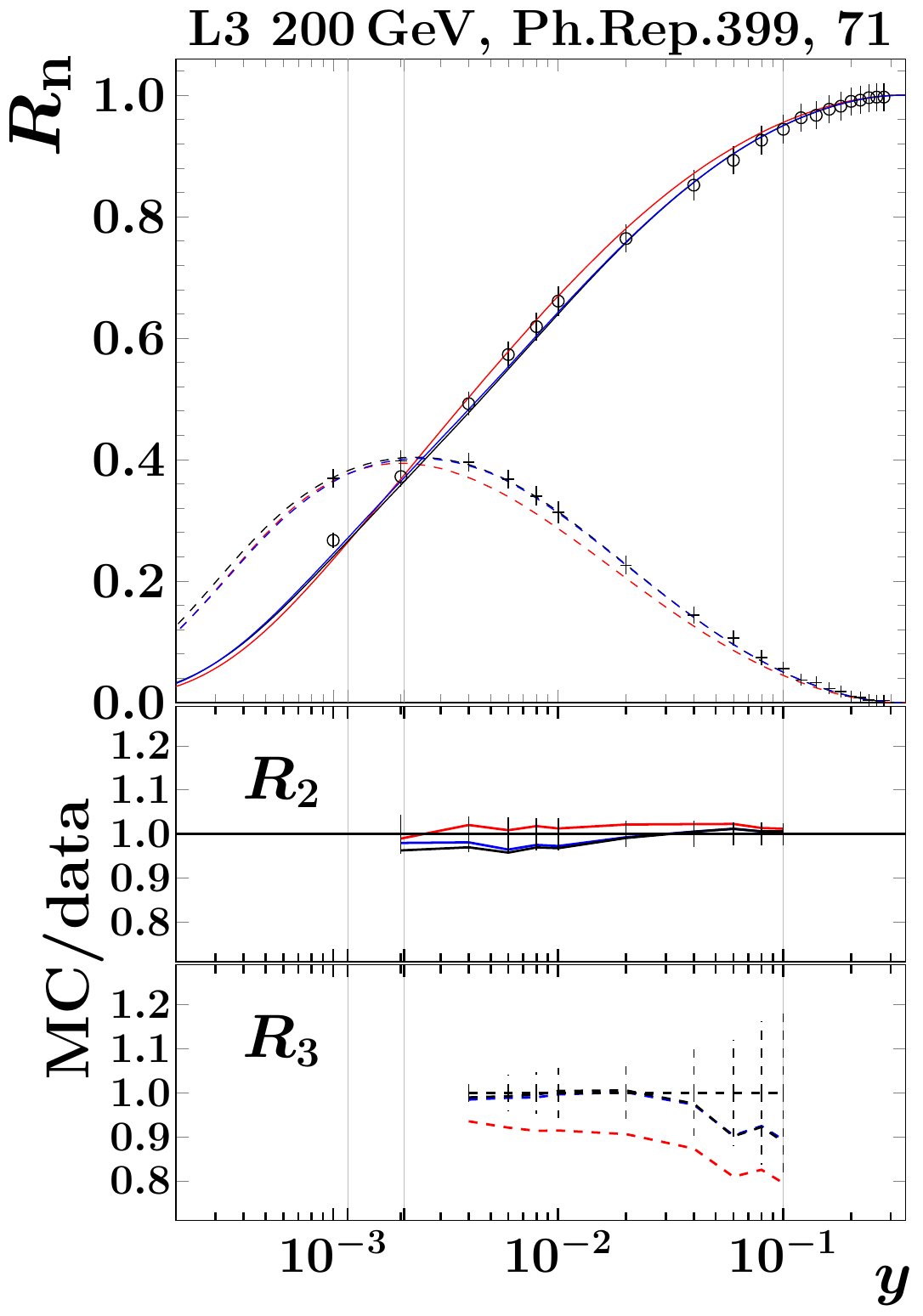}\epjcbreak{}\draftbreak{}\arxivbreak{}\includegraphics[width=\FIGWIDTH,height=\FIGHEIGHT]{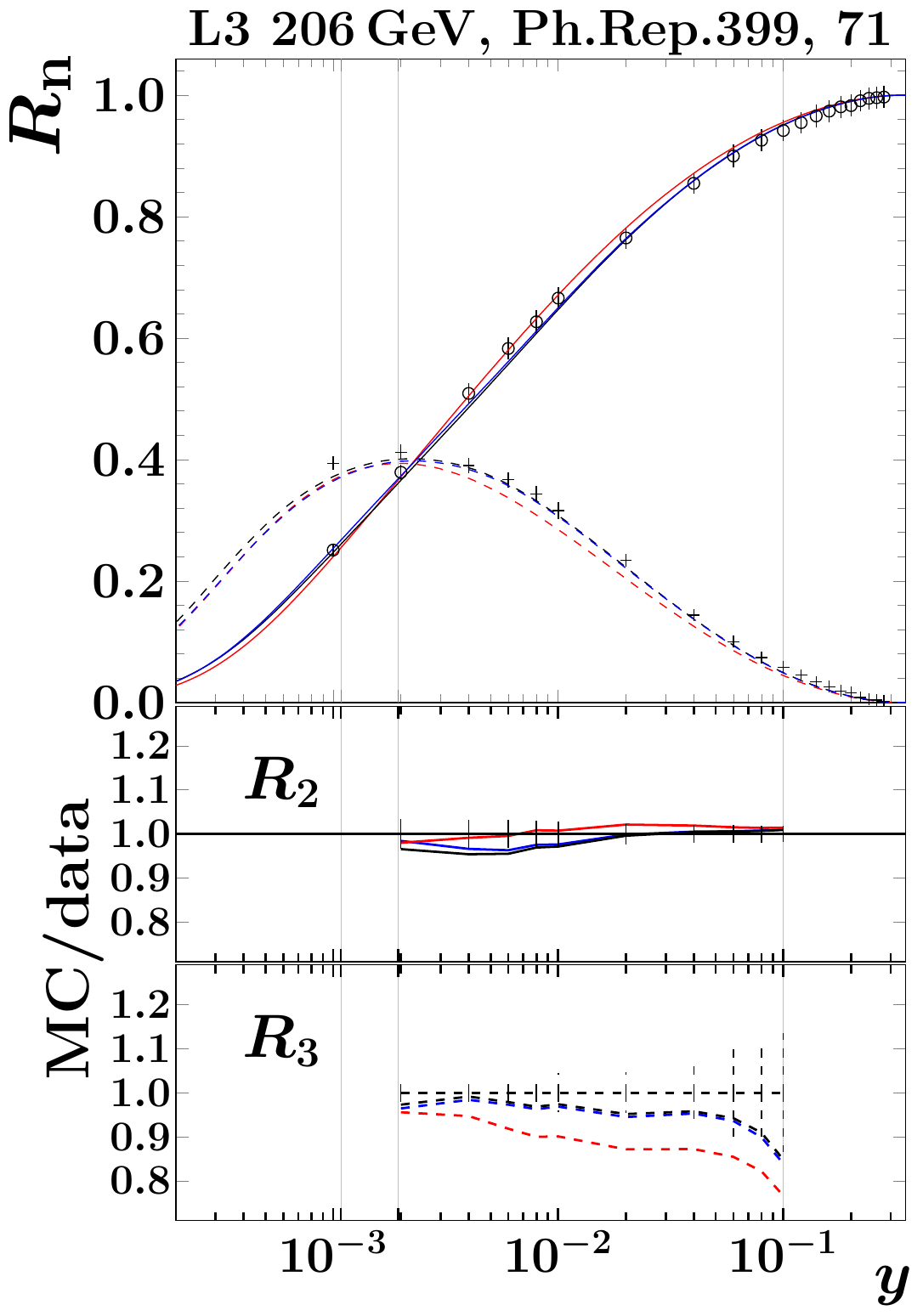}\includegraphics[width=\FIGWIDTH,height=\FIGHEIGHT]{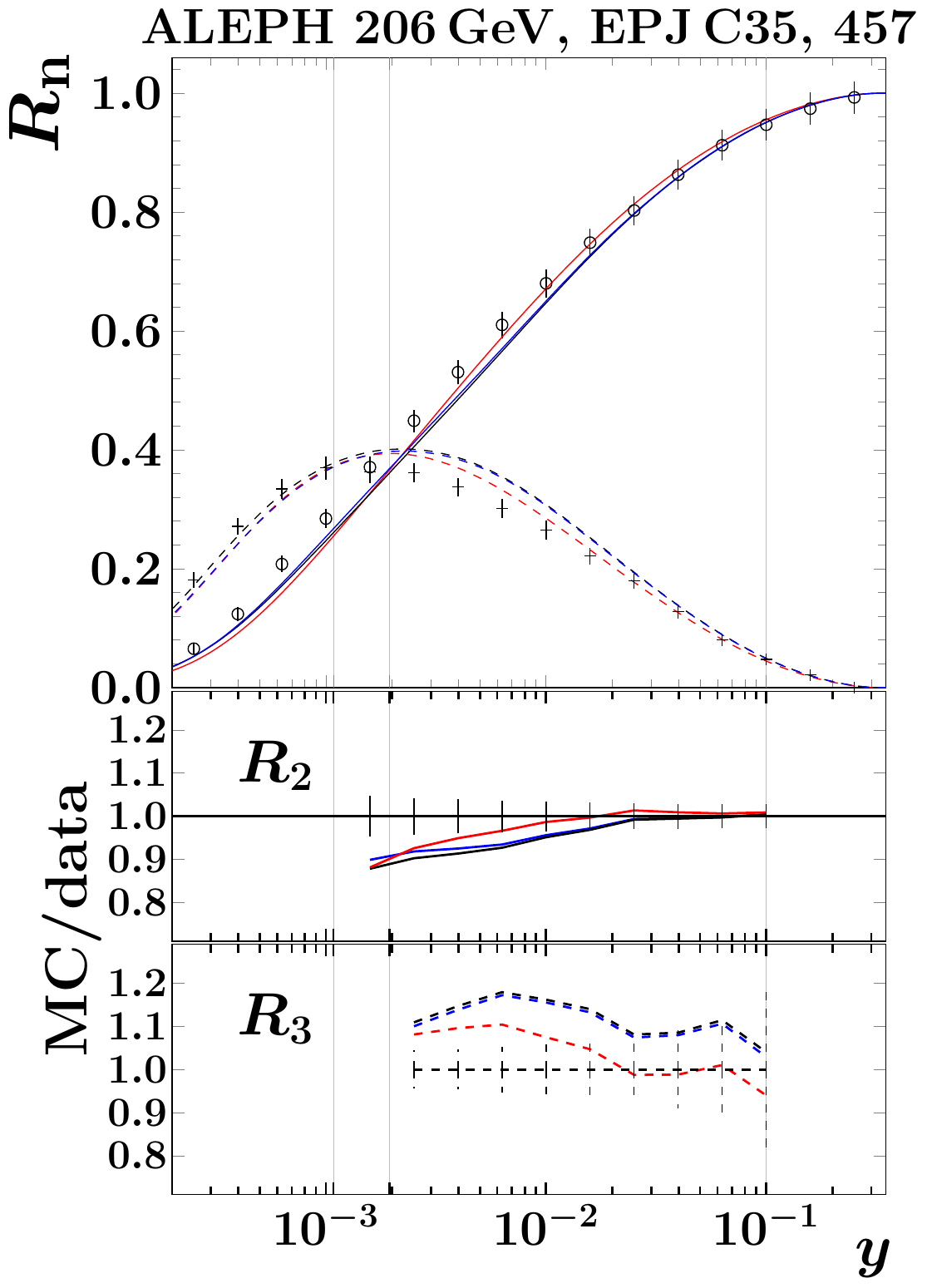}\includegraphics[width=\FIGWIDTH,height=\FIGHEIGHT]{Figures/fighadronsfoo-figure0.pdf}\\
\caption{Predictions  obtained  with $S^{C}$, $H^{C}$ and $H^{L}$ MC setups at  hadron level.}
\label{fig:hadrons:three}
\end{figure}

The figures in this section contain data measurements and predictions
for jet rates calculated in very fine binning at parton
(Fig.~\ref{fig:partons:all}) and hadron
(Figs.~\ref{fig:hadrons:one}, \ref{fig:hadrons:two} and
~\ref{fig:hadrons:three}) levels using different MC setups. The plots
with ratios of hadron level predictions to data measurements are
calculated with the binning used in the measurements.  The error bars
in the plots are calculated from the total uncertainty of the data
points. The vertical lines in the plots show the fit ranges used in
this work for fits of the $R_2$ and $R_3$ observables.

\FloatBarrier 
\section{Hadronization corrections}
\label{app:had}
\begin{figure}[htbp]\centering
\includegraphics[width=\FIGWIDTH,height=\FIGHEIGHT]{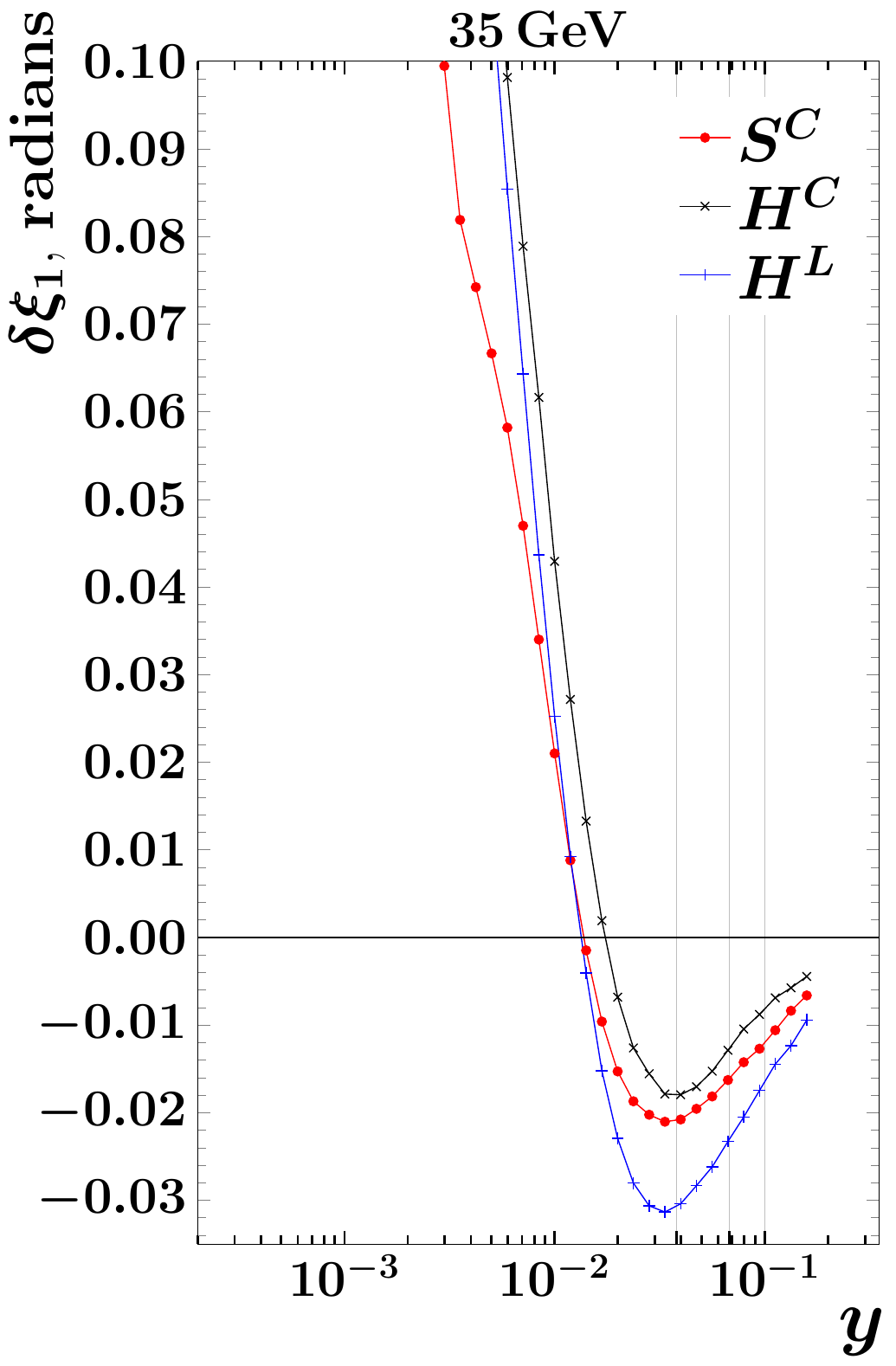}\includegraphics[width=\FIGWIDTH,height=\FIGHEIGHT]{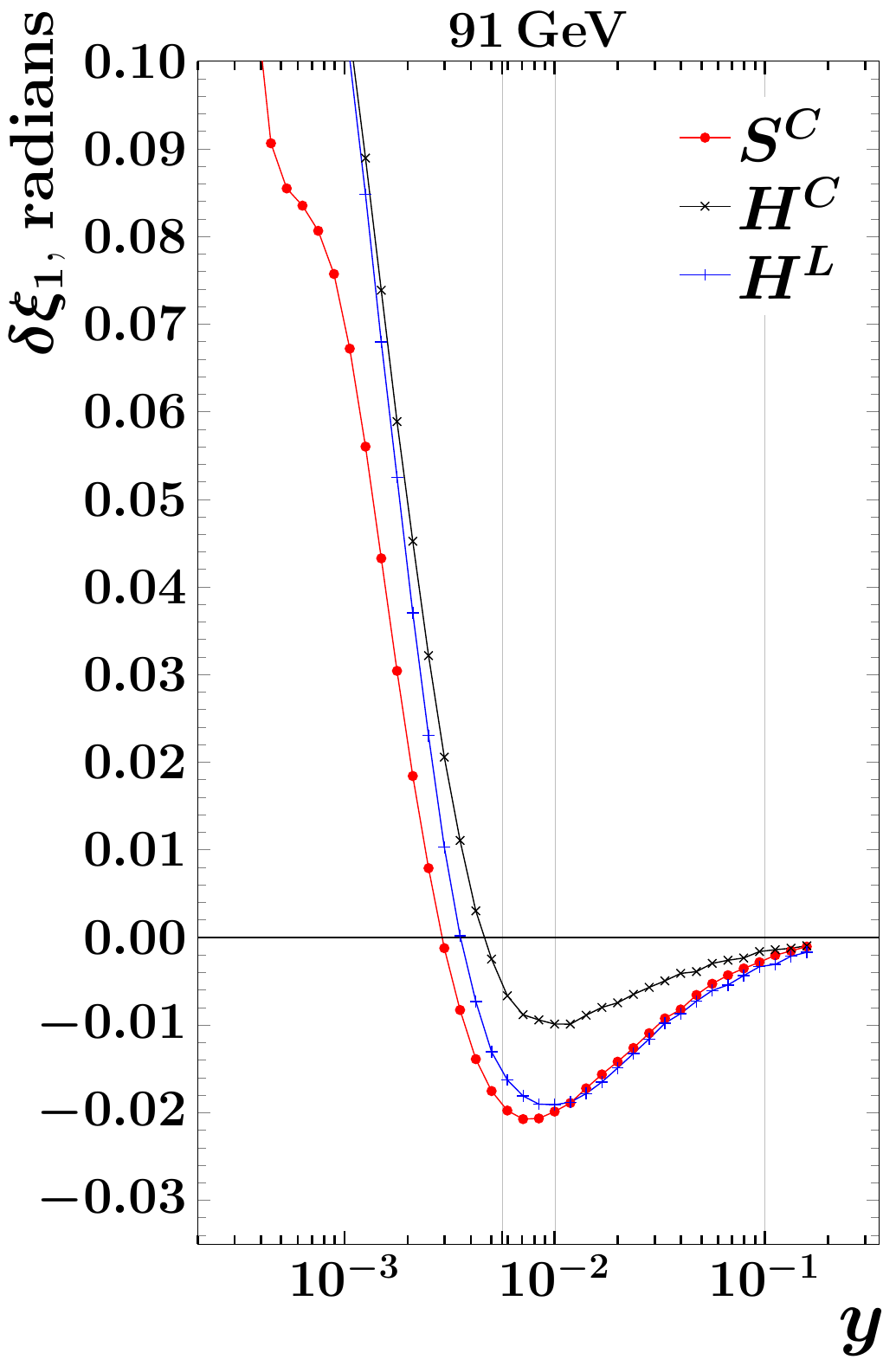}\epjcbreak{}\includegraphics[width=\FIGWIDTH,height=\FIGHEIGHT]{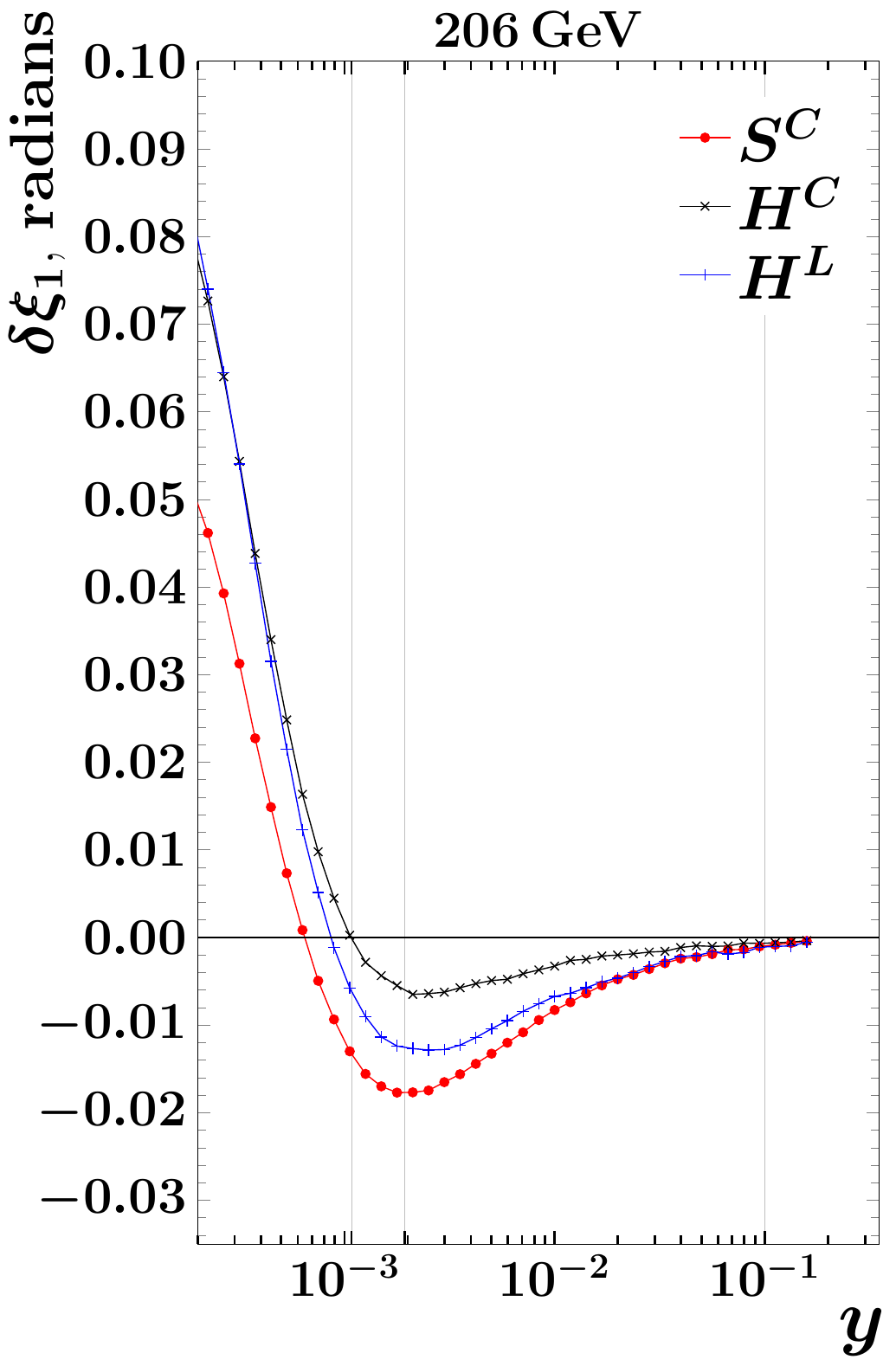}\draftbreak{}\arxivbreak{}\includegraphics[width=\FIGWIDTH,height=\FIGHEIGHT]{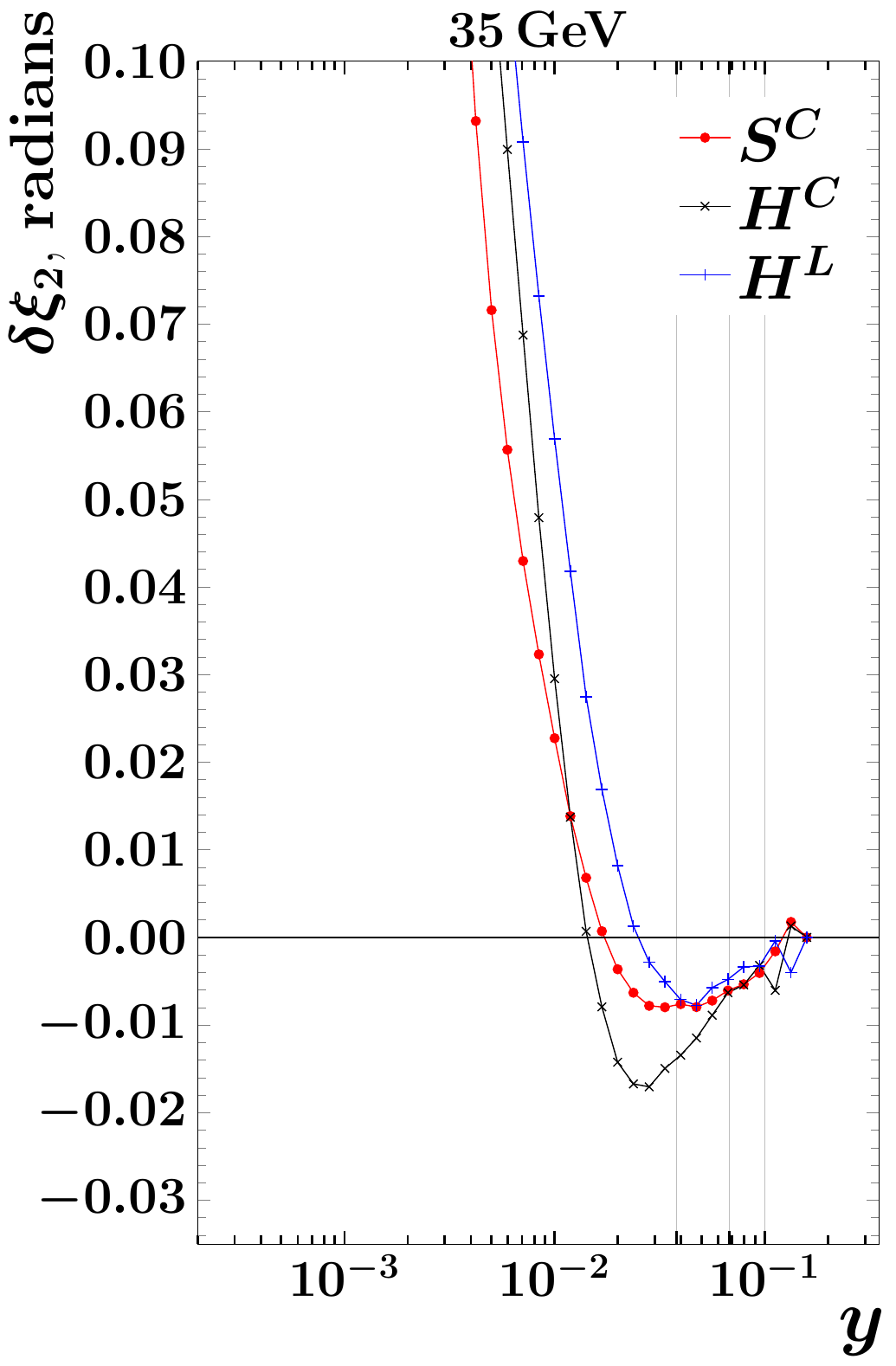}\epjcbreak{}\includegraphics[width=\FIGWIDTH,height=\FIGHEIGHT]{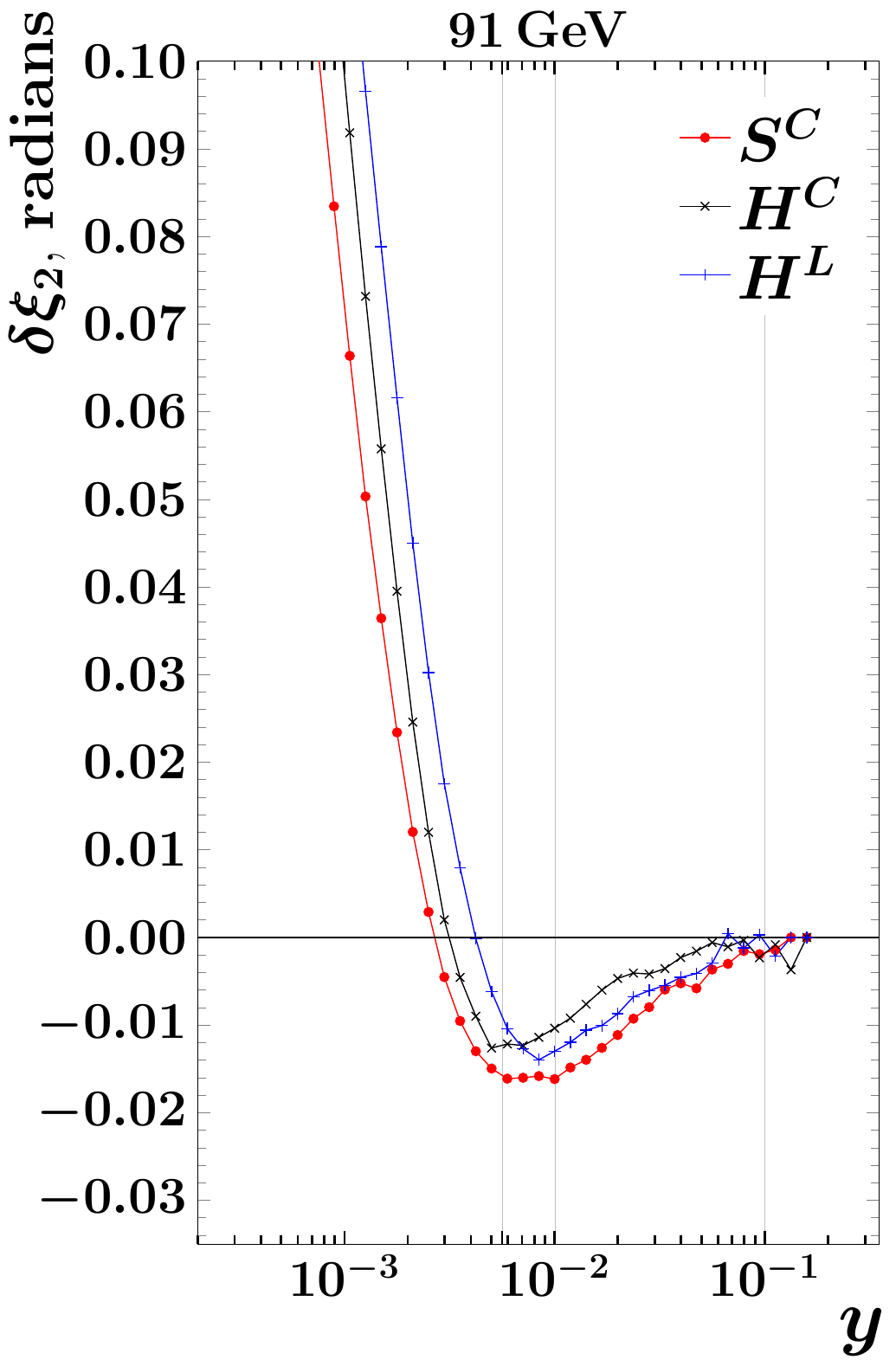}\includegraphics[width=\FIGWIDTH,height=\FIGHEIGHT]{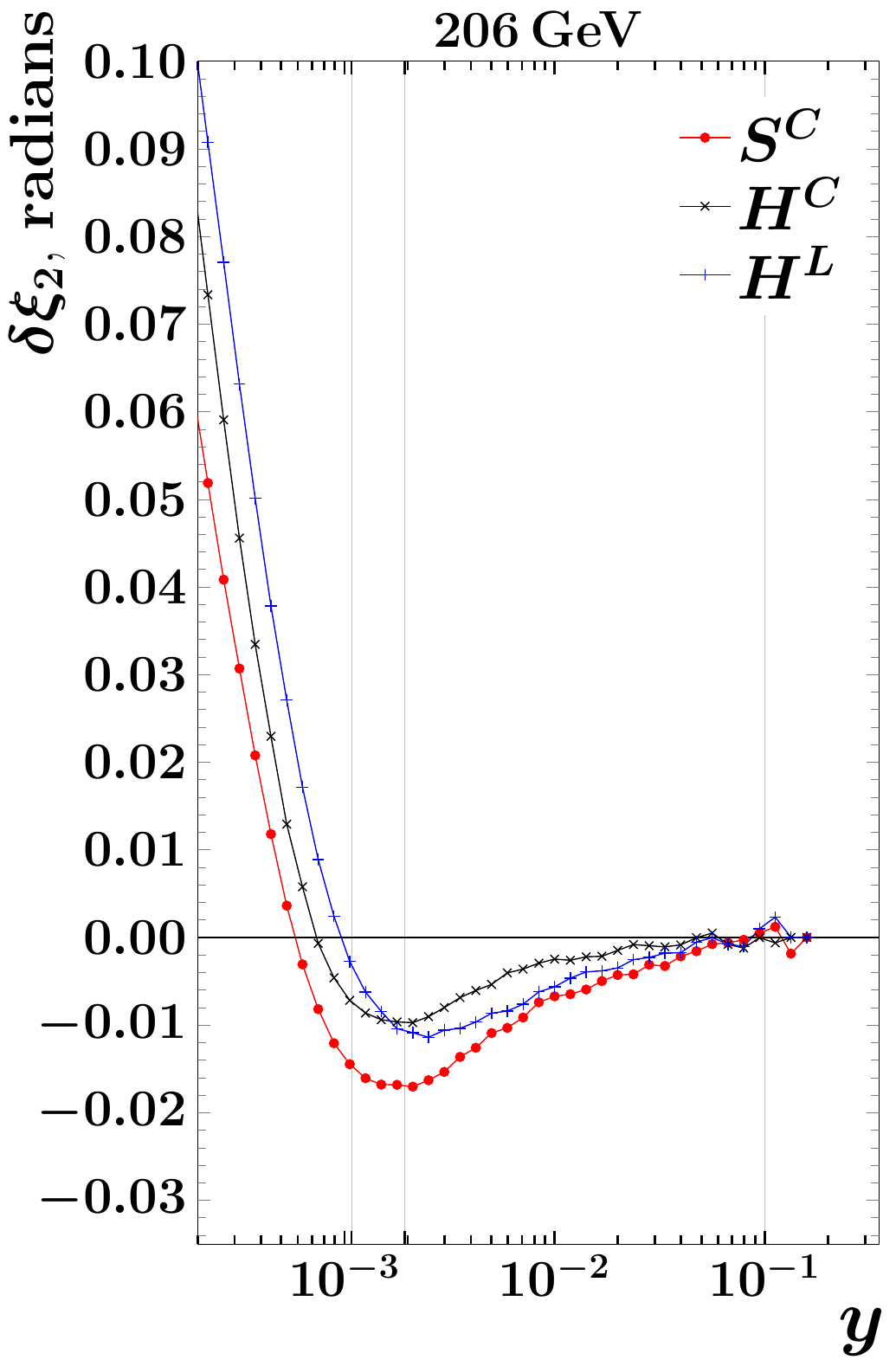}\\
\caption{Hadronization corrections  obtained  with $S^{C}$, $H^{C}$ and $H^{L}$ hadronization models.}
\label{fig:mchadr:all}
\end{figure}

\begin{figure}[htbp]\centering
\includegraphics[width=\FIGWIDTH,height=\FIGHEIGHT]{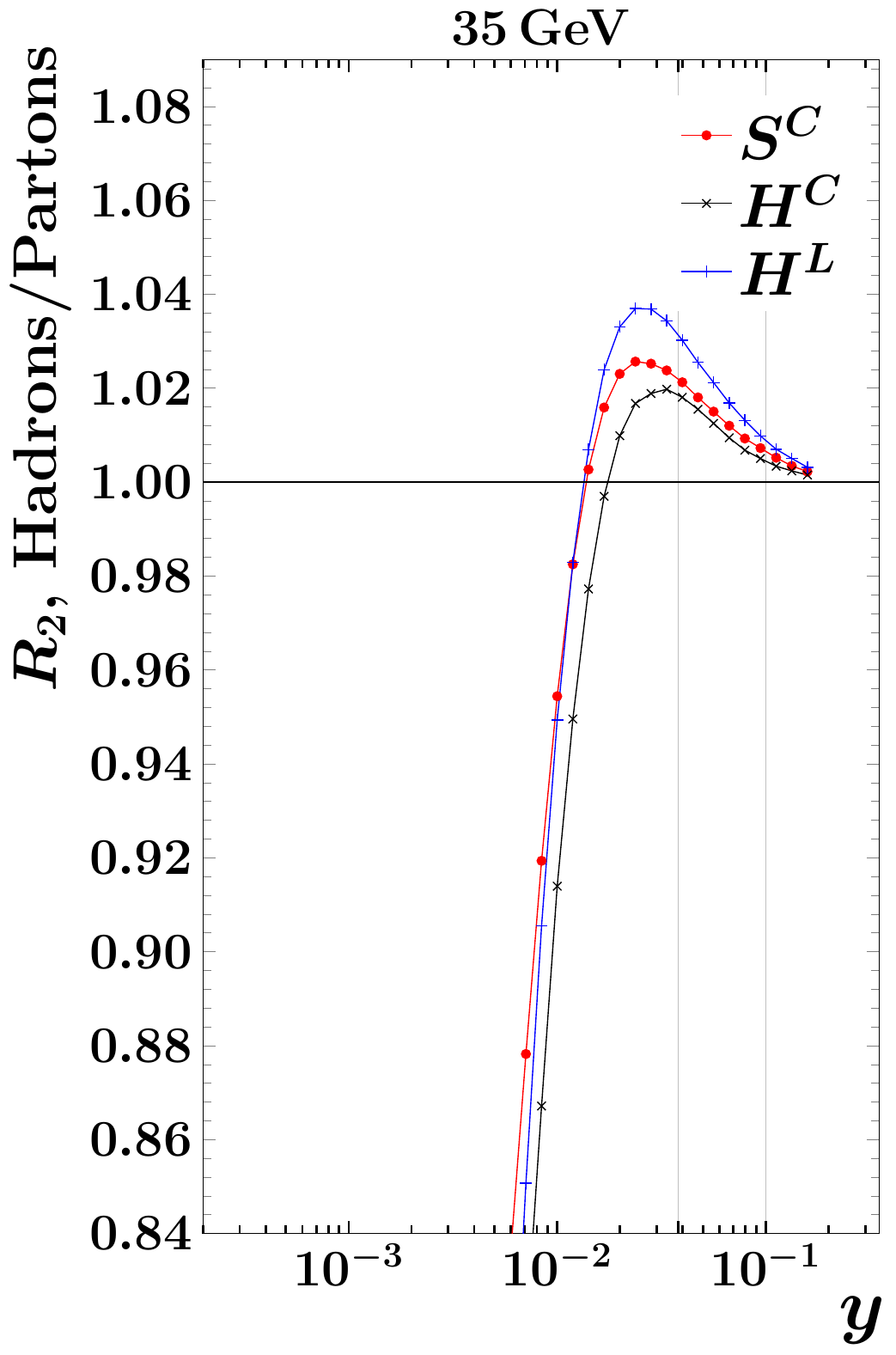}\includegraphics[width=\FIGWIDTH,height=\FIGHEIGHT]{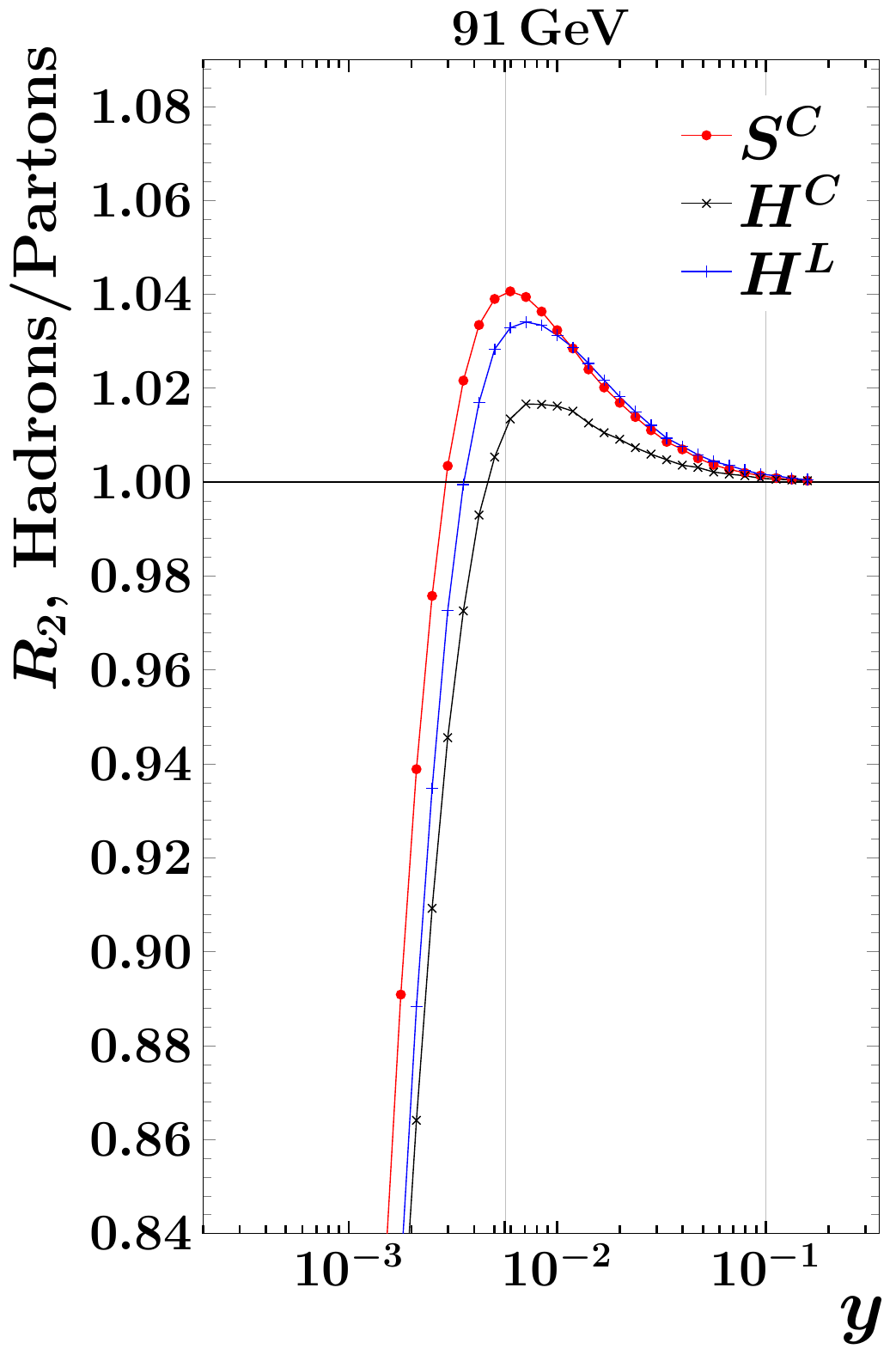}\epjcbreak{}\includegraphics[width=\FIGWIDTH,height=\FIGHEIGHT]{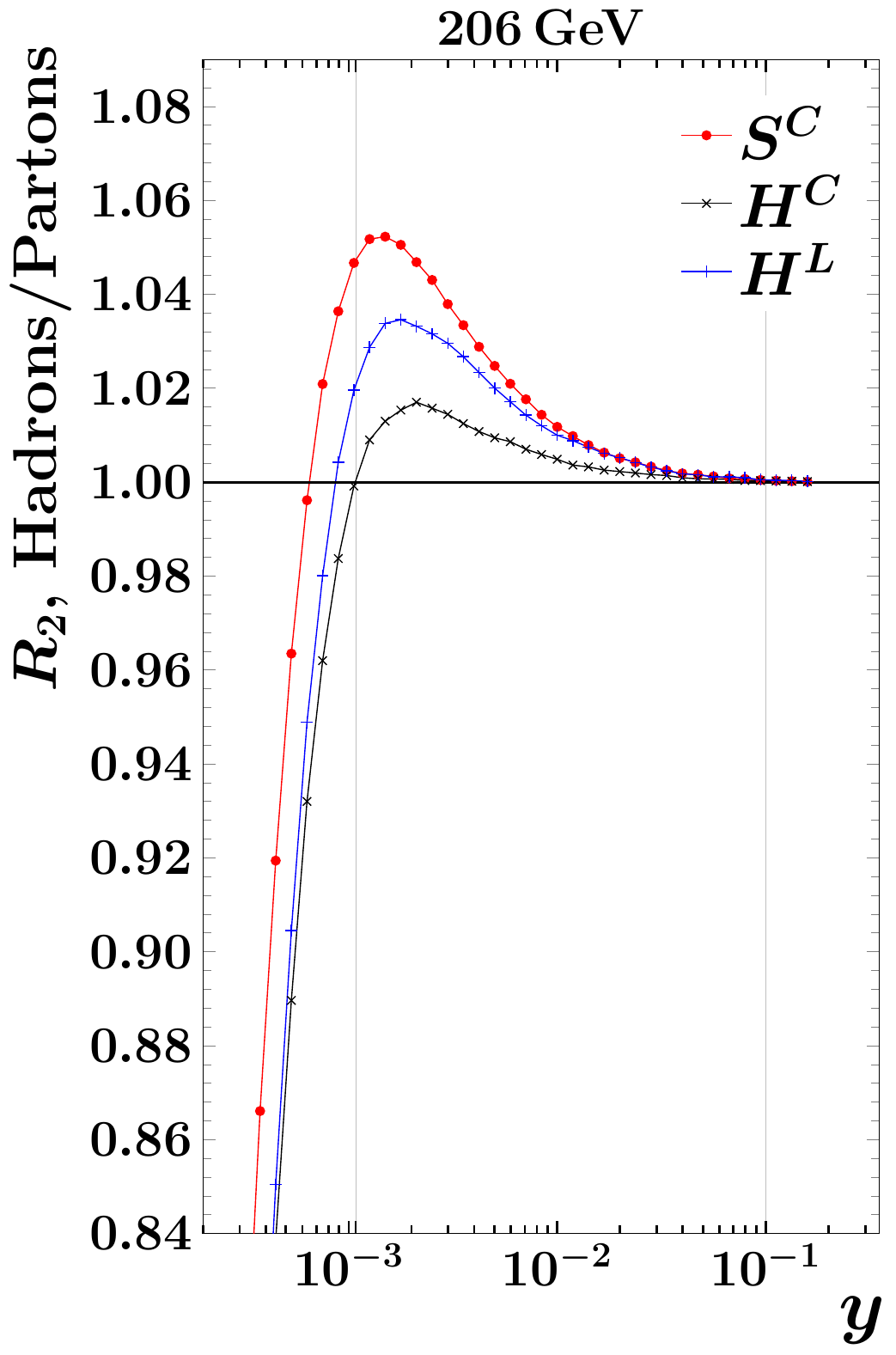}\draftbreak{}\arxivbreak{}\includegraphics[width=\FIGWIDTH,height=\FIGHEIGHT]{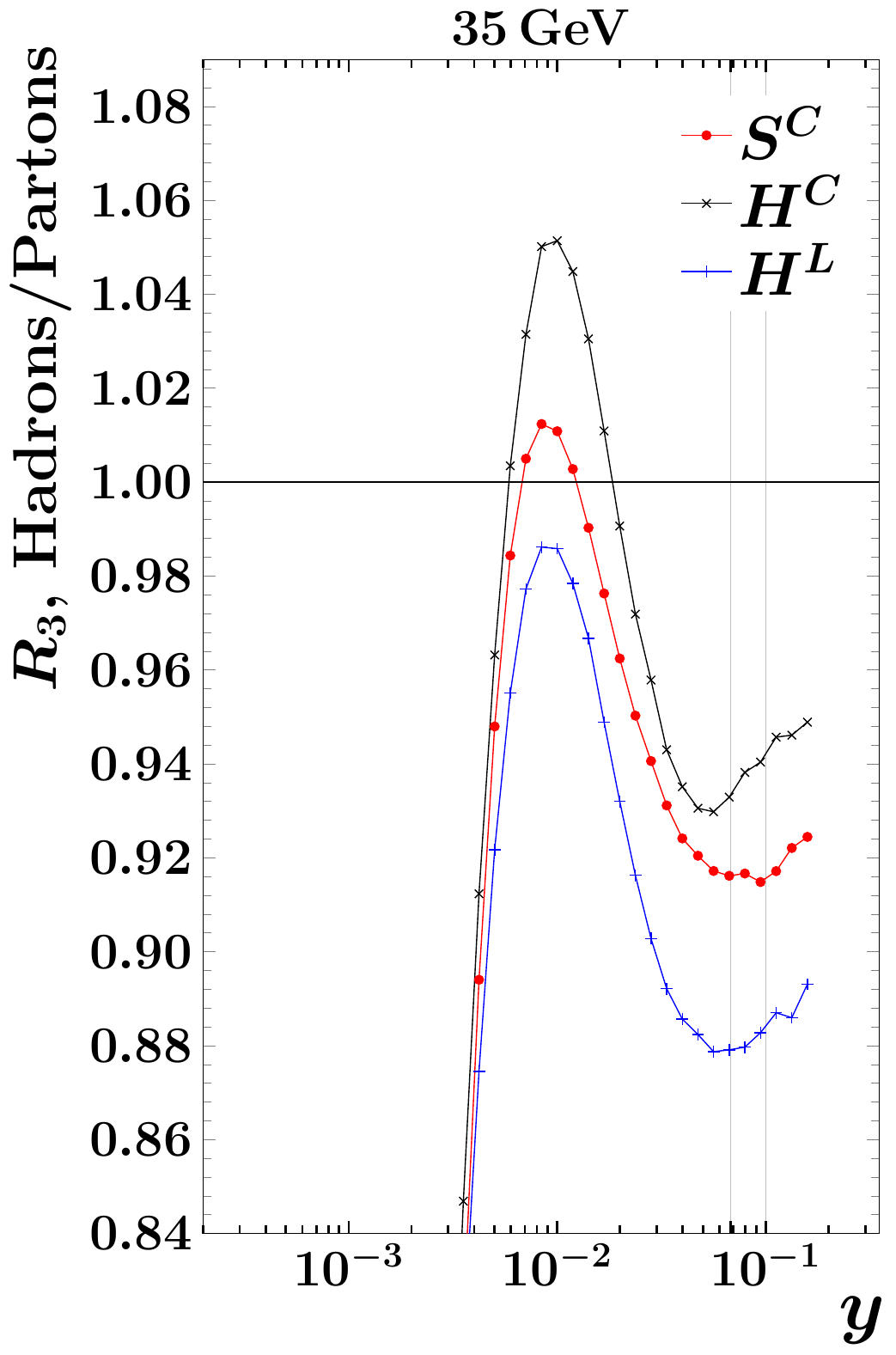}\epjcbreak{}\includegraphics[width=\FIGWIDTH,height=\FIGHEIGHT]{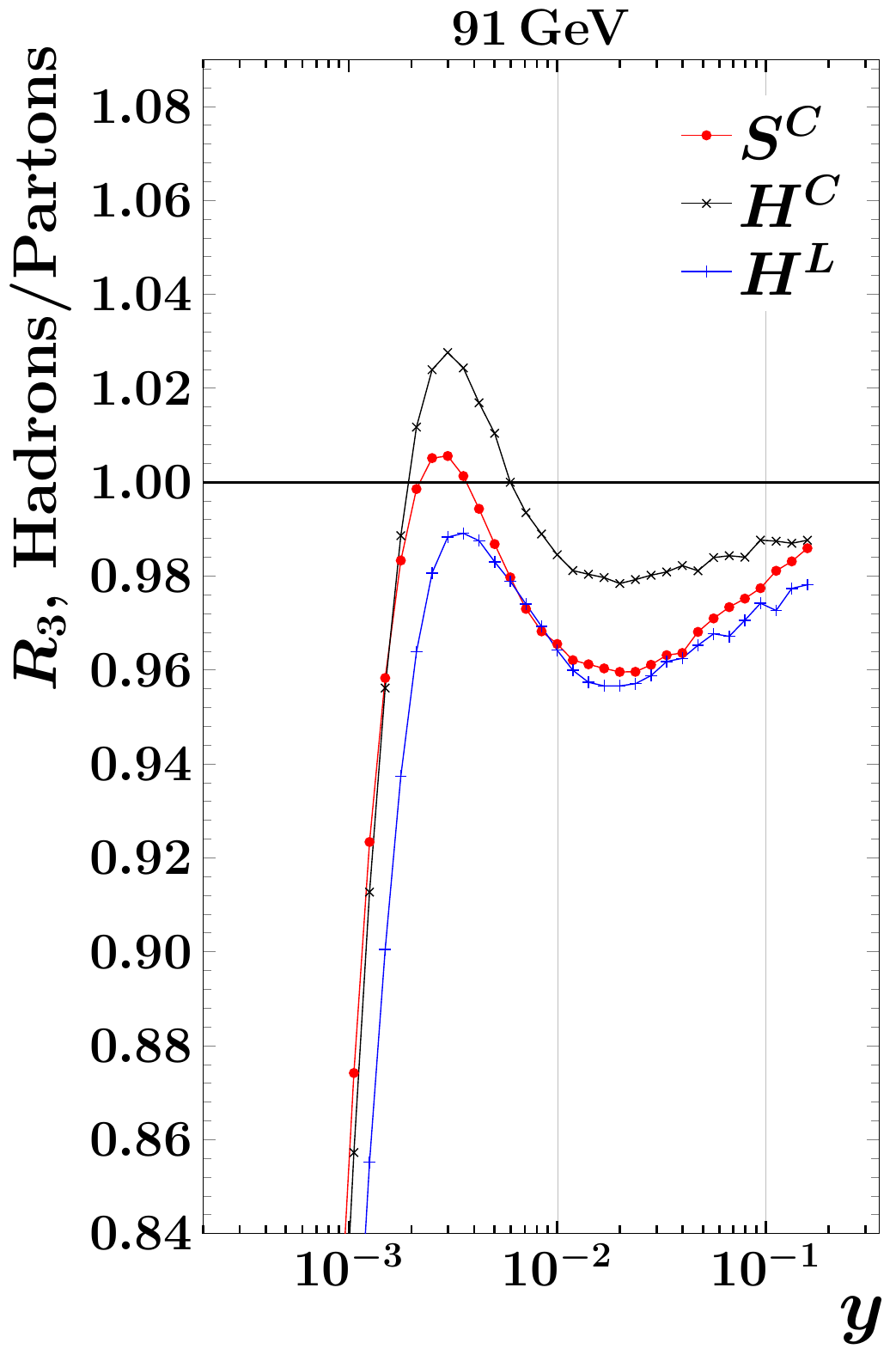}\includegraphics[width=\FIGWIDTH,height=\FIGHEIGHT]{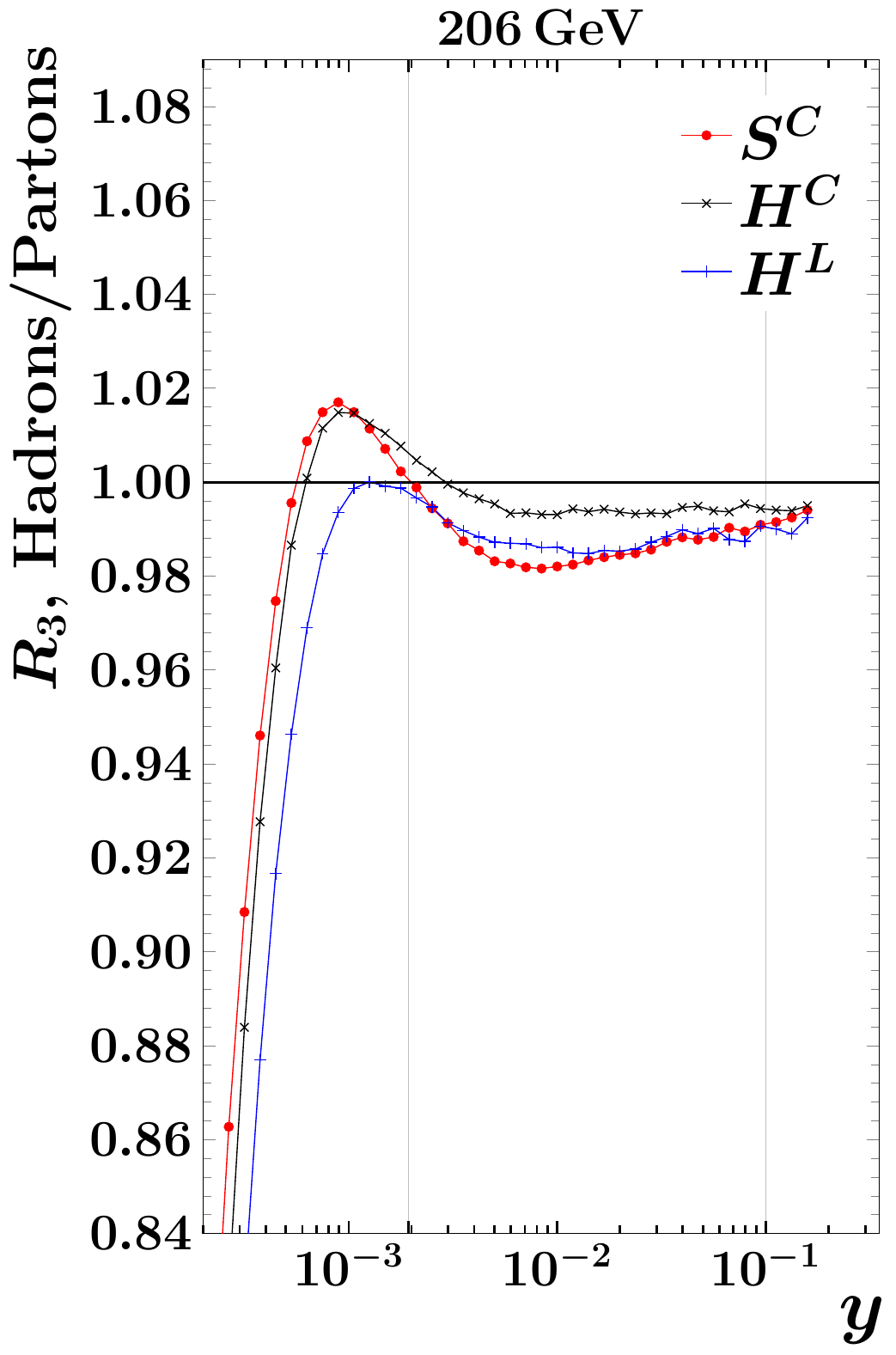}\\
\caption{Hadronization corrections  obtained  with $S^{C}$, $H^{C}$ and $H^{L}$ hadronization models.}
\label{fig:mchadrr:all}
\end{figure}

Fig.~\ref{fig:mchadr:all} shows the $\delta \xi_1$ and $\delta \xi_2$
distributions used to model hadronization corrections of jet rates at
different centre-of-mass energies.  As before, the vertical lines in
the plots show the fit range for reference fits of the $R_2$ and $R_3$
observables. Fig.~\ref{fig:mchadrr:all} shows the size of
hadronization corrections obtained from the $\delta \xi_1$ and
$\delta \xi_2$ values using eqs.~\eqref{eq:xiparton}
and~\eqref{eq:xihadron}.

\newpage

\newpage
\section{Additional fits}
\label{app:special}
In this section we present results of additional fits that we
have performed. 

Table~\ref{tab:result:rtwothree} shows a simultaneous fit of
$\alpha_s(M_Z)$ from experimental data for $R_2$ and $R_3$ obtained
using N$^3$LO and N$^3$LO+NNLL predictions for $R_2$ and NNLO
predictions for $R_3$. Three different hadronization models are used
and four different choices of the fit range are shown. 
The reported uncertainty is only the statistical uncertainty of
the fit, as given by \prog{MINUIT2}.
%
We note that these results show a significant dependence on the fit range
used. 

\renewcommand{\arraystretch}{1.0}
\begin{table}[htbp]\centering
\addtolength{\tabcolsep}{-3pt}
\begin{tabular}{|c||c|c|}\hline
Fit ranges, $\log y$&  N$^3$LO, NNLO                  & N$^3$LO+NNLL, NNLO               \\
Hadronization      &   $\chi^{2}/ndof$                    & $\chi^{2}/ndof$ \\\hline\hline
\arxivonly{
\JRTtabularresultNNLOandNNLORES
}
\draftonly{
\JRTtabularresultNNLOandNNLORES
}
\epjconly{
\JRTtabularresultNNLOandNNLORESepjc
}
\end{tabular}
\caption{
Simultaneous fit of $\alpha_s(M_Z)$ from experimental data for $R_2$ and $R_3$ obtained using
N$^3$LO and N$^3$LO+NNLL predictions for $R_2$ and NNLO predictions for $R_3$. Three different hadronization
models are used and four different choices of the fit range are shown, as given in the
brackets, with ${\cal L} = \log_{10}(M_Z^2/Q^2)$. The fit range for $R_{2}$  is given in the first pair of brackets and the 
the fit range for $R_{3}$ in the second.
The reported uncertainty is the fit 
uncertainty only, as given by \prog{MINUIT2}.
}
\label{tab:result:rtwothree}
\end{table}

We also show in Tab.~\ref{tab:result:rthree} a fit of $\alpha_s(M_Z)$
from experimental data for $R_3$ only obtained using NNLO predictions,
three different hadronization models and four different choices of the
fit range. As before, the reported uncertainty is the fit uncertainty
only, as given by \prog{MINUIT2}. These results can be compared
directly with
results of similar analyses, such as the study of
Ref.~\cite{Dissertori:2009qa}. Altogether we find good agreement
between the fitted values reported in Tab.~\ref{tab:result:rthree} and
the results of Ref.~\cite{Dissertori:2009qa}.

\renewcommand{\arraystretch}{1.0}
\begin{table}[htbp]\centering
\addtolength{\tabcolsep}{-3pt}
\begin{tabular}{|c||c|}\hline
Fit ranges, $\log y$&  NNLO                          \\
Hadronization      &   $\chi^{2}/ndof$     \\\hline\hline
\JRTtabularresultNNLOthree
\end{tabular}
\caption{
Fit of $\alpha_s(M_Z)$ from experimental data for $R_3$ obtained using
NNLO predictions, three different hadronization
models and four different choices of the fit range, as given in the
brackets, where ${\cal L} = \log_{10}(M_Z^2/Q^2)$.
The reported uncertainty is the fit 
uncertainty only, as given by \prog{MINUIT2}.
}
\label{tab:result:rthree}
\end{table}


\newpage
{\bibliographystyle{./JHEP}{\raggedright\bibliography{JRT.bib}}}\vfill\eject
\clearpage
\end{document}